\documentclass[a4paper,fleqn,usenatbib]{mnras}

\usepackage{newtxtext,newtxmath}
\usepackage[T1]{fontenc}
\usepackage{ae,aecompl}

\usepackage{graphicx}	% Including figure files
\usepackage{amsmath}	% Advanced maths commands
\usepackage{amssymb}	% Extra maths symbols
\usepackage{color}
\usepackage{gensymb}

\def \Jbhv{$\vec{J}_{\rm BH}\,$}
\def \Jdv{$\vec{J}_{\rm d}\,$}
\def \fedd{${f}_{\rm Edd}\,$}

\definecolor{myorange}{RGB}{220,100,40}
\definecolor{myblue}{rgb}{0.0, 0.5, 0.69}

\def \LGalaxies{\texttt{L-Galaxies}\,}

\title[Nuclear and wandering supermassive black holes]{From galactic nuclei to the halo outskirts: tracing supermassive black holes across cosmic history and environments}

\author[D. Izquierdo-Villalba et al.]{
David Izquierdo-Villalba,$^{1}$\thanks{E-mail: dizquierdo@cefca.es}
Silvia Bonoli,$^{2,3}$
Massimo Dotti,$^{4,5}$
Alberto Sesana,$^{4}$ \newauthor \space
Yetli Rosas-Guevara,$^{2}$
Daniele Spinoso$^{1}$
\\
$^{1}$ Centro de Estudios de F\'{\i}sica del Cosmos de Arag\'{o}n (CEFCA), Plaza San Juan 1, Planta-2, Teruel, 44001, Spain.\\
$^{2}$ Donostia International Physics Centre (DIPC), Paseo Manuel de Lardizabal 4, 20018 Donostia-San Sebastian, Spain\\
$^{3}$ IKERBASQUE, Basque Foundation for Science, E-48013, Bilbao, Spain\\
$^{4}$ Dipartimento di Fisica ``G. Occhialini'', Universit\`a degli Studi di Milano-Bicocca, Piazza della Scienza 3, 20126 Milano, Italy\\
$^{5}$ INFN, Sezione Milano–Bicocca, Piazza della Scienza 3, I-20126 Milano, Italy
}

\date{Accepted XXX. Received YYY; in original form ZZZ}

\pubyear{2015}

\begin{document}
\label{firstpage}
\pagerange{\pageref{firstpage}--\pageref{lastpage}}
\maketitle

\begin{abstract}

We study the mass assembly and spin evolution of supermassive black holes (BHs) across cosmic time as well as the impact of gravitational recoil on the population of nuclear and wandering black holes (wBHs) by using the semi-analytical model \LGalaxies run on top of \texttt{Millennium} merger trees. We track spin changes that BHs experience during both coalescence events and gas accretion phases. For the latter, we assume that spin changes are coupled with the bulge assembly. This assumption leads to predictions for the median spin values of $z=0$ 
BHs that depend on whether they are hosted by pseudobulges, classical bulges or ellipticals, being $\overline{a}\,{\sim}\,0.9$, $0.7$ and $0.4$, respectively. The outcomes of the model display a good consistency with $z\,{\leq}\,4$ quasar luminosity functions and the $z\,{=}\,0$ BH mass function, spin values and black hole-bulge mass correlation. Regarding the wBHs, we assume that they can originate from both the disruption of satellite galaxies (\textit{orphan} wBH) and ejections due to gravitational recoils (\textit{ejected} wBH). The model points to a number density of wBHs that increases with decreasing redshift, although this population is always  $\rm {\sim}\, 2\, dex$ smaller than the one of nuclear black holes. At all redshifts, wBHs are typically hosted in $\rm M_{halo}\,{\gtrsim}10^{13} \, M_{\odot}$ and $\rm M_{stellar}\,{\gtrsim}10^{10} \, M_{\odot}$, being \textit{orphan} wBHs the dominant type. Besides, independently of redshift and halo mass, \textit{ejected} wBHs inhabit the central regions ($\rm {\lesssim}\,0.3R_{200}$) of the host DM halo, while \textit{orphan} wBHs linger at larger scales ($\rm{\gtrsim}\,0.5R_{200}$). Finally, we find that gravitational recoils cause a progressive depletion of nuclear BHs with decreasing redshift and stellar mass. Moreover, ejection events lead to   changes in the predicted local black hole-bulge relation, in particular for BHs in pseudobulges, for which the relation is flattened at $\rm M_{bulge}\,{>}\,10^{10.2}\,M_{\odot}$ and the scatter increase up to $\rm {\sim}\,3\, dex$.

\end{abstract}


%%%%%%%%%%%%%%%%%%%%%%%%%%%%%%%%%%%%%%%%%%%%%%%%%%

%%%%%%%%%%%%%%%%% BODY OF PAPER %%%%%%%%%%%%%%%%%%

\section{INTRODUCTION}

Since the discovery of the first quasar \citep{Schmidt1963} our understanding of supermassive black holes (BHs), their growth and relation with the host galaxies  has substantially increased, although a complete and exhaustive picture on their formation, evolution and impact on the environment is still missing. A couple of decades ago, observational results confirmed both the existence of BHs at the center of most massive galaxies and the presence of a correlation between their masses and the properties of their hosts \citep{Soltan1982,Haehnelt1993,Faber1999,ODowd2002,HaringRix2004,Kormendy2013,Savorgnan2016}. Their effect on shaping the  star formation of their hosts has also been extensively studied \citep{Silk1998,Birzan2004,Diamond-StanicandRieke2012,Mullaney2012a,Mullaney2012b,Eisenreich2017}. Observational studies are finally finding direct evidence of AGN-driven winds \citep{Tombesi2014,Tombesi2015,Cresci2015,Bischetti2017}.
%\citep{Hardcastle2006,Best2007,Georgakakis2011,Harrison2012,IzquierdoVillalba2018}.\\

From a theoretical point of view, it is well established that black holes grow via gas accretion and coalescence with other black holes \citep{Marulli2008,Bonoli2016,Hirschmann2012,Hirschmann2014,Fanidakis2011,Dubois2012}. While the former seems to be the main contributor to the total mass-assembly budget  \citep{Soltan1982}, the latter appears to have just a sub-dominant importance \citep{SmallBlandford1992,Fanidakis2011,Dubois2014}. Concerning gas accretion,  several physical processes are believed to be responsible for driving gas towards the black hole surroundings. One of these processes is galaxy mergers, whose gravitational torques can modify the galaxy cold gas structure and trigger gas inflows towards the nuclear region \citep{Hernquist1989,BarnesHernquist1991,MihosandHernquist1996,DiMatteo2005,Springel2005,Li2007}. Secular processes such as the formation of bar structures or the redistribution of gas after supernovae explosions may have similar effects \citep{Shlosman1989,Hopkins2009c,HopkinsQuataert2010,Fanali2015,Dubois2015,Du2017}. However, no clear observational conclusions about these two latter feeding mechanisms have been reached up to date. After being transported towards the nuclear parts, the gas has to lose a significant amount of its angular momentum to reach closer orbits around the BH to be ultimately accreted by it. Even though this process is not trivial and probably does not have an unique answer, many works have pointed out that both \textit{local viscous stress} and \textit{gravitational torques} (such as bars-within-bars, spiral waves or large-scale magnetic stress) can be very effective in transporting gas close to the BH (${\lesssim}\,0.01-0.1\,\rm pc$) before being transformed into stars \citep{ShlosmanBegelman1989,Balbus1998,Thompson2005}.\\

As a result of gas accretion onto the BHs, a large amount of energy can be released prompting the birth of a \textit{quasar} or an  \textit{Active Galactic Nucleus} \citep[][]{Soltan1982,Hopkins2007,Ueda2014}. The amount of energy released depends on the mass of the black hole, the geometry of the accretion disk and on the spin of the black hole \citep{ShakuraSunyaev1973,MerloniANDHeinz2008}. The dimensionless \textit{spin} parameter, $a$, is defined as: $a\,{=}\,cJ_{\rm BH}/G \rm M_{BH}^2 \,{\lesssim} 1$ where $c$ is the speed of light, $G$ is the gravitation constant and $J_{\rm BH}$ is the black hole angular momentum. Its specific value plays an important role in the BH assembly and feedback. While highly-spinning BHs can convert up to ${\sim}\,40\%$ of the accreted matter into radiation, slowly spinning ones can only reach up to ${\sim}\,5\%$. This translates into potentially very bright radiative emission and relatively slow mass-growth of the first ones, as opposed to the generally high growth-rates and low luminosity of the latter ones
%While highly spinning BHs slow down their growth converting up to ${\sim}\,40\%$ of the accreted matter into radiation, making them more luminous, low spinning black holes speed their assembly converting only the ${\sim}\,5\%$
\citep{Bardeen1970,NovikovThorne1973,Page1974}. Moreover, the production and strength of AGN jets   might be related to the value of the BH spin \citep{Blandford1977}. How BHs acquire their spin is not fully understood yet. Up to date, all the proposed theories have assumed that a BH reaches its final spin after repeated gas accretion episodes and BHs coalescence that modify the initial rotation \citep{Fanidakis2011,Volonteri2013,BertiVolonteri2008,Barausse2012}. On the gas accretion side, %\cite{King2005} showed that the ratio between the accretion disk and the BH angular momentum determines if the gas consumption can spin-up or spin-down the BH. Based on that rule and physical processes such as torques due to the \textit{Lense Thirring precesion} \citep{BardeenPetterson1975,Bardeen1975} 
several theoretical scenarios have been proposed. For instance, \cite{King2005} suggested the \textit{chaotic} scenario where the gas accretion always proceeds via consumption of self-gravitating accretion disks in uncorrelated directions. In this scenario, BHs tend to spin-down towards a typical final value of $a\,{\lesssim}\,0.2$ \citep{King2008}. On the other side, \cite{Volonteri2007} showed that a \textit{prolonged} gas accretion in the form of a constant and coherent gas consumption, efficiently spins-up the BHs up to $a\,{\sim}\,1$. These two can be seen as extreme cases, and intermediate scenarios have been explored \cite[see e.g.][]{Dotti2013,Fiacconi2018,Bustamante2019}. Over the last years, some works have attempted to measure BHs spin by using X-ray spectroscopy \citep{Brenneman2006,Reynolds2013}. However, the challenges and biases in the observations make it difficult to draw solid conclusions. In the near future, with the upcoming experiments such as \textit{Laser Interferometer Space Antenna} \citep[LISA,][]{Lisa2017} or \textit{Advanced Telescope for High-ENergy Astrophysics} \citep[ATHENA,][]{Nandra2013} it will be possible to reach robust observational conclusions ruling out many theoretical models of BH spin evolution.\\

Even though BH-BH coalescences seem to play a sub-dominant role in the evolution of both BH mass and spin, they can have an important effect on the co-evolution between black hole and galaxy \citep{Redmount&Ress1989}. Because of conservation of linear momentum, the propagation of gravitational waves imparts a recoil (or \textit{kick}) in the remnant BH \citep{Bekenstein1973,Baker2008,Lousto2012}. Small kicks could simply lead to  a small displacement of the remnant BH from the galaxy center whose observational counterpart would be an offset AGN \citep{Madau&Quataert2004,Loeb2007,Blecha&Loeb2008,Guedes2011}. At the other extreme, if the recoil velocity is larger than the escape velocity of its host galactic center, the black hole can be \textit{kicked out} of the galaxy becoming  a \textit{wandering} black hole. This last case could be more probable at high-$z$, where the galaxy potential wells are smaller and relatively small recoil kicks can effectively expel the BH from the galaxy and possibly even from the halo potentials. %All the works performed to compute analytically the strength of the recoil agree in the fact that its magnitude correlates with the properties of the merging BH, such as the BH mass ratio, the magnitude of the spins and their orientations \cite[see eg.][]{Baker2008,Gonzalez2007,Lousto2008,vanMeter2010}. 
Such BH ejections might have some important consequences to be taken into account when drawing a complete black hole-galaxy co-evolution picture.  For instance \cite{Blecha2011}, by using hydrodynamical simulations, found that the bulge velocity dispersion and the BH mass relation changes (a shift in the velocity dispersion axis) when the simulation is run with and without the inclusion of recoil velocities. This deviation is due to the fact that the ejection interrupts the BH growth while the galaxy continue its evolution (see similar results of \citealt{Blecha2008,Gerosa&Sesana2015}). In the same work, \cite{Blecha2011} discussed the halting of AGN feedback as a consequence of the BH ejection. The outcome is an extended period of active star formation in the galaxy which yields a dense and massive nuclear stellar cusps. This type of nuclear regions could be a target property which would point out galaxies which underwent a BH ejection. Indeed, from an observational point of view, some galaxies have been tagged as candidates of having undergone a BH recoil (see eg. BCG in A2261 cluster \citep{Postman2012}, SDSS J1056+5516 \citep{Kalfountzou2017}, CID-42 \citep{Civano2012}, NGC 1399, 4261, 4486, IC 429 \citep{Lena2014}, 3C 186 radio quasar \citep{Chiaberge2017} and \cite{Komossa2012} for a review). On top of ejection via gravitational recoil, the complex dynamics of the two BHs before the final coalescence might cause the formation of a wandering black hole. If the two black holes spend enough time orbiting around each other \citep{Begelman1980,Quinlan1997} a third BH could reach their sphere of influence. As a consequence, the three BHs are likely to undergo a complex 3-body scattering interaction with the final outcome of the ejection of the less massive BH and the shrinking of the separation between the two remnants BHs \citep{Volonteri2003,Volonteri2005,Bonetti2018a}. Even though this type of interactions might no have a main role in the population of wandering black holes \cite[see e.g.][]{Volonteri2003,Volonteri2005} they seem to play importance in speeding up the final BH-BH coalescence \citep{Bonetti2018}.\\

Wandering black holes can also be a consequence of the disruption of a satellite galaxy during the infall onto a larger system.
%there is another mechanism by which a BH can be hosted in a dark matter halo without being surrounded by its hots. Right after the merger of two dark matter halos, their host galaxies start to sink towards the inner parts of the new halo where they eventually will merge \citep{BinneyTremine1987}. During this process, galaxies undergone 
Environmental processes such as ram pressure or tidal interactions, in fact, can  gradually strip away satellites before they merge with the central galaxy \citep{Moore1999,Springel2001}. 
%If one of these galaxies is not compact enough its final dissolution in the dark matter halo will take place, losing the possibility of any kind of merger. Such dissolution provokes the incorporation of the host galaxy supermassive BH in the dark matter halo as a \textit{wandering black hole}. 
The frequency of satellite galaxy disruptions is still not well constrained. Some recent works have suggested that dwarf galaxies hosted in clusters are the ones which experience these events the most \citep{Faltenbacher2005,Martel2012}. Naively, the BHs hosted by the disrupted satellites are expected to be incorporated in the halo of the central galaxy and, thus, contribute to the population of wandering BHs. Even though such population of BHs is interesting because it would give an idea of the galaxy disruption rate, it has been mainly studied only through hydrodynamical simulations \citep{Bellovary2010,Miki2014,Tremmel2018,Pfister2019}. However, the current limits in the resolution achievable in simulations,  make it difficult to  well-resolve the fate of  the cusp of stars and/or gas around the BH after the disruption and to properly assess the evolution of such ``orphan'' BHs.\\

Given the  observational limitations in detecting wandering black holes, more theoretical work is needed to constrain where and how these black holes can be found  with current and upcoming instruments. Several works have attempted to determine the orbits and time spent by a BH in a wandering orbit. For instance, by using 3D hydrodynamical simulation \cite{Blecha2011} found that at a fixed ratio between recoil and escape velocity, the recoil trajectories have a substantial variation with the galaxy gas fraction. On the other hand, under an analytic perspective, \cite{Choksi2017} explored different parameters and contributions in the orbital damping of wandering BHs finding a longer wandering phase in lighter halo and galaxy systems. These authors found that in many circumstances the wandering phase may be larger than the Hubble time.\\% at the starting point of the wandering phase.

In this work we use the \LGalaxies semi-analytical model in the version presented in \cite{IzquierdoVillalba2019} to study the statistics and environment of black hole growth and spin evolution and to explore the population of wandering black holes across cosmic time. We include both galaxy mergers and galaxy disk instabilities as physical processes that can lead to BH growth. For tracking spin evolution we follow the approach introduced by \cite{Sesana2014}, where spin changes are coupled to the bulge assembly. We also introduce a time-delay between galaxy and BH mergers that depends on the BHs mass ratio and the galaxy gas fraction. Recoil velocities after BH coalescences are calculated on the basis of the progenitors properties while the trajectories of kicked BHs depend on the physical properties of the host halos and galaxies. We also follow the trajectory of \textit{orphan} black holes from stripped satellites to complete the picture of wandering black holes.\\  

The paper is organised as follows: in section~\ref{sec:SAM_MILL} we describe the main characteristics of the semi-analytical model and the merger trees used in this work. In section~\ref{sec:BHmodelling} we describe all the physics included in the semi-analytical model to properly follow the BH mass growth, spin evolution and  wandering phases. In section~\ref{sec:results_nuclear_BHs} we presents our main findings regarding nuclear black holes (i.e., black holes in the center of galaxies). In section~\ref{sec:res_wandering_BHs} we explore the properties of wandering black holes across cosmic time. In section~\ref{sec:grav_rec_impr} we present the effects of gravitational recoils in both BH occupation fraction and bulge-black hole mass relation. Finally, in section~\ref{sec:conclusion} we summarize our main findings.
A lambda cold dark matter $(\Lambda$CDM) cosmology with parameters $\Omega_{\rm m} \,{=}\,0.315$, $\Omega_{\rm \Lambda}\,{=}\,0.685$, $\Omega_{\rm b}\,{=}\,0.045$, $\sigma_{8}\,{=}\,0.9$ and $\rm H_0\,{=}\,67.3\, \rm km\,s^{-1}\,Mpc^{-1}$ is adopted throughout the paper \citep{PlanckCollaboration2014}. 

\section{Galaxy formation model} \label{sec:SAM_MILL}

In this section we briefly describe the \LGalaxies semi-analytical model (SAM) and the dark matter simulations used in this work. The bulk of the model is the one presented in  \citet{Henriques2015}, with the  modifications detailed in \cite{IzquierdoVillalba2019} which contribute to a better description of  galaxy morphologies and radii for both the \texttt{Millennium} and \texttt{MillenniumII} merger trees.

\subsection{Merger trees}
%\subsubsection{Dark matter simulations}

The version of \LGalaxies used is this work is built on top of the subhalo merger trees of the \texttt{Millennium} \citep[hereafter MS,][]{Springel2005} and \texttt{Millennium II} \citep[hereafter MSII,][]{Boylan-Kolchin2009} dark matter (DM) $N$-body simulations.  While MS follows the cosmological evolution of $2160^3$ DM particles with a mass of $8.6 \times 10^8\, \mathrm {M_{\odot}}/h$ within a periodic cube of 500 ${\rm Mpc}/h$ on a side,  MSII tracks the same number of particles  in  a box  125 times smaller in volume ($\rm 100\,Mpc/\textit{h}$  on a side) but with 125 times better mass resolution ($ \mathrm{ 6.885\,{\times}\,10^6\,M_{\odot}}/h$). Both simulations were run with \texttt{GADGET} code \citep{Springel2001,Springel2005} by using WMAP1 \& 2dFGRS concordance cosmology: $\Omega_{\rm m}\,{=}\,0.25$, $\Omega_b\,{=}\,0.045$, $\Omega_{\rm \Lambda}\,{=}\,0.75$, $h\,{=}\,\rm 0.73 \,km\,s^{-1}\,Mpc^{-1} $, $n\,{=}\,1$, $\sigma_{8}\,{=}\,0.9$ \citep{Colless2001}. However, the latest \LGalaxies version was tuned on a re-scaled versions of MS and MSII simulations \citep{AnguloandWhite2010} to match the cosmological parameters obtained by Planck first-year data \citep{PlanckCollaboration2014}: $\Omega_{\rm m} \,{=}\,0.315$, $\Omega_{\rm \Lambda}\,{=}\,0.685$, $\Omega_{\rm b}\,{=}\,0.045$, $\sigma_{8}\,{=}\,0.9$ and $h\,{=}\,0.673\, \rm km\,s^{-1}\,Mpc^{-1}$.\\

%After re-scaling, the particle mass corresponds to $\mathrm{m_p}\,{=}\,\rm  1.43 \,{\times}\,10^9\, M_{\odot}/\textit{h}$ and $\mathrm{m_p}\,{=}\,7.68\,{\times}\,10^{6}\, \mathrm{M_{\odot}}/\textit{h}$ for MS and MSII, respectively.\\

Particle data of MS and MSII simulations were stored respectively at 63 and 68 epochs or \textit{snapshots} spaced approximately logarithmically  in  time  at $z\,{>}0.7$
and  linearly  at  $z\,{<}0.7$ \citep[see][]{Boylan-Kolchin2009}. Dark matter halos were selected within these snapshots by using a friend-of-friend (FOF) group-finder.  Subhalos structures, i.e locally overdense, self-bound particle groups formed inside the DM halos, were identified
with \texttt{SUBFIND} algorithm \citep{Springel2001}. By applying \texttt{L-HALOTREE} \citep{Springel2005} all halos and subhalos were arranged in merger trees structures allowing to follow the evolutionary path of any DM (sub)halo in the simulations. Although these (subhalo) merger trees structures are the skeleton of the SAM, the time resolution given by their 63 and 68 snapshots are not enough to properly trace the baryonic physics. Thus, to accurately follow the galaxy evolution, the SAM does an internal time discretization between two consecutive snapshots with approximately $\rm{\sim}\,5{-}20 \,Myr$ of time resolution. These extra-temporal subdivisions used by the SAM are called \textit{sub-steps}. Even though this work is based on the merger trees of the \texttt{Millennium} simulation, we combine them with the ones of \texttt{Millennium II}  (as detailed in Section~\ref{sec:init_galaxy}) to improve the galaxy and black hole initialization.

\subsection{Baryonic physics}\label{sec:barPhys}

We briefly describe here the baryonic processes included in \LGalaxies that lead to the build up of the disk and bulge components.

\subsubsection{Galaxy disk}

In the current understanding of how cosmological structures form, it is accepted that the initial density fluctuations of the early universe act as the birthplaces of dark matter halos \citep{PressANDSchechter1974}. As soon gravitational instability triggers the collapse of these density fluctuations, part of the diffuse baryonic gas present in the universe is attracted and collapsed within them. Following this scenario, \LGalaxies starts associating a baryonic matter to each newly resolved DM subhalo in a form of diffuse, pristine, spherical and quasi-static \textit{hot} gas atmosphere. Gradually, a fraction of this atmosphere is allowed to condensate and migrate towards the center of the subhalo, leading to the formation of a cold, rotationally supported gas component \citep{WhiteandRees1978}. If the mass of the cold gas structure is large enough, star formation (SF) episodes are prompted, leading to the build-up of the stellar disk component. Shortly after the SF event, a fraction of the new stars explode as supernovae, enriching the environment with newly formed heavy elements, reheating a fraction of cold gas via energy injection and, eventually, expelling part of the hot gas atmosphere beyond the subhalo virial radius, $R_{\rm 200c}$ \citep{Guo2011}. At later times, this ejected hot gas is allowed to be reincorporated into the gas atmosphere, being available once again for cooling onto the galactic disk. The reincorporation of this ejecta at later times helps to regulate the low-$z$ star formation, especially in low-mass galaxies \citep{Henriques2015}. Besides, to prevent the stellar component of massive galaxies from overgrowth, the model introduces feedback from central supermassive black holes (BHs) as an additional mechanism to regulate star formation at low redshifts (see Section~\ref{sec:BHmodelling}).

\subsubsection{Galaxy bulge} \label{sec:bulgesSAM}

Galaxies are allowed to assembly a bulge component via \textit{galaxy mergers} and \textit{disk instabilities}. While the former is related with the hierarchical growth of the DM subhalos, the latter takes more importance in isolated galaxies with a quiet merger history.\\

%Galactic encounters happen after the merger of two halos in a time-scale given

Galaxy encounters follow the merger of the two parent subhalos on a time-scale given by the dynamical friction presented in \citet{BinneyTremine1987}. According to the baryonic (gas plus stars) mass ratio of the two involved galaxies, $\rm m_R$, the model differentiates between \textit{major} ($\rm m_R\,{>}\,m_R^{th}$) and \textit{minor} ($\rm m_R\,{<}\,m_R^{th}$) interactions ($\rm m_R^{th}$ is set to 0.2 as in \citealt{IzquierdoVillalba2019}). Major  interactions are assumed to completely destroy the disks of the two interacting galaxies, generating a pure spheroidal remnant which undergoes a \textit{collisional starburst}. On the other hand, during minor interactions the disk of the larger galaxy survives and experiences a burst of star formation, while its bulge integrates the entire stellar mass of the satellite \citep[see][]{Guo2011}. Besides, the model used in this work includes the prescription of \textit{smooth accretion} in order to deal with the physics of extreme minor mergers \citep{IzquierdoVillalba2019}. During those events it is expected that the stellar remnant of the satellite (bulge and disk) gets diluted  inside the disk of the central galaxy before being able to reach the nucleus, thus, losing the possibility of make the bulge of the primary to grow. \\

On the other hand, in \LGalaxies a disk instability (DI) event refers to the process by which the stellar disk becomes massive enough to be inclined to non-axisymmetric instabilities which eventually lead to the formation of a central ellipsoidal component trough the buckling of nuclear stellar orbits \citep[see references in][]{MoWhite2010}. During this process, the result is the formation of a \textit{bar} and/or \textit{pseudobulge} structure \citep{Combes&Sanders1981,Pfenniger1990,Athanassoula2005,Sellwood2016,spinoso2017}. The \LGalaxies model accounts for galactic DI with an analytic stability criterion based on the \cite{Efstathio1982} and \cite{MoMaoWhite1997} simulations:
\begin{equation}\label{eq:instability_criterium}
\rm  \frac{V_{max}}{\left(G M_{\star,d}/R_{\star,d}\right)^{1/2}}\,{\leqslant}\, \epsilon^{\rm DI}, 
\end{equation} 
where $\rm V_{max}$ is the maximum circular velocity of the host dark matter, $\rm R_{\star,d}$ and  $\rm M_{\star,d}$ are respectively the length and stellar mass of the stellar disk and $\epsilon^{\rm DI}$ a parameter which determines the importance of the disk self-gravity (set to $1.5$ as in \citealt{IzquierdoVillalba2019}). If the stability criterion of Eq.~\eqref{eq:instability_criterium} is met, an amount
\begin{equation}\label{eq:Delta_M_stars_DI}
 \rm \Delta M_{stars}^{DI} \,{=}\, M_{\rm \star, d} - \left( V_{max}^2 R_{\star,d}/G \epsilon_{\rm DI}^{2}\right) \, ,
\end{equation}
of the disk stellar mass is transferred to the bulge in order to restore the disk (marginal) stability. Following the history and the physical conditions of the galaxy in which a DI takes place, the model distinguishes between instabilities that are \textit{merger-induced} and the ones that are a consequence of the slow, \textit{secular} evolution of galaxies. On one hand, \textit{merger-induced} DIs are produced as a consequence of the fast increase of stellar disk mass after the collisional starburst or \textit{smooth} satellite galaxy accretion. On the other hand, \textit{secular} DIs result from the slow, but continuous, mass growth of the disk, taking a major importance in galaxies evolving in isolation. Under the assumption that bars and pseudobulges are a consequence of the secular evolution of galaxies \citep{Debattista2006,MendezAbreu2010,Kormendy2013,Moetazedian2017,Zana2018A,Zana2018B,RosasGuevara2019} the model of \cite{IzquierdoVillalba2019} links \textit{secular} DIs with the formation of galactic bars and pseudobulges, whereas \textit{merger-induced} DIs contribute to the formation of classical bulges. For more details we refer the reader to \cite{IzquierdoVillalba2019}.\\

Finally, in order to distinguish between bulge morphologies in the SAM, we follow the definition of \cite{IzquierdoVillalba2019}. In brief, we assume that a galaxy hosts a pseudobulge when the fraction of bulge formed via  \textit{secular induced} DI is at least 2/3. Galaxies display a classical bulge or elliptical structure when the fraction of bulge formed via \textit{secular induced} DIs is smaller than 2/3 of the total bulge mass and its \textit{bulge-to-total} stellar fraction ($\rm B/T$) is respectively $\rm 0.01\,{<}\,B/T\,{<}\,0.7$ and $\rm B/T\,{>}\,0.7$. Finally, galaxies with $\rm B/T\,{<}\,0.01$ are considered bulgeless galaxies.

\subsubsection{Improving the galaxy initialization in \texttt{Millennium} trees} \label{sec:init_galaxy}

As we discussed, the SAM populates each new resolved subhalo with an amount of hot gas in form of a diffuse atmosphere. From that moment, different analytical recipes are included to deal with the baryonic physical processes that lead to a mature galaxy. The subhalo mass resolution of the \texttt{Millennium} simulation used in this work ($\rm M_{halo}^{Res}\,{\sim}\,10^{10}M_{\odot}/\mathit{h}$) imposes a clear limit on the galaxy and BH evolutionary pathway that can be followed. Nevertheless, new resolved subhalos could already host a mature galaxy and its evolved BH. To improve this aspect we follow the approach of \cite{Bonoli2014} and \cite{Angulo2014} initializing each MS subhalo with random galaxies (hosted in the same redshift and subhalo mass) extracted from a run of \LGalaxies on top of the high resolution \texttt{Millennium II} merger trees ($\rm M_{halo}^{Res}\,{\sim}\,10^{8} M_{\odot}/\mathit{h}$). The galaxy \textit{grafting} includes all the properties of the galaxy and its respective central black hole (including wandering BHs). We stress that this does not imply for our model to be able to follow subhalos below the resolution limit of MS, as we do not have access to their merger history and cosmological evolution. The aim of this procedure is to better match the DM mass of the newly resolved subhalo to the properties of the newly initialised host galaxy. %{\bf Silvia - Maybe we should say this earlier, when we first talk about initialization, and then we refer to it when we talk about the seeds}
%\subsubsection{Environmental processes}
%Regarding the \textit{large scale} effects, the model includes \textit{environmental processes} such as ram pressure or tidal interactions which can completely remove the hot gas atmosphere around satellite galaxies and eventually destroy their stellar and cold gas components \citep[we refer to][for more details]{Guo2011,Henriques2015}. {\bf Silvia: this is very short, and not related to the bulge. Either we put it somewhere else, or we cut it}

\section{A comprehensive model of black hole growth} \label{sec:BHmodelling}
Astrophysical black holes can be fully determined by two quantities: \textit{mass} and \textit{spin}. In this section we present the implemented physic to track their evolution along the cosmic history.

\subsection{Black hole seeding and spin initialization} \label{sec:seeding_and_spin_init}

Even though the existence of BHs is now well established, their origin is still an open issue. Up to date, several theories have been proposed and explored \citep[see][for a review]{Volonteri2010,MayerBonoli2019}. The most studied one is the \textit{light seed scenario} where BHs of mass between $\rm 10\,{-}\,100 \, \rm M_{\odot}$ were formed at $z\,{\gtrsim}\,20$ as remnants of Population III stars \citep[Pop III,][]{Madau2001,Heger2002}. Another complementary formation channel is the \textit{heavy seed scenario} where much more massive black holes ($\rm {\sim}10^4\,{-}\,10^5 M_{\odot}$) are formed after the direct collapse of massive gas clouds, either in pristine primordial halos or in gas-rich merging galaxies \citep{Koushiappas2004,VolonteriStark2011,Mayer2010,Bonoli2014}. In this work we simply assume that each newly resolved subhalo (independently of redshift and halo properties) is seeded with an initial BH mass of $\rm 10^{4} M_{\odot}$. This initial seed mass is a reasonable assumption given the minimum mass of new resolved subhalos in MS and MSII (${\sim}\,10^{10}\rm M_{\odot}$ and ${\sim}\,10^8 \rm M_{\odot}$, respectively). In future works, a proper BH seeding in \LGalaxies will be tackled (Spinoso et al. in prep.).\\

Besides the initial mass, the BH angular momentum, $|\vec{J}_{\rm BH}|$, has to be set. The exact value of $|\vec{J}_{\rm BH}|$ can be expressed as: 
\begin{equation} 
|\vec{J}_{\rm BH}| \,{=}\,  \rm \frac{1}{\sqrt{2}} M_{\rm BH} \mathit{a} \left(\mathit{G} M_{\rm BH} \mathit{R}_{\rm Sch}\right)^{1/2} = \frac{\mathit{a}\mathit{G} M_{\rm BH}^2}{\mathit{c}}\, ,
\end{equation}
where $\rm \mathit{R}_{Sch}\,{=}\, 2 \, \mathit{G} M_{BH}/\mathit{c}^2$ is the Schwarzchild radius, $G$ the gravitational constant, $c$ the light speed, and $a$ the dimensionless \textit{spin} parameter, whose value can be in the range  $0\,{<}\,|a|\,{<}\,0.998$ \citep{Thorne1974}. While $\rm M_{BH}$ and $\rm \mathit{R}_{Sch}$ are fully determined by the seed mass, the value of $a$ does not have to depend on the initial mass. Indeed, no observational constraints and theoretical predictions about the initial BH spin exist to date \citep[see e.g.][]{Batta2017,Fuller2019,Roulet2019,Zackay2019}. Here we assume that the initial value of $a$ is a random number selected between $0\,{<}\,|a|\,{<}\,0.998$. We have tested that the initial selection of $a$ does not affect the results presented in this work. In fact, a BH loses any memory of its initial spin when it has accreated in a coherent way a gas mass comparable to its own \citep[see for instance][Figure 1]{Fanidakis2011}.

\subsection{Black hole gas accretion}\label{sec:BHaccretion_model}

In this section we summarize the different mechanisms that feed central black holes with gas: (i) galaxy mergers, (ii) galaxy disk instabilities and, (iii) gas accretion from the hot gas atmosphere.

\subsubsection{Cold phase accretion}\label{sec:quasar_accretion}

The cold gas accretion is triggered by two different processes: galaxy mergers and disk instabilities.

\begin{enumerate}
\item[i)] \textit{Gas accretion after mergers}. Simulations have shown that gravitational torques during galaxy mergers are able to drive cold gas towards the galaxy inner regions, triggering the black hole accretion \citep{DiMatteo2005,Springel2005,Hopkins2009c}. Following \cite{KauffmannandHaehnelt2000}, \cite{Croton2006} and \cite{Bonoli2009} we assume that the fraction of cold gas accreted by the BH after a galaxy merger is:
\begin{equation}\label{eq:QuasarMode_Merger}
\rm   \Delta {M}_{BH}^{gas} \,{=}\,\mathit{f}_{BH}^{merger} (1+\mathit{z}_{merger})^{3/2} \frac{m_{R}}{1 + (V_{BH}/V_{200c})^2}\, M_{\rm gas}^{\rm Cold},
\end{equation}
where $\rm \Delta {M}_{BH}^{gas}$ includes a dependence on the galaxy baryonic merger ratio, $\rm m_{R}\,{<}\,1$, the virial velocity of the host DM subhalo, $\rm V_{200c}$, the redshift of the merger, $z_{\rm merger}$, and two adjustable parameters $\rm V_{BH}$ and $\rm \mathit{f}_{BH}^{\rm merger}$ set to $\rm 280 \, km/s$ and $0.034$, respectively.\\ %The redshift dependence is based on \cite{MoMaoWhite1997} work which pointed out that at high-$z$ galaxy disks are smaller and with higher surface density than the ones at $z\,{=}\,0$. Therefore, any galaxy process (mergers or galaxy instabilities) would feed the BH more efficiently at high-$z$ than at $z\,{=}\,0$ \citep[notice that this approach was already used by][]{Croton2006,Bonoli2009}.\\

\item[ii)] \textit{Gas accretion after disk instabilities}. Non-axisymmetric instabilities of galactic disks have a deep impact on the morphology of the nuclear parts. They are able to modify the gas disk structure via gravitational torques \citep[see][]{ShlosmanBegelman1989,HopkinsQuataert2010}. These processes produce strong gas inflows that lead to nuclear bursts of star formation  and can drive part of the  gas also in the BH surroundings \citep{Shlosman1989,Hopkins2009c,HopkinsQuataert2010,Fanali2015,Du2017,spinoso2017}. Consequently, each time a galaxy undergoes a disk instability episode we allow the central black hole to accrete an amount of gas given by:
\begin{equation}\label{eq:QuasarMode_DI}
\rm    \Delta {M}_{BH}^{gas} \,{=}\, \mathit{f}_{BH}^{DI} (1+\mathit{z}_{DI})^{3/2} \frac{\Delta M_{stars}^{DI}}{{1 + (V_{BH}/V_{200c})^2}},
\end{equation}
where $\rm V_{\rm BH}$ and $\rm V_{200c}$ are the same parameters described before, $\rm \mathit{z}_{DI}$ the redshift in which the DI takes place and $\rm \Delta M_{\rm stars}^{DI}$ the mass of stars that triggers the stellar disk instability (see Eq.~\ref{eq:Delta_M_stars_DI}). $\rm \mathit{f}_{BH}^{DI}$ is a free parameter that takes into account the gas accretion efficiency. As we discussed in Section~\ref{sec:bulgesSAM}, we assume DIs to  be the consequence of either the galaxy secular evolution or of galaxy minor mergers. In the former case, we set the efficiency $\rm \mathit{f}_{BH}^{DI} \,{=}\, 0.008$. Instead, for the case of merger-induced DIs, which are causally connected with the last galaxy minor interaction (\rm ${\lesssim}\,10\mathit{t}_{\rm dyn}^{\star}$, \citealt{IzquierdoVillalba2019}) we assume $\rm \mathit{f}_{BH}^{DI}\,{=}\,\mathit{f}_{BH}^{merger}$, i.e since the system is not relaxed the torque causing the gas inflow during the DI has the same efficiency that the one of the merger \citep{HopkinsQuataert2010}. %Instead, for DI merger induced we assume $\rm \mathit{f}_{BH}^{DI}$ to be the same as the one of mergers: $\rm \mathit{f}_{BH}^{DI}\,{=}\, \mathit{f}_{BH}^{merger}$. 
The exact values of $f_{\rm BH}$ for both mergers and DI have been chosen to match the observed correlation between the bulge and the BH mass. \\
\end{enumerate}

%\subsubsection{Accretion rate during quasar phase}
In this work, we assume that all the cold gas accreted by the BH after a merger or disk instability is stored in a reservoir, $\rm M_{Res}$. Motivated by numerical simulations, the reservoir is progressively consumed according to a \textit{light curve} composed by two different phases \citep{Hopkins2005,Hopkins2006a}. While the first one is characterized by a rapid growth truncated at Eddington limit, the second one is a quiescent regime of low accretion rates which starts when the BH reaches a certain threshold of accreted mass. Previous works have shown that this average AGN light curve is able to match successfully the faint end of the AGN luminosity functions \citep{Marulli2008,Bonoli2009,Hirschmann2012,Volonteri2013}. Note that if a galaxy undergoes a new merger or DI while the central BH is still accreting mass from a previous event, the new cold gas driven in the BH surroundings is added to the previous remnant gas reservoir and the light curve re-starts under the new initial conditions.\\

To characterize both phases we use the \textit{Eddington factor}, \fedd, defined as:
\begin{equation} \label{eq:lbolandledd}
L_{\rm bol}(t) \,{=}\,  f_{\rm Edd}(t) \, L_{\rm Edd}(t),
\end{equation}
where $L_{\rm bol}$ and $ L_{\rm Edd}$ are respectively the black hole bolometric and Eddington luminosity. These quantities can be expressed as:
\begin{equation} \label{eq:Leddington}
L_{\rm Edd}(t)\,{=}\, \frac{4\pi G\; m_p\; c}{\sigma_T} \mathrm{M_{\rm BH}}(t)\,{=}\,\frac{\mathrm{M_{\rm BH}}(t)c^2}{t_{\rm Edd}},
\end{equation}
\begin{equation} \label{eq:Luminosity}
L_{\rm bol}(t) \,{=}\, \frac{\epsilon(t)}{1-\eta(t)}  \dot{\rm M}_{\rm BH} \rm c^2,
\end{equation} 
where $\sigma_T$ is the Thomson scattering cross-section for the electron, $m_p$ the proton mass, $t_{\rm Edd}\,{=}\, \sigma_T c/(4\pi m_P G) \,{=}\, 0.45 \, \rm Gyr$ and $\eta$ and $\epsilon$ the black hole accretion and radiative efficiency, respectively. Notice that, since we want to obtain a general equation that describes the growth of supermassive black holes, in Eq.~\eqref{eq:Luminosity} we have distinguished between $\eta$ and $\epsilon$. We refer to other works such as \cite{Pacucci2015} or \cite{VolonteriSilkDubus2015} in which the same distinction is performed.\\

By definition of Eddington limit, the first phase is characterized by $f_{\rm Edd}(t)\,{=}1\,$. The BH grows in this phase for the time necessary to reach the mass $\rm M_{\rm BH,peak}$ defined by the moment in which the BH consumed a fraction $\gamma$ (free parameter set to 0.7) of its whole gas mass reservoir \citep[see][]{Marulli2008,Bonoli2009,Hirschmann2012}. Once this mass is reached, the black hole enters in a self-regulated or quiescent growth regimen characterized by small $f_{\rm Edd}$ values. Following  \cite{Hopkins2006} we assume that $\rm \mathit{f}_{Edd}$ in this phase is parametrized as:
        \begin{equation}\label{equation:m_dot}
                \rm \mathit{f}_{Edd}(\mathit{t}) \,{=}\,\frac{1}{\left[ 1 + (\mathit{t}/\mathit{t}_\mathit{Q})^{1/2}\right]^{2/\beta} },
        \end{equation}
where $t_Q \,{=}\, t_0\,\xi^{\beta}/(\beta \ln 10)$, being $t_0 \,{=}\, 1.26{\times}10^8 \, \rm yr $, $\beta \,{=}\, 0.4$ and $\xi \,{=}\, 0.3$ \citep[see the discussion of these values in][]{Hopkins2006}. We highlight that $t$ is relative to the moment in which the BH finishes the Eddington phase.\\

As pointed out by \cite{Hopkins2006} and \cite{MerloniANDHeinz2008} we also take into account the fact that the radiative efficiency has a dependence with the accretion efficiency and the nature of the accretion flow:
    \begin{equation} \label{equation:epsilon_dep}
    \epsilon \left(\eta, f_{\rm Edd} \right) \rm \,{=}\, \left \{ \begin{matrix} \eta & \rm \mathit{f}_{Edd} \,{>}\, \mathit{f}_{Edd}^{\rm crit} \\ \eta \left( \frac{\rm \mathit{f}_{Edd}}{\rm \mathit{f}_{Edd}^{crit}} \right) &
    \rm \mathit{f}_{Edd} \leq \mathit{f}_{Edd}^{\rm crit}  \end{matrix}\right.
    \end{equation}
where $\rm \mathit{f}_{Edd}^{crit}$ \citep[set to 0.03]{MerloniANDHeinz2008,Volonteri2013} is the \textit{Eddington factor} by which below it the accretion disc becomes radiatively inefficient.\\

Finally, the black hole mass evolution during any accretion event can be obtained from the black hole growth rate, $ \dot{\rm M}_{\rm BH} \rm \,{=}\,  d M_{BH}/d\mathit{t}$.  By using Eq.~\eqref{eq:lbolandledd}, Eq.~\eqref{eq:Leddington} and Eq.~\eqref{eq:Luminosity}, $\dot{\rm M}_{\rm BH}$ can be computed as:
\begin{equation} \label{eq:growth_rate}
\dot{\rm M}_{\rm BH} \rm \,{=}\, \mathit{f}_{Edd}(\mathit{t})\left(\frac{1-\eta(\mathit{t})}{\epsilon(\mathit{t})}  \right)  \frac{M_{BH}(\mathit{t})}{\textit{t}_{Edd}},
\end{equation}
As we will see in Section~\ref{sec:Spin_evolution}, $\epsilon$ and $\eta$ experience a time variation given their dependence with the black hole spin, $a$, which varies with time as a consequence of BH-BH mergers and gas accretion. In order to ease notation, instead of denoting the accretion and radiative efficiency as $\eta(a(t))$ and $\epsilon(a(t))$ we will use just the notation $\eta$ and $\epsilon$. %, keeping in mind that these two values have a spin dependence and, therefore, their value can vary with time. 
%In Section~\ref{sec:Spin_evolution} we will show the values we obtain for the radiative efficiency. 
%Finally, with knowing the values of \fedd, $\eta$ and $\epsilon$ at any moment, we can compute the growth rate of the black hole and its bolometric luminosity. As we will see in Section XXX $\epsilon$ and $\eta$ have a time variation given their dependence with the black hole spin $a$ which varies with time as a consequence of BH-BH mergers and gas accretion. In order to clarify the notation, instead of denoting the accretion and radiative efficiency as $\eta(a(t))$ and $\epsilon(a(t))$ respectively we will use the notation $\eta$ and $\epsilon$, keeping in mind that this two values has a spin dependence and therefore a time variation.\\

\subsubsection{Hot phase accretion}
This mechanism of black hole growth is triggered by the accretion of the hot gas which surrounds the galaxy. Following \cite{Henriques2015} the BH mass accretion rate during this mode is determined by:
\begin{equation}\label{eq:Radio_mode}
\dot{\rm M}_{\rm  BH} \rm \,{=}\, \mathit{k}_{AGN} \left( \frac{M_{hot}}{10^{11}M_{\odot}} \right) \left( \frac{M_{BH}}{10^{8}M_{\odot}}\right),
\end{equation}
where $\rm M_{hot}$ the total mass of hot gas surrounding the galaxy and $\rm \mathit{k}_{AGN}$ is a free parameter set to $\rm 3.5\,{\times}\,10^{-3} M_{\odot}/yr$ to reproduce the turnover at the bright end of the galaxy luminosity function.

As pointed out by \cite{Marulli2008}, the accretion rate of Eq.~\eqref{eq:Radio_mode} is orders of magnitude below the BH Eddington limit. Therefore the contribution of this \textit{radio mode} in the black hole mass growth is minimal. However, the AGN feedback generated during this phase is essential to inject enough energy into the galaxy hot atmosphere  to decrease or even stop the gas cooling rate in the galaxy \citep{Croton2006,Bower2006}.  

%\subsection{Black hole accreiton rate and AGN luminosity during the \textit{quasar mode}} \label{sec:accretionrate}

\subsection{Tracing the BH spin evolution} \label{sec:Spin_evolution}
In this section we present the main equations implemented in \LGalaxies to properly follow the evolution of BH spin. Note that we have applied the spin model presented here to both modes of gas accretion (hot and cold). However, since the hot gas accretion is characterized by small values of {\fedd}, we do not expect a large variation of the spin during these accretion events. 

\subsubsection{Spin evolution during gas accretion} \label{sec:gas_spin}

%Tracking the black hole spin evolution during its feeding episodes is not an easy task. 
During an accretion event, the gas settles in a disk which may not lie in the equatorial plane of the black hole. The disk could have a random orientation with an angular momentum \Jdv, misaligned with respect to the one of the  black hole \Jbhv. %(see Fig~\ref{fig:BlackHoleCartoon}). 
When this misalignment happens, the rotating BH induces on the disk the so called \textit{Lense-Thirring precession} \citep{BardeenPetterson1975}. %, characterized by an angular velocity $\vec{\omega}_{\rm LT} \,{=}\, (2\,G/c^2) \vec{J}_{\rm BH}/R^3$. 
For large-viscosity disks\footnote{This property is met in the $\alpha-$disk model which describes the accretion disk around black holes \citep{ShakuraSunyaev1973}.} this misalignment configuration is unstable making the orbital plane of the inner parts of the disk align (\textit{prograde} orbit)  or counter-align (\textit{retrograde} orbit) to the black hole angular momentum. This process is called \textit{Bardeen-Petterson effect} \citep{Bardeen1975}. %This (anti)alignment of the inner parts results in a warped disk which is maximally deformed at radii \Rw, corresponding to the distance where the warp-diffusion timescale $\tau_{\rm warp} \,{=}\,  R^2/\nu_2(R)$ (with $\nu_2$ the vertical shear viscosity) is comparable with the Lense-Thirring precession time $\tau_{\rm LT} \,{=}\, 2\pi/\omega_{\rm LT}$. This warp region determine if the inner parts of the disk are aligned (\textit{prograde} orbit) or antialigned (\textit{retrograde} orbit) respect to \Jbhv.\\
The criterion of counter-alignment was established by \cite{King2005} and it takes into account the ratio between the black hole and disk angular momentum  and the angle $\theta$ %($\rm -1 \leq cos\,\theta \leq 1$) 
formed between them:\begin{equation} \label{eq:condition_alig}
\mathrm{cos}\,\theta\,{<}\,-\frac{|\vec{J}_{\rm d}|}{2\,|\vec{J}_{\rm BH}|}.
\end{equation}
Thus, for most of the accretion event, the inner part of the accretion disk is aligned if i) $|\vec{J_{d}}|\,{>}\,2|\vec{J}_{\rm BH}|$ or ii) $|\vec{J_{d}}|\,{<}\,2|\vec{J}_{\rm BH}|$ and $\theta\,{<}\,\pi/2$. On the contrary, when $|\vec{J_{d}}|\,{<}\,2|\vec{J}_{\rm BH}|$ and $\theta\,{>}\,\pi/2$, the (anti-)alignment depends if the ratio of $|\vec{J}_{\rm d}|/2\,|\vec{J}_{\rm BH}|$ is small enough with respect to exact value of $\rm cos\,\theta$.\\%Notice that aligned accretions spin up the BH while anti-alignments spin down it.\\

%\begin{figure}
%	\centering
%    \includegraphics[width=1.\columnwidth]{figures/Black_Hole_physics.pdf}
%    \caption{Scheme of the accretion physics of a black hole. }
%\label{fig:BlackHoleCartoon}
%\end{figure} 

We model this in our semi-analytical framework following the spin model presented in \cite{Dotti2013} and \cite{Sesana2014}. The model assumes that the gas available during an accretion event (given by $\dot{\rm M}_{\rm BH}$, see Section~\ref{sec:BHaccretion_model}) is accreted in chunks of transient accretion disks limited by self-gravity, instead of being consumed in a single episode \citep{King2008}. In an $\alpha$-disk the self-gravity radius, $\rm \mathit{R}_{disk}^{sg}$, is determined by the distance to the BH at which the Toomre parameter $Q$ is less than unity: $Q \,{\sim}\, c_s \Omega/\pi G \Sigma \,{\lesssim}\,1$, where $\Omega$ is the Keplerian angular velocity, $c_s$ sound speed and $\Sigma$ the gas surface density. Thus, the value of $\rm \mathit{R}_{disk}^{sg}$ is:
\begin{equation}\label{eq:self-gravity-radius}
\rm \frac{\mathit{R}_{disk}^{sg}}{\mathit{R}_{Sch}} {\approx}\, 10^5  \mathit{f}_{Edd}^{\,-22/45} \left( \frac{\alpha}{0.1}\right)^{28/45} \left( \frac{M_{BH}}{10^6\,M_{\odot}}\right)^{-52/45}  \left( \frac{\eta}{0.1}\right)^{22/45},
\end{equation}
where $\rm \mathit{R}_{Sch}$ is the Schwarzchild radius and $\alpha$ the radial shear viscosity parameter (set from hereafter to 0.1). Concerning the self-gravity mass, $\rm M^{sg}$, the value is expressed as:
\begin{equation}
\rm M^{sg} \,{\approx}\, 2 \times 10^4  \mathit{f}_{Edd}^{\,4/45} \left( \frac{\alpha}{0.1}\right)^{-1/45} \left( \frac{M_{BH}}{10^6\,M_{\odot}}\right)^{34/45} \, M_{\odot} \, ,
\end{equation} 
which was computed by solving the expression $\rm M^{sg}\,{=}\,\int_0^{\mathit{R}_{disk}^{sg}} 2 \pi \Sigma(\mathit{R}) \mathit{R}^2 d\mathit{R}$ with a surface density:
\begin{equation}
\begin {split}
\rm \Sigma(\mathit{R})\,{=}\,7{\times}10^7\mathit{f}_{Edd}^{\,7/10} {\left( \frac{\alpha}{0.1}\right)^{-4/5}}   &{\left( \frac{\rm M_{BH}}{10^6\,\mathrm{M_{\odot}}}\right)^{19/20}}\\
& {\left( \frac{\eta}{0.1}\right)^{-7/10}} \rm {\left( \frac{\mathit{R}}{\mathit{R}_{Sch}}\right)^{-3/4}} {g \, cm^{-2}}.
\end{split}
\end{equation}
%($\rm M_{BH} \,{>}\,M_{BH}^{crit}$)($\rm M_{BH} \,{<}\,M_{BH}^{crit}$)
\noindent To avoid unrealistic accretion episodes, \cite{Dotti2013} set the maximum mass of the transient accretion disk to the minimum between $\rm M^{sg}$ and a fixed molecular cloud mass, $\rm M^{cld}$, set in our case to $\rm 7\,{\times}\,10^3\,M_{\odot}$. As pointed out by \cite{Sesana2014}, fixing the value of $\rm M^{cld}$ is an idealized case, as molecular clouds might be characterized by a broad mass spectrum ($\rm {\lesssim}\,10^{5} M_{\odot}$) eventually dependent on the host galaxy properties. However, since the cloud mass function is not trivial, we follow \cite{Dotti2013} keeping $\rm M^{cld}$ constant over the whole BH life. The continuous increase of the BH mass during accretion episodes with constant $\rm M^{cld}$ implies the formation of accretion disks which carry less angular momentum relative to the BH. In this way, the larger is the BH mass, the larger is the number of retrograde accretions possible given that the condition  {$|\vec{J_{d}}|/2|\vec{J}_{\rm BH}|\,{<}\,1$} holds for a significant fraction of the duration of each accretion event. Typically, the number of retrograde accretions exceeds the prograde ones at $\mathrm{M_{BH}} \,{=}\,\rm 4\,{\times}\,10^5(0.1\mathit{f}_{Edd}/\eta)^{-2/27}( M^{cld}/10^4 M_{\odot})^{45/34} M_{\odot}$. %\textbf{The value of $\rm M^{cld}$ plays a relevant role in the model, given that the chances of BH-disk alignment or anti-alignment depend on the disk angular momentum ($\vec{J_{d}}$, see Eq.~\ref{eq:condition_alig}). When $\rm M^{sg}\,{=}\,M^{cld}$, which translates to a BH mass of $\rm 4\,{\times}\,10^5(0.1\mathit{f}_{Edd}/\eta)^{-2/27}( M^{cld}/10^4 M_{\odot})^{45/34} M_{\odot}$, the condition $|\vec{J_{d}}|/2|\vec{J}_{\rm BH}|\,{<}\,1$ holds and retrograde accretions are possible for a significant fraction of the duration of each accretion event.} %The value of $\rm M^{cld}$ plays also a relevant role in the model, determining the mass above which a BH \textbf{can undergo a significant accretion event with $|\vec{J_{d}}|\,{<}\,2|\vec{J}_{\rm BH}|$.
%changes from maximally to not maximally spinning. 
%Specifically, this happens when $\rm M^{sg}\,{=}\,M^{cld}$ with a critical BH mass of $\rm 4\,{\times}\,10^5(0.1\mathit{f}_{Edd}/\eta)^{-2/27}( M^{cld}/10^4 M_{\odot})^{45/34} M_{\odot}$ \citep{Dotti2013}. 
For a given cloud mass, the radius of the accretion disk is \footnote{The value of $\rm \mathit{R}_{disk}^{cld}$ is computed by solving $\rm M^{cld} \,{=}\, \int_0^{\mathit{R}_{disk}^{cld}} 2 \pi \Sigma(\mathit{R}) \mathit{R}^2 d\mathit{R}$.}:
\begin{equation}
\begin {split} 
\rm \frac{\mathit{R}_{disk}^{cld}}{\mathit{R}_{Sch}} \,{\approx}\, 4 {\times} 10^4 {\mathit{f}_{Edd}^{\,-14/25}} &{\left( \frac{\rm M_{cloud}}{10^4\,\mathrm{M_{\odot}}}\right)^{4/5}} {\left( \frac{\alpha}{0.1}\right)^{16/25}} \\\ 
& {\left( \frac{\rm M_{BH}}{10^6\,\mathrm{M_{\odot}}}\right)^{-44/25}} {\left( \frac{\eta}{0.1}\right)^{14/25}}.
\end{split}
\end{equation}

\noindent Each transient accretion disk has associated an angular momentum $|\vec{J}_{\rm d}|$ determined by \cite{Perego2009}:
\begin{equation}\label{eq:disk-angular-momentum}
|\vec{J}_{\rm d}| \,{=}\, \rm \frac{8}{21} \frac{\mathit{R}_{d}^{7/4} (G M_{BH})^{1/2}}{A_{\nu}} \dot{M}_{BH} \, ,  
\end{equation}
where \begin{equation}
\rm A_{\nu}\,{=}\, 9.14\,{\times}\,10^6 {\mathit{f}_{Edd}^{3/10}} {\left(\frac{\alpha}{0.1}\right)^{4/5}} {\left( \frac{M_{BH}}{10^6 \mathrm{M_{\odot}}}\right)^{1/20}} {\left( \frac{\eta}{0.1}\right)^{-3/10}}  {cm^{5/4}/s} \, , 
\end{equation} and i) $\rm \mathit{R}_{d}\,{=}\, \rm \mathit{R}_{disk}^{cld}$ if $\rm M^{cloud} \,{<}\, M^{sg}$ or ii) $\rm \mathit{R}_{d}\,{=}\, \rm \mathit{R}_{disk}^{sg}$ if $\rm M^{sg} \,{<}\, M^{cld}$.\\

\noindent Therefore, the BH spin evolution during any gas accretion episode is followed by using Eq.\eqref{eq:self-gravity-radius}-\eqref{eq:disk-angular-momentum} and checking Eq.\eqref{eq:condition_alig}. If $|\vec{J_{d}}|\,{>}\,2|\vec{J}_{\rm BH}|$, the fraction of transient disks consumed in a prograd accretion, $n_{Pa}$, is always equal to 1. On contrary, when $|\vec{J}_{d}|\,{<}\,2|\vec{J_{\rm BH}}|$ the scenario is more complex and the exact value of $\rm cos\, \theta$ is needed to check the (anti)alignment. To determine $\rm cos\, \theta$ in the regime of $|\vec{J}_{d}|\,{<}\,2|\vec{J_{\rm BH}}|$, previous works have assumed the \textit{chaotic accretion} presented in \cite{King2005,King2008}, consisting in choosing randomly the value of $\rm cos\,\theta$ \citep{Fanidakis2011,Volonteri2013}. This selection gives the same probability of having an alignment and counter-alignment, i.e $n_{p}\,{\approx}\, 0.5$ \citep{King2005}. Since the gas accretion in a counter-rotating orbit spins down the BH more efficiently than the spin-up due to a co-rotating accretion, the net result of the \textit{chaotic scenario}
is a low-spinning BH population with a typical $a$ value oscillating around $0.2$ \citep[see][]{King2008,BertiVolonteri2008}. Here, instead, following the approach of \cite{Dotti2013} and \cite{Sesana2014}, we assume a much more general scenario, allowing an asymmetry in the $n_{Pa}$ value. According to \cite{Sesana2014} the fraction of transient disks accreted in a prograde accretion when $|\vec{J}_{d}|\,{<}\,2|\vec{J_{\rm BH}}|$ is given by:
\begin{equation} \label{eq:npa}
n_{Pa} = F \,{+}\, \frac{|\vec{J_{d}}|}{2|\vec{J}_{\rm BH}|}(1\,{-}\,F) \, ,  
\end{equation}
where $F$ is an \textit{isotropy parameter} which takes into account the fraction of accretion events with an initial angular momentum with $\theta\,{<}\,\pi/2$\footnote{Notice that by fixing $F\,{=}\,0.5$ the model presented here becomes the standard \textit{chaotic scenario} of \cite{KingPringle2006}.}.\\

%\noindent \textbf{The particular conditions of the gas inflow at subparsec scales that determine the value of $F$ are still unknown. \cite{HopkinsQuataert2010} studied the gas dynamic at different scales by analyzing  nested hydrodynamics simulations. The authors showed that gas inflow at $\rm {\sim}\, pc$ scales displays a correlation with the  non-axysimetric disturbances that act on galactic scales such as mergers or dynamical instabilities of self-gravitating disks. Indeed, they reported that strongly disk-dominated systems show characteristic inflow rates at $\rm {\sim}\,pc$ scales up to 3 orders of magnitude higher than strongly bulge-dominated systems (THIS ARE INFLOW RATES, not alignemnt). On the observational side, \cite{Smethurst2019} compared a sample of nearby disk-dominated AGN to a sample of AGN with merger-dominated histories. The authors found out that the former displays larger black hole accretion rates (${\sim}\,5$ times), suggesting that the geometry of the smooth, planar inflow in a secular dominated system would spin-up the black holes, increasing their accretion efficiency and reducing the feedback responsible for lowering the accretion rate \citep[see also][]{Nayakshin2012}. Motivated by all these results, here we decide to follow the \textit{assumption} of \cite{Sesana2014} which links the value of $F$ to the bulge morphology. With this connection we presume that gas driven to circumnuclear scales retains certain memory of both the event which triggered its nuclear in-inflow and the bulge assembly.}
\noindent The particular conditions of the gas inflow at subparsec scales that determine the value of $F$ are still unknown. \cite{HopkinsQuataert2010}, by analyzing  nested hydrodynamical simulations, found out that the gas inflow at $\rm {\sim}\, pc$ scales displays a correlation with the  non-axysimetric disturbances that act on galactic scales such as mergers or dynamical instabilities of self-gravitating disks. As discussed previously, \cite{King2007} suggested that a chaotic BH feeding at small scales ($F\,{=}\,1/2$) would offer a promising explanation of the growth of the most BHs. However, the large level of anysotropy during the chaotic accretion leads to a population of low spinning BHs which makes it difficult to reconcile the model with observations. In this work, to obtain a better agreement with the spin observations, we relaxed the chaotic scenario by following the the \textit{assumption} of \cite{Sesana2014} which links the value of $F$ to the bulge morphology. With this connection we presume that the gas driven to circumnuclear scales retains certain memory of both the event which triggered its nuclear in-inflow and the bulge assembly (see the recent observational work of \cite{Smethurst2019} which would suggest this). Specifically, in \cite{Sesana2014} the exact value of $F$ is associated to the ratio between the bulk rotation velocity of the bulge, $v$, and its velocity dispersion, $\sigma$. This ratio gives an idea of the coherence motion of the galactic bulge. If $v/\sigma\,{>}\,1$ then $F\,{=}\,1$, if $v/\sigma \,{=}\,0$ then $F\,{=}\,0.5$ and if $ 0 \,{<}\, v/\sigma \,{<}\, 1$ then $F \,{=}\, (1 \,{+}\, v/\sigma)/2$. Thus, bulges characterized by a large coherent motion (e.g. bars) feed the BHs with a larger number of transient chunks of co-rotating accretion disks than bulges dominated by velocity dispersion (e.g. ellipticals).\\

\noindent Since the exact value of $v/\sigma$ is shaped by a complex galaxy dynamics which \LGalaxies does not address, we decide to take the $v/\sigma$ values from the observed probability distribution function (PDF) of pseudobulges, classical bulges and ellipticals presented in Eq. 14-18 of \cite{Sesana2014}. In brief, while elliptical structures have an asymptotic PDF around $0$, classical bulges have a PDF centred at ${\sim}\,0.55$ (see Eq.15 and 16 of \citealt{Sesana2014}). For pseudobulges, we follow the \textit{hybrid} model of \cite{Sesana2014} assuming a log-normal PDF with $\rm 0.34\,dex$ of standard deviation and an average value given by its Eq.14. Following \cite{IzquierdoVillalba2019} we assume that in \LGalaxies, DI \textit{secular} events build-up the pseudobulge/bar structure, minor mergers and DIs \textit{merger-induced} assembly the classical bulge component, and major mergers shape an elliptical galaxy. Therefore, each time that a galaxy undergoes one of these events we extract a random number from its $v/\sigma$ out-coming bulge PDF and we perform a mass-weighted average between that value and the one which characterizes the bulge at that moment. In the case of major mergers, we re-set the value of $v/\sigma$ with a value randomly selected from the elliptical-bulge PDF. For the special case of \textit{smooth accretion} events which bring gas onto the BH but do not build-up a bulge structure, we assume that the $v/\sigma$ of the gas brought towards the BH surroundings is linked to the kinematics of the galaxy gaseous disk. For that, as we did with the pseudobulges, we assume a log-normal PDF with $\rm 0.34\,dex$ of standard deviation and an average value given by Eq.14 of \cite{Sesana2014}.\\

\noindent Following \cite{Thorne1974} we can determine the evolution of the BH spin by:
\begin{equation}\label{eq:spin_vaiation_general}
\dot{\mathit{a}} = \left[ L_{\rm ISCO}(a) - 2a E_{\rm ISCO}(a)\right] \left(\frac{ \dot{\rm M}_{\rm BH}}{\rm M_{BH}}\right),
\end{equation}
where $\rm M_{BH}$ is the black hole mass, $\dot{\rm M}_{\rm BH}$ the accretion rate onto the black hole and $L_{\rm ISCO}$, $E_{\rm ISCO}$ quote respectively the specific angular momentum and energy in the \textit{innermost stable circular orbit} ($\rm ISCO$). Since an accretion event is composed by a fraction of $n_{Pa}$ prograde and $(1-n_{Pa})$ retrograde orbits, Eq.~\eqref{eq:spin_vaiation_general} can be rewritten as \citep{Sesana2014,Barausse2012}:
\begin{eqnarray}
 \dot{a} &{=}& \left[ \, \left(n_{Pa}\, L_{\rm ISCO}^{\rm pro}(a) + (1{-}n_{Pa})L_{\rm ISCO}^{\rm retro}\right)\right . \nonumber\\
 & {-} &\left . 2a\left( n_{Pa}\, E_{\rm ISCO}^{\rm pro}(a) + (1{-}n_{Pa})E_{\rm ISCO}^{\rm retro}(a)\right) \right]\, \frac{ \dot{\rm M}_{\rm BH}}{\rm M_{\rm BH}}, 
\end{eqnarray}
where $L_{\rm ISCO}^{\rm pro}$ ($L_{\rm ISCO}^{\rm retro}$) and $E_{\rm ISCO}^{\rm pro}$ ($E_{\rm ISCO}^{\rm retro}$) are, respectively, the specific angular momentum and energy in a prograde (retrograde) ISCO \citep{Barde1972}.\\ %These quantities can be computed as follows \citep{Barde1972}:
%\begin{equation}
%(L_{\rm ISCO})^{\rm pro}_{\rm retro} \,{=}\, \pm \frac{(r_{\rm ISCO} \, \mp 2a + a^2)}{r_{\rm ISCO} (r_{\rm ISCO} - 3 \pm 2a r_{\rm ISCO}^{-1/2})^{1/2}},
%\end{equation}
%\begin{equation}
%(E_{\rm ISCO})^{\rm pro}_{\rm retro} \,{=}\, \pm \frac{(r_{\rm ISCO}^{3/2} \, - 2r_{\rm ISCO} \pm a)}{r_{\rm ISCO} (r_{\rm ISCO} - 3 \pm 2a r_{\rm ISCO}^{-1/2})^{1/2}},
%\end{equation}
%the upper signs corresponds to the prograd orbits while the lower ones to the retrograd orbits. $r_{\rm ISCO}$ is the radius of the innermost stable circular orbit (ISCO) which depends in the spin value and the (anti)alignment \citep{Barde1972,Barausse2012_2}\footnote{To guide the reader, for non spinning or Schwarzschild BHs ($a\,{=}\,0$) the innermost stable circular orbit (ISCO) has a value $r_{\rm ISCO} \,{=}\, 6$. While for maximum rotating black hole ($a\,{=}\,1$) with a co-rotating accretion event $ 1 \,{\leq}\, r_{\rm ISCO} \,{<}\, 6 $. In the latter case, if the accretion event is counter-rotating the ISCO span in the range $ 6 \,{<}\, r_{\rm ISCO} \,{\leq}\, 9$.}:
%\begin{equation}
%	r_{\rm ISCO} \,{=}\, 3 + Z_2 \mp a\sqrt{(3-Z_1)(3 + Z_1 + 2Z_2)} 
%\end{equation}
%where the upper and lower sign refers to a prograde and retrograde orbits and the vales of $Z_1$ and $Z_2$ are determined by:
%\begin{equation}
%\begin{split}
%& Z_1 \,{=}\, 1 + (1-a^2)^{1/3} \left[(1+a)^{1/3} + (1-a)^{1/3} \right] %\\ 
%& Z_2 \,{=}\, \sqrt{3a^2 + Z_1^2}
%\end{split}
%\end{equation}

\noindent Finally, the accretion efficiency, $\eta$, it is computed as \citep{Bardeen1970}:\begin{equation} \label{eq:eta}
\eta \,{=}\, 1 - E_{\rm ISCO} = 1 - \sqrt{1-\frac{2}{3} \frac{1}{r_{\rm ISCO}}} \, , 
\end{equation}
where $r_{\rm ISCO}$ is the radius of the ISCO (in units of  $\rm \mathit{R}_{Sch}/2$) %the innermost stable circular orbit 
and its exact value depends in the spin value and the (anti)alignment of the transient accretion disk \citep{Barde1972,Barausse2012_2}:%\footnote{To guide the reader, for non spinning or Schwarzschild BHs ($a\,{=}\,0$) the innermost stable circular orbit (ISCO) has a value $r_{\rm ISCO} \,{=}\, 6$. While for maximum rotating black hole ($a\,{=}\,1$) with a co-rotating accretion event $ 1 \,{\leq}\, r_{\rm ISCO} \,{<}\, 6 $. In the latter case, if the accretion event is counter-rotating the ISCO span in the range $ 6 \,{<}\, r_{\rm ISCO} \,{\leq}\, 9$.}:
\begin{equation}
	r_{\rm ISCO} \,{=}\, 3 + Z_2 \mp a\sqrt{(3-Z_1)(3 + Z_1 + 2Z_2)} \, , 
\end{equation}
where the upper and lower sign refers respectively to a prograde and retrograde orbits and the vales of $Z_1$ and $Z_2$ are determined by:
\begin{equation}
\begin{split}
& Z_1 \,{=}\, 1 + (1-a^2)^{1/3} \left[(1+a)^{1/3} + (1-a)^{1/3} \right] , \\ 
& Z_2 \,{=}\, \sqrt{3a^2 + Z_1^2} \, , 
\end{split}
\end{equation}
Therefore, the final value of $\eta$ after a fraction of prograde $n_{Pa}$ and (1-$n_{Pa}$) of retrograde orbits is: \begin{equation}\label{eta_Sesana}
\eta(a_{\rm BH}) \,{=}\, n_{Pa}\,\eta_{\rm pro}(a) + (1-n_{Pa})\,\eta_{\rm retro}(a) \, ,
\end{equation}
where $\eta_{\rm pro}$ and $\eta_{\rm retro}$ are the values of accretion efficiency computed from Eq~\eqref{eq:eta} assuming a $r_{\rm ISCO}$ radius in a prograde and retrograde orbit, respectively.

%\begin{figure*} 
%	\centering
%    \includegraphics[width=1.\textwidth]{figures/disruption_WBH.pdf}
%    \caption{Scheme of \textit{orphan} black holes formation. When two dark matter halos merger, their host galaxies do it as well. The smallest galaxy via dynamical friction starts to skin, with a velocity $\rm \mathit{V}_{infall}$, towards the central potential of the halo where the main galaxy is hosted. During this process the satellite galaxy is disrupted before reaching the halo center and merge with the central galaxy. In this case, the central black hole of the satellite galaxy is deprived of its galaxy and is incorporated in the central dark matter halo. We assume that the \textit{orphan} black hole loses as well its accreiton disk.}
%\label{fig:WNHS_orphan_cartoon}
%\end{figure*} 

\subsubsection{Spin evolution after a BH-BH coalescence}
The last stage of a BH binary system is the coalescence. The final spin of the remnant BH is shaped by the initial spin configuration of its progenitors and their initial masses.
%After a galaxy merger both BHs are sink towards the inner parts of the newly formed galaxy, forming a binary black hole. Due to dynamical friction with gas and three body interaction with nuclear stars, the system is able to shrink its orbit and eventually merger.
In this work we follow the results of \cite{BarausseANDRezzolla2009} which found that the spin modulus of the remnant black hole, $|a_f|$, is determined by:
\begin{equation}\label{BH_spin_merger1}
\begin {split}
 |a_{f}| = \frac{1}{(1+q)^2} & \left[ |a_1^2| + |a_2^2|\,q^4 +  2\,|a_2||a_1|q^2\,\rm{cos}\,\alpha \right.\\
& \left. +\,  2(|a_1|\rm{cos}\,\beta + \mathit{|a_2|}q^2 \rm{cos}\,\gamma)|\ell| +  |\ell|^2q^2 \right]^{1/2},
\end{split}
\end{equation}
with 
\begin{equation}\label{BH_spin_merger2}
\begin{split}
|\ell| \,{=}\, & 2\sqrt{3} + t_2\nu + t_3\nu^2 + \\ & \frac{s_4}{(1+q^2)^2} (\, |a_1|^2 + |a_2|^2q^4 + 2|a_1||a_2|q^2 \rm{cos} \,\alpha) \, \\ &  + \left( \frac{s_5 \nu + t_0 + 2}{1+q^2}\right)  (\, |a_1| \rm{cos}\,\beta + \mathit{|a_2|}q^2\rm{cos}\,\gamma) ,
    \end{split}
\end{equation}
where $q \,{=}\, \rm M_{BH,2}/M_{BH,1} \,{\leq}\, 1$ is the binary mass ratio, $\nu \,{=}\, q/(1+q)^2$ the symmetric mass ratio, $|a_1|$ and $|a_2|$ are the initial spin magnitudes, $\alpha$ is the angle formed between the two spins and $\beta$, $\gamma$ the angle formed between the spin (1 and 2) with the orbital angular momentum, respectively. The other parameters are set to the following values: $s_4 = 0.1229\,{\pm}\, 0.0075$, $s_5\,{=}\,0.4537 \pm 0.1463$, $t_0 \,{=}\, -2.8904\,{\pm}\,0.0359$, $t_3 \,{=}\, 2.5763\,{\pm}\,0.4833$ and $t_2 \,{=}\, -3.51714\,{\pm}\,0.1208$ \citep{BarausseANDRezzolla2009,Barausse2012}.\\

Given that in the SAM we only track the evolution of the BH spin modulus we have to establish the direction between the two black hole spins and their directions with respect to the orbital angular momentum, i.e the values of $\alpha$, $\beta$ and $\gamma$ of Eq.\eqref{BH_spin_merger1} and Eq.\eqref{BH_spin_merger2}. To do that, we use the standard approach of dividing the BH-BH mergers in two types \citep{Barausse2012,Volonteri2013}: \textit{wet} and  \textit{dry}. While the former is characterized by $\rm (M_{\rm Res,1}+M_{\rm Res,2})\,{>}\,(M_{\rm BH,1} + M_{\rm BH,2})$, %where $M_{\rm Res}$ is the reservoir of gas around the two black holes. 
the latter is characterized by $\rm (M_{\rm Res,1}+M_{\rm Res,2})\,{<}\,(M_{\rm BH,1} + M_{\rm BH,2})$. In the case of wet mergers, it has been shown that the dissipative dynamics between black holes and the massive gas disc produces a torque with the effect of aligning both spins to the orbital angular momentum. However, \cite{Dotti2010} showed that this alignment is not perfect and there is a residual offset in the spin direction relative to the orbital angular momentum at the level of 10\degree. Therefore for the wet BH mergers we assume that $\cos{\alpha}$, $\cos{\beta}$ and $\cos{\gamma}$ are random numbers between [$\cos{0\degree}\,{-}\,\cos{10\degree}$]. For dry mergers, we assume random values of $\cos{\alpha}$, $\cos{\beta}$ and $\cos{\gamma}$.\\

\subsection{Black hole coalescence} \label{section:BH_delay_merger}

\subsubsection{Black hole binaries: merger delay}

In the standard \textit{hierarchical} paradigm galaxies grow mainly through mergers with smaller or comparable mass companions \citep{WhiteandRees1978}. This fact, together with recent observational results confirming the existence of BHs being hosted in the center of galaxies \citep{Haehnelt1993}, suggest that binary BH (BBH) systems have been formed and merged during the whole lifetime of the universe. The evolutionary pathway of BBHs is described by three different stages \citep{Begelman1980}: i) an initial \textit{pairing} phase in which dynamical friction exerted by the galaxy gas and stars drives both BHs individually to the nucleus of the merger remnant eventually forming a binary; ii) a following \textit{hardening} phase in which the BBH orbital separation shrinks due to three-body slingshots and/or interaction with a massive gaseous disk; iii) a final \textit{gravitational wave inspiral} phase which drives the binary to the final coalescence \citep[see][and references therein for a review]{Colpi2014}. Building on this scenario, several studies have tackled the evolution of BBHs \citep[see for instance][]{Quinlan1997,Sesana2013,Kelley2016,Kelley2017,Bonetti2018,Bonetti2018a}. On the semi-analytical modelling side, up to date, most of the studies presented in the literature have assumed an instantaneous coalescence right after a galaxy-galaxy merger \cite[see the few exceptions of][]{Volonteri2003,Tanaka2009,Antonini2015,Bonetti2018a,Bonetti2018}. This might be an acceptable assumption at high-$z$, where galaxy interactions are gas rich and  dynamical friction can efficiently  reduce the angular momentum of both BHs and speed up the hardening and final coalescence \citep{Mayer2013,delValle2015}. However, at low-$z$  galaxies are generally poorer in gas and the timescales for merger can become very large.\\

In this work we relax the assumption of instantaneous BH-BH coalescence by allowing for a time delay between the galaxy merger and the BH binary final coalescence. Therefore, we introduce the possibility for galaxy-merger remnants of hosting a binary BH system for a given time interval $t_{delay}^{\rm BH}$. Following the results of several dedicated hydrodynamical simulations \citep{Callegari2011a,Callegari2011b,Fiacconi2013,Mayer2013,delValle2015}, we implement a simple prescription to determine the lifetime of a BH binary system:
\begin{equation}\label{eq:BHBHdelay}
    t_{delay}^{\rm BH} \, {\approx}\, 0.01 \left( \frac{0.1}{q} \right) \, \left( \frac{0.3}{\mathit{f}_{\rm gas}}  \right) \mathcal{F}(e) \, \, \rm Gyr,
\end{equation}
where $q$ (${<}\,1$) is the mass ratio between the two BHs at the moment of the galaxy-galaxy merger, $f_{\rm gas}$ is the galaxy gas fraction, $\mathcal{F}(e)$ is a factor which takes into account the dependence of $t_{delay}^{\rm BH}$ with the binary eccentricity $e$ (for simplicity, we assume $\mathcal{F}(e)\,{\approx}\,1$) and the $\rm 0.01 \, Gyr$ amplitude is based on the hydrodynamical simulations of \cite{Dotti2007}, \cite{Fiacconi2013} and \cite{delValle2015} who found that the BH separation in a binary system decays, typically, at parsecs scales in few $\rm Myrs$. We clarify that the $t_{delay}^{\rm BH}$ presented in Eq.~\ref{eq:BHBHdelay} refers only to the time spent by the BBH system in the hardening and gravitational wave inspiral phase. This is counted starting from the  galaxy merger time computed by the \LGalaxies, which  marks the moment in which the two galactic nuclei merge. We have checked that the black hole mass function, spin distribution and BH-Bulge scaling relation presented in this work do no significantly change if we run the model with a delay $100\,t_{delay}^{\rm BH}$. %\footnote{\textbf{We have explored how the results presented in this work change if we increase a factor of 100 $t_{delay}^{\rm BH}$. The black hole mass function, spin distribution and scaling relations are almost unchanged. The main differences is shown in the ejection rate of BHs with mass ${>}\,10^{7}\, M_{\odot}$ where the model with $100{\times}t_{delay}^{\rm BH}$ display ${\sim}\,$4 times less ejections.}}.
Even though the hydrodynamical simulations of \cite{Escala2004,Escala2005}, \cite{Dotti2007} and \cite{Cuadra2009} showed that dense gaseous regions are very effective in hardening BBHs, promoting their coalescence in less than $\rm {\lesssim}\,10^7\,yr$, the dependence with the gas fraction might be much more complex. For instance, \cite{Tamburello2017} showed that the final BH-BH coalescence can be speeded-up or delayed depending on the clumpiness of the circumnuclear discs around the BHs. In a future work we plan to explore how different assumptions for the merger delays affect the global BH population.\\ %\textbf{Finally, we have checked that the black hole mass function, spin distribution and BH-Bulge scaling relation presented in this work do no significantly change if we run the model with a delay $100\,t_{delay}^{\rm BH}$. The main difference is found on the ejection number density of $\rm M_{BH}\,{>}\,10^7 \,M_{\odot}$ which is  ${\sim}\,5$ times smaller in the run with the largest time delay. This translates in a slightly larger occupation fraction of BHs in elliptical galaxies (${\sim}95\%$ in comparison with the ${\sim}90\%$ predicted by the Eq.\ref{eq:BHBHdelay}).}\\

\begin{figure}
	\centering
	\includegraphics[width=1.0\columnwidth]{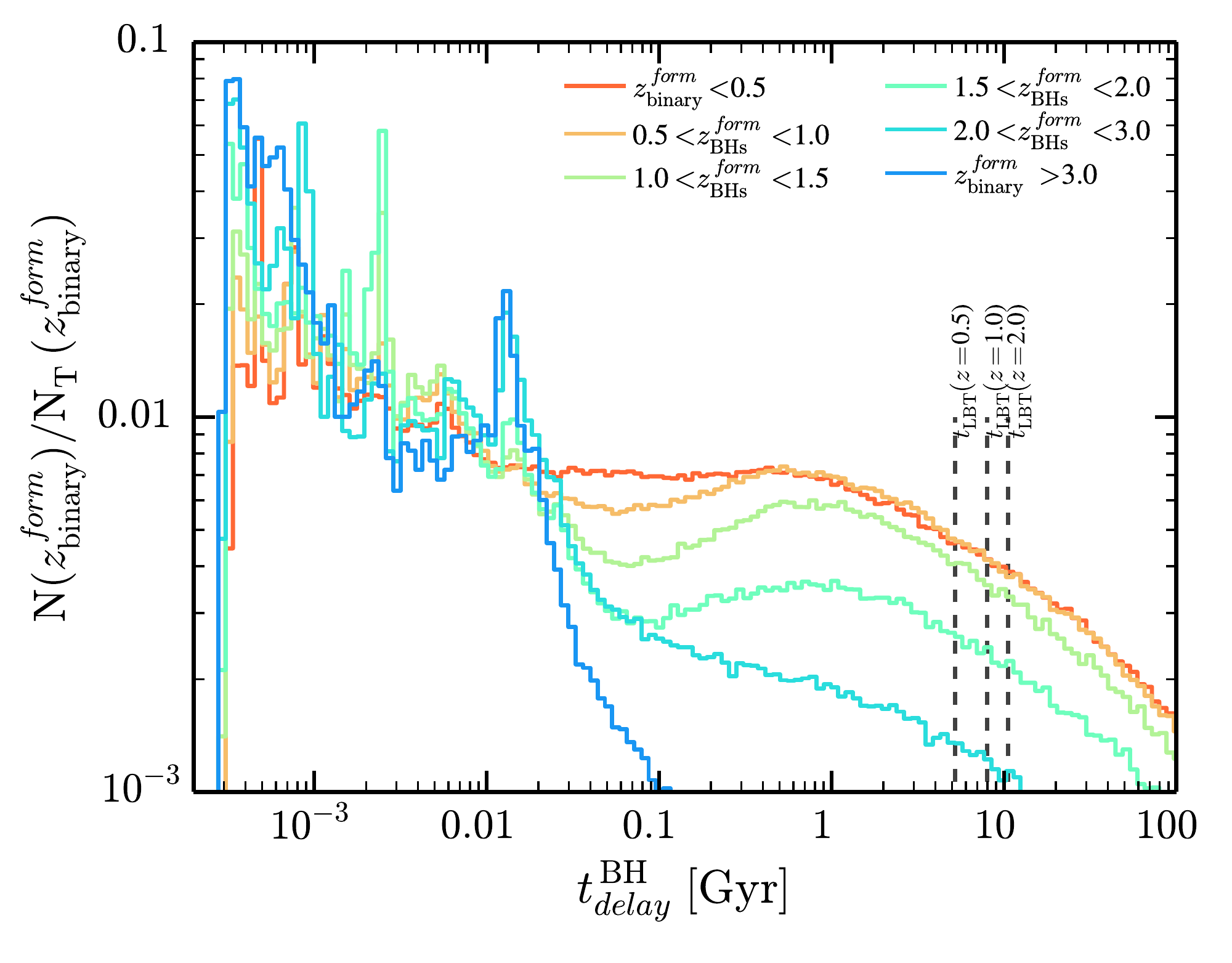}
    \caption{Distribution of BBH merger delay timescales, $t_{\rm delay}^{\rm BH}$, for different formation redshifts ($z_{\rm BHs}^{\rm form}$, as described in the legend). Vertical lines represents the lookback time ($t_{\rm LBT}$) at $z\,{=}\,0.5,1.0,2.0$. Binary systems whose $z_{\rm BHs}^{\rm form}$ is $0.5,1.0,2.0$ and are beyond these lines will not merge by $z\,{=}\,0$. The increase of the binary system lifetime towards low $z$ is caused principally by the dependence of the gas fraction in Eq.~\eqref{eq:BHBHdelay}.}
\label{fig:tdelay_distr}
\end{figure} 

We allow both BHs to accrete their respective gas reservoir carried before the merger while they are in a binary system. We also assume that any new amount of gas accreted after the galaxy merger (see Section~\ref{sec:quasar_accretion})  is added to the reservoir of the most massive BH. Still, the life of a galaxy is rather complicated and it can undergo multiple mergers in short time scales. If a merger happens in a moment in which the galaxy is still hosting a BBH, a third BH (or even another BBH) will perturb the system \citep{Hut1992}. The final configuration is not trivial and few studies have addressed this \citep[see e.g.][and references therein]{Bonetti2018a,Bonetti2018}. Here we use a simplified approach allowing only the two most massive BHs to remain in the galaxy \citep[see][]{Volonteri2003,Volonteri2005}. The lightest BH suffers a scattering event and it is ejected from the galactic nucleus. We highlight that we do not follow the wandering phase of these scattered BHs since we do not track the binary system separation (as, for instance, \citealt{Volonteri2003} do) and therefore we cannot set the initial velocity of the ejection (see Section~\ref{sec:wandering_BHs}). We emphasize that our treatment of 3-body scattering is a simplified approach. Recent works of \cite{Hoffman2007} and \cite{Bonetti2018} pointed out that triple interaction can lead to the coalescence of stalled binaries. Even though this effect should be taken into account to have a complete picture of the evolution of BBHs, we do not expect significant modifications to our final results by neglecting it. For instance, \cite{Bonetti2018} found that triple interactions drive to coalescence \textit{only}  the $\sim30\%$ of stalled binaries and, for those, the merger timescale is dictated by the dynamical friction timescale of the incoming intruder which is usually few hundreds of Myrs. In Appendix~\ref{Appendix:Ejection_Recoil_Three_Body} the number density of both gravitational and 3-body scattering ejections is compared. We show that at any redshift and subhalo mass, the former is $\rm {\gtrsim}\,2\,dex$  more frequent. %Note that \cite{Volonteri2003} and \cite{Volonteri2005} showed that the number of wandering BHs formed in this way is relatively low (${\sim}\,3$ in Milky Wasy size halo) {\bf Silvia: we need to discuss this last sentence}.
Finally, if the two initial BHs forming the binary system are the most massive ones, we eject the intruder without changing the $t_{\rm delay}^{\rm BH}$ value. Conversely, if there was an exchange of BHs in the system, we reset the value of $t_{\rm delay}^{\rm BH}$ with the new initial conditions. The distribution of BH-BH merger delays, $t_{\rm delay}^{\rm BH}$, is presented in Fig.~\ref{fig:tdelay_distr}. As shown, the earlier is the redshift of binary formation, $z^{\rm form}_{\rm BHs}$, the shorter is the time spent by the two BHs in a binary system. While at $z^{\rm form}_{\rm BHs}\,{>}\,3$ BBHs merge in $\rm {\lesssim}\,100 \, Myr$, at $z^{\rm form}_{\rm BHs}\,{<}\,3$ a considerable fraction of BHs spend more than $\rm 1 \, Gyr$ orbiting each other before the final coalescence. Besides, at  $z^{\rm form}_{\rm BHs}\,{<}\,3$ a sizeable fraction of BBHs survive for times larger than the lookback time corresponding to their $z^{\rm form}_{\rm BHs}$, and therefore still inhabit galaxy nuclei today. The increase of the binary system lifetime towards low $z$ is caused principally by the dependence of the gas fraction in Eq.~\eqref{eq:BHBHdelay}. While at high-$z$ galaxy mergers are mainly gas-rich, sinking effectively both BHs until the final coalescence, at low-$z$ galaxies run out of gas stalling the two BHs in a binary system for a long time.\\

\begin{figure*} 	
	\centering
    \includegraphics[width=1.\textwidth]{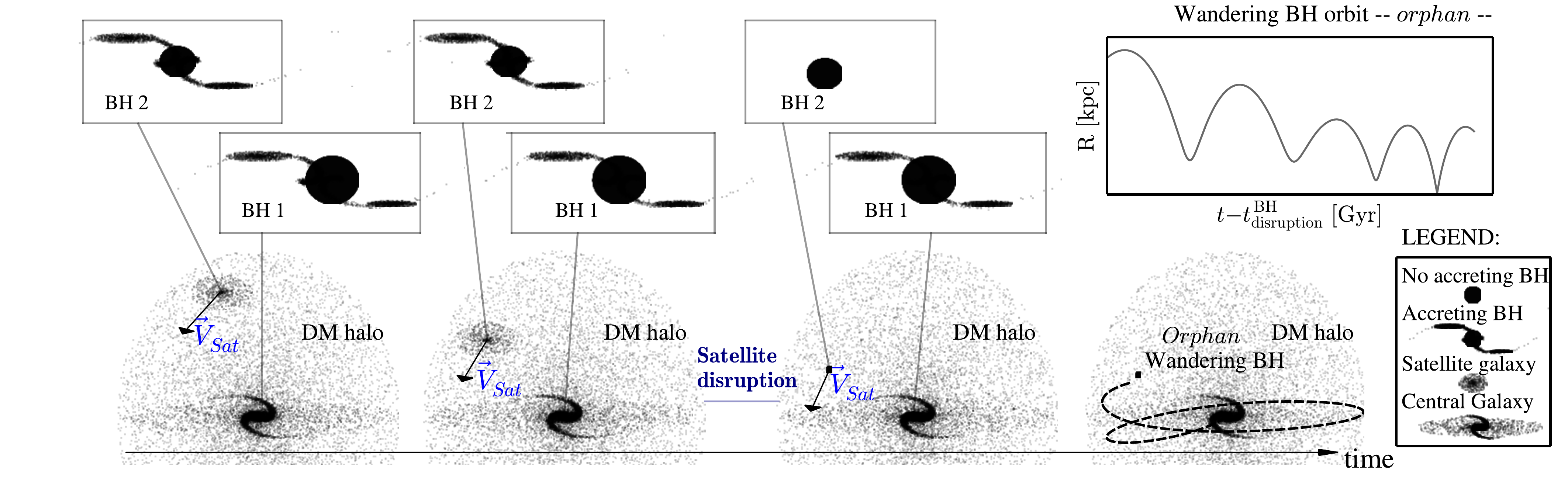}
    \includegraphics[width=1.\textwidth]{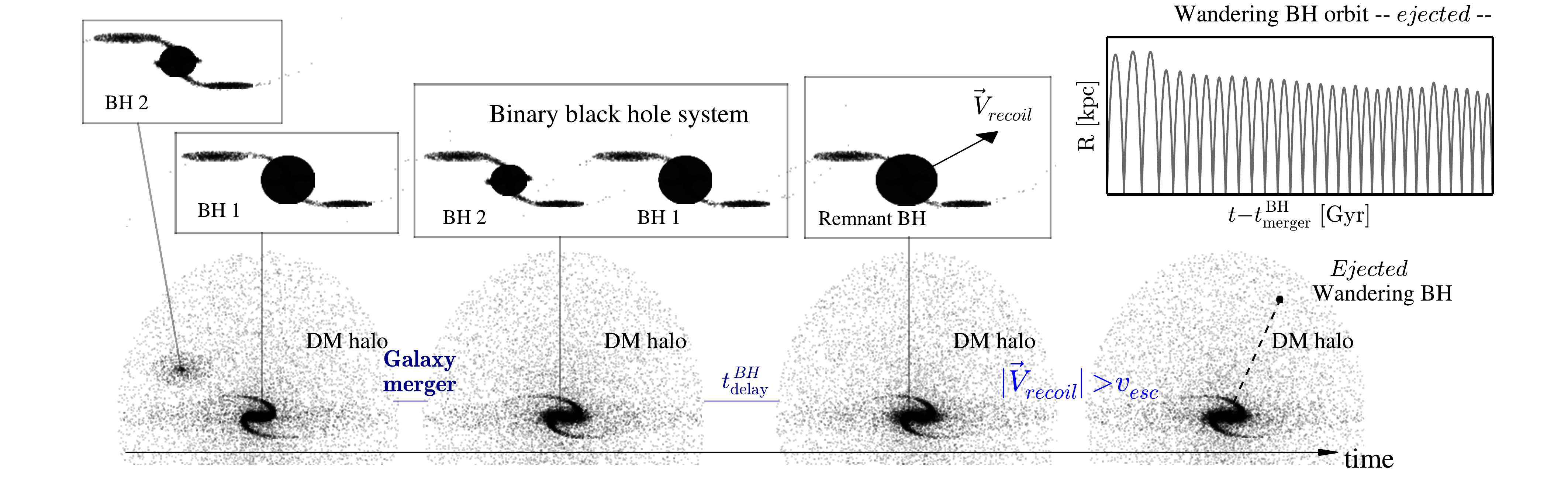}
    \caption{A schematic view of the two pathways that lead to wandering BHs. \textbf{Upper panel}: \textit{Orphan} wandering black holes originate from the complete disruption of the host galaxy during its infall towards the central galaxy. Once the black hole is deprived of its host galaxy, we assume that its accretion disk (in case it was in an accreting phase), is also removed. The initial orbital speed of the new wandering BH is assumed to be the 3D velocity of the satellite galaxy at the disruption time $\vec{V}_{\rm  Sat}$ (see Section~\ref{sec:OrphanBHs}). We highlight that an accreting BH is presented surrounded by a warped accretion disk. This is the result of the Bardeen-Petterson effect which makes the orbital plane of the inner parts of the disk align  or counter-align to the black hole angular momentum. \textbf{Lower panel}:  An \textit{ejected} black hole is the result of the coalescence of a binary system  (after a $\rm \mathit{t}_{delay}^{BH}$ subsequent to the merger of the two host galaxies) and the resulting gravitational recoil. The kick velocity, $\vec{\rm V}_{\rm recoil}$, depends on the properties of two progenitors.  If the modulus of the recoil velocity is larger than the escape velocity of the galaxy ($v_{\rm esc}$) the kicked BH is ejected from the host galaxy, starting its wandering phase.}
\label{fig:WNHS_ejected_cartoon}
\end{figure*}

\subsubsection{Final mass after a coalescence}
Since gravitational waves carry away part of the energy of the system, the final mass of the remnant black hole, $\rm M_{BH}^f$, is not the sum of the two progenitor black hole masses. According to  \cite{TichyandMarronetti2008}, the final mass of a black holes can be computed as:\begin{equation}\label{eq:BHrem2}
\begin {split}
\rm M_{\rm BH}^f{=}(M_{\rm BH,1} + M_{\rm BH,2}) \,  \left[ 1 + 4\nu\right.&\left(\mathit{m}_0-1\right) \, + \\ 
& \left. 16\mathit{m}_1\nu^2\left(\mathit{a}_1 \cos{\beta} + \mathit{a}_2 \cos{\gamma}\right)\right] \, ,
\end{split}
\end{equation} where $m_0\,{=}\,0.9515\pm0.001$, $m_1\,{=}\,-0.013 \pm 0.007$.
%In the case of equal-mass black holes with spins (anti)aligned with the orbital angular momentum, \cite{Reisswig2009} established that the mass of the final BH can be parameterized as:\begin{equation}\label{eq:BHrem1}
%\rm M_{\rm BH}^f = (M_{\rm BH,1} + M_{\rm BH,2})\, \left[1-\left(\mathit{p}_0 + 2\,\mathit{p}_1\,\mathit{a} + 4\,\mathit{p}_2\,\mathit{a}^2\right)\right]
%\end{equation}
%where $p_0 \,{=}\, 4.826\times10^{-2}$, $p_1 \,{=}\,1.559\times 10^{-2}$, $p_2 \,{=}\, 0.485\times10^{-2}$ and $a\,{=}\, (a_1 + a_2)/2$. In the case of non aligned spins or non equal mass black holes we follow \cite{TichyandMarronetti2008} formalism, assuming that:
%\begin{equation}\label{eq:BHrem2}
%\rm {M_{\rm BH}^f}{=}{(M_{\rm BH,1} + M_{\rm BH,2})}{[1 + 4\nu(\mathit{m}_0-1)+{16\mathit{m}_1\nu^2(\mathit{a}_1 %\cos{\beta} + \mathit{a}_2 \cos{\gamma})}]}
%\end{equation} where $m_0\,{=}\,0.9515\pm0.001$, $m_1\,{=}\,-0.013 \pm 0.007$.

\subsection{The population of wandering black holes} \label{sec:wandering_BHs}

Black holes outside of a galaxy in bound orbits within the dark matter halo are defined as wandering black holes (wBHs). In this section we study two types of events responsible for the formation of the wBH population: (i) \textit{disruption of satellite galaxies} and (ii) \textit{gravitational recoils} after a BH-BH merger.
%Indeed, any process that is able to deposit a black hole in the galaxy dark matter halo contribute to the wandering black hole population.

\subsubsection  {Orphan black holes: a disruption event of its host galaxy} \label{sec:OrphanBHs}

In \LGalaxies, as soon as two dark matter subhalos merge their host galaxies do it as well in a time scale given by the dynamical friction formula of \cite{BinneyTremine1987}. The smallest galaxy starts to sink towards the inner parts of its new host subhalo. During this process, if the galaxy is not compact enough it can be disrupted before reaching the subhalo center, losing the possibility to merge with the central galaxy. When this happens, the central black hole (or binary system) of the satellite galaxy is deprived of its host and is incorporated in the dark matter subhalo as a wBH. The disruption process implemented in {\LGalaxies} is presented in \cite{Guo2011}. We follow the 3D orbit of the black hole using the formalism presented in Section~\ref{section:orbit_WBH}. As initial conditions, we consider the position in which the satellite galaxy is disrupted ($\vec{\rm p}_{\rm Sat}\,{=}\,\rm(x_{\rm Sat},y_{\rm Sat},z_{\rm Sat})\,$) and the satellite galaxy velocity ($\vec{V}_{\rm Sat}\,{=}\,(V_{\rm Sat}^{\rm x},V_{\rm Sat}^{\rm y},V_{\rm Sat}^{\rm z})\,$) evaluated at the same instant. Additionally, we assume that the black hole completely loses its accretion disk. From here onwards, we tag this type of wBHs as \textit{orphan black holes}. In the upper panel of Fig.\ref{fig:WNHS_ejected_cartoon} we present a cartoon showing the formation of the orphan black hole population.

%As initial conditions, we consider the position in which the satellite galaxy is disrupted and the infalling velocity, assumed to be radial, of its destroyed host galaxy. Besides, we assume that the black hole loses its accretion disk. From here on, we tag this type of wBHs as \textit{orphan black holes}. In the upper panel of Fig.\ref{fig:WNHS_ejected_cartoon} we present a cartoon showing the formation of the orphan black hole population. 

\subsubsection{Ejected black holes: Kicks due to gravitational recoils} \label{section:ejectedWBHs}

During a BH merger, the binary system emits gravitational waves able to carry away energy and angular momentum. Due to conservation of linear momentum, in the instant of the merger, the emission of gravitational waves imparts a kick to the remnant black hole. The velocity characterizing this kick is called \textit{recoil velocity}, $\vec{V}_{\rm recoil}$. Using numerical simulations of general relativity, it has been shown that recoil velocities can take a wide range of values: from small ones which just displace the remnant BH from the galaxy center, to larger values big enough to exceed the escape velocity of the host galaxy or even of the host halo \citep{Baker2008,Lousto2008,vanMeter2010}. Here we use the fitting formula established by \cite{Lousto2012}:
\begin{equation} \label{eq:recoil_velocity}
\begin{split}
& \vec{V}_{\rm recoil} \,{=}\, v_m\boldsymbol{\hat{e}}_1 + v_{\perp}\left( \cos \zeta \boldsymbol{\hat{e}}_1 + \sin \zeta \boldsymbol{\hat{e}}_2 \right) + v_{\parallel} \boldsymbol{\hat{n}}_{\parallel} \, , \\
& v_{m} \,{=}\, A_m \nu^2 \frac{(1-q)}{(1+q)} \left[ 1 + B_m \nu \right] \, , \\
&  v_{\perp} \,{=}\, H \frac{\nu ^2 }{(1+q)} \left( a^{\parallel}_2 - q a^{\parallel}_ 1 \right) \, , \\
& v_{\parallel} \,{=}\, \frac{16 \nu^2}{(1+q)} \left[ V_{1,1} + V_{A} \tilde{S}_{\parallel} + V_{B}\tilde{S}^2_{\parallel} + V_{C} \tilde{S}^3_{\parallel}\right] |\vec{a}^{\perp}_2  - q\vec{a}^{\perp}_1| \cos(\phi_{\Delta} - \phi_1) \, , 
\end{split}
\end{equation}
where, $v_m$ is the velocity due to the \textit{mass-asymmetry contribution}\footnote{since it does not depend on the spin it disappears in equal binary mass mergers} and, $v_{\parallel}$, $v_{\perp}$ are \textit{spin contribution} producing kicks parallel and perpendicular to the orbital angular momentum respectively.  As in the previous section, $q \,{=}\, \rm M_{BH}^2/M_{BH}^1 \,{\leq}\, 1$ is the binary mass ratio, $\nu \,{=}\, q/(1\,{+}\,q)^2$ the symmetric mass ratio, $|a_i|$ is  the initial spin magnitude of the black hole $i$, $\parallel$ and $\perp$ refer respectively to components parallel and perpendicular to the orbital angular momentum. $\boldsymbol{\hat{e}}_1$  and $\boldsymbol{\hat{e}}_2$ are orthogonal unit vectors in the orbital plane. $\zeta = 145^{o}$ is the angle between the unequal mass and spin contribution \citep{Gonzalez2007,Lousto2008}, $\vec{\tilde{S}} \,{=}\, 2 (\vec{a}_2 \,{+}\, q^2\vec{a}_1)/(1\,{+}\,q)^2$. $\phi_{\Delta}$ is the angle between the in-plane component $\vec{\Delta}^{\perp} \,{=}\, (M_{\rm BH}^1 \,{+}\, M_{\rm BH}^2)(\vec{S}_2^{\perp}/M_{\rm BH}^2 \,{-}\, \vec{S}^{\perp}_ 1/M_{\rm BH}^1)$ and the infall direction at merger. The values of the other coefficients are obtained numerically: $A_m \,{=}\, 1.2 \times 10^4$, $B_m \,{=}\, -0.93$, $H \,{=}\, 6.9\times 10^3$ \citep{Gonzalez2007,Lousto2008} and $V_{1,1}\,{=}\, 3677.76\, \rm km/s$, $V_A \,{=}\, 2481.21 \,\rm  km/s$, $V_B \,{=}\, 1792.45\, \rm km/s$ and $V_C \,{=}\, 1506.52 \, \rm km/s$ \citep{Lousto2012}.\\

If the modulus of this velocity is larger than the escape velocity of its host galaxy, the black hole is kicked from it and incorporated in the DM halo as a wBH. Indeed, if the recoil velocity is large enough, the BH can even escape from the halo. This happens more frequently at high-$z$, where the halos are smaller and their potential wells shallower. This scenario is summarized in the lower cartoon displayed in Fig.\ref{fig:WNHS_ejected_cartoon}. Hereafter, we call this type of wBHs \textit{ejected black holes}. When an \textit{ejected} wBH is kicked out of its host galaxy we assume a \textit{purely} radial motion for the orbital integration (i.e, along the $\hat{r}$ coordinate without movement in the $\hat{\theta}$ and $\hat{\varphi}$ ones, see Section~\ref{section:orbit_WBH}). However, when an \textit{ejected} wBH loses its host halo and is incorporated to a larger structure, after a galaxy merger, we start treating the \textit{ejected} wBH as an \textit{orphan} wBH, and we start tracking its 3D orbit according to the formalism presented in Section~\ref{section:Orbits_during_mergers}. In what follows, we will call \textit{inborn} the \textit{ejected} wBHs that are still orbiting the galaxy from whose nucleus they were ejected, while we will use the term  \textit{acquired} to refer to \textit{ejected} wBHs that have lost their host galaxy and have been incorporated by the halo of a larger one.
%Even though this is a reasonable assumption when the \textit{ejected} wBH and its host halo evolve in isolation, the scenario might change when the latter merges with a larger galaxy. In this case, the motion of the \textit{ejected} wBH inside the new host is unlikely to be purely radial. Hence, after an subhalo-subhalo merger we start tracking the 3D orbit according to the formalism presented in Section~\ref{section:Orbits_during_mergers} for \textit{orphan} BHs. We tag as \textit{inborn} \textit{ejected} wBHs  the \textit{ejected} wBHs that are still around the galaxy from which they were ejected, thus still on a radial orbit. On the other hand, \textit{ejected} wBHs that were accreted by another subhalo and follow 3D orbits, are labeled as \textit{acquired} \textit{ejected} wBHs. When it is not specify between \textit{inborn} and \textit{acquired}, we will refer to both populations of \textit{ejected} wBHs.}

Finally, we neglect the fact that the ejected BH can retain an accretion disk. It is not clear how massive is the disk that can be retained and subsequently accreated by a BH after a recoil \citep[see][]{Blecha2011}. \cite{Loeb2007} showed that in an $\alpha$-disk the maximum amount of the disk material that a recoiled BH can carry out is, at most $\rm {\sim}M_{BH}$ which accounts for a relatively small AGN lifetime ($\rm {\sim} \,4.5\,{\times}\,10^6 \mathit{f}_{Edd}/\epsilon\, yr$, \citealt{VolonteriMadau2008}). In this work we neglect the accretion disk of  ejected BHs, but we plan to include it in a future work, focused on the observability of AGN pairs.

%keeping in mind that in the best scenario the BH would only double its mass in the wandering phase and it would be visible as an AGN during a very shot period ($\rm {\sim}\,10^7\,yr$ at \fedd${=}\,1$ and $\epsilon\,{=}\,0.1$).

\subsubsection{Tracking the orbits of wandering black holes}\label{section:orbit_WBH}

Once in the wandering phase, the BH starts to orbit within the subhalo potential (i.e. around its centre $r_0$\footnote{Here we assume that the galaxy position is always at the center of the host subhalo, i.e $r_0$}). The equation of motion can be expressed as:
\begin{equation}\label{eq:oscilationsWBH}
\ddot{\vec{u}} \,{=}\, \rm \left( - \frac{\mathrm{G}\,{\mathrm{M}\left({<}\left(\mathit{r}\,{-}\,\mathit{r}_0\right)\right)}}{\left(\mathit{r}\,{-}\,\mathit{r}_0\right)^2} + \mathrm{\mathit{a}_{df}^{cless}} + \mathrm{\mathit{a}_{df}^{gas}}\right)\, \mathit{\hat{u}},
\end{equation}
where the first term refers to the gravity acceleration caused by a mass  $\mathrm{M}\left({<}\left(\mathit{r}\,{-}\,\mathit{r}_0\right)\right)$ of dark matter, stars and gas (in both galaxy and inter-cluster medium) within a radius $\mathit{r}\,{-}\,\mathit{r}_0$. The second and third terms are related to the dynamical friction caused by collisionless (stars and dark matter) and gas components, respectively. Since in this work we are interested in the position of wBHs, we track their orbital evolution by solving numerically Eq.~\eqref{eq:oscilationsWBH} using a $\rm 4^{th}$ order \textit{Runge-Kutta} integrator with time step of $\rm 1\,Myr$\footnote{Note that the \textit{sub-steps} that \LGalaxies does between DM \textit{snapshots} are $\rm {\sim} 2\,{-}\,20\, Myr$, depending on redshift}. We have also used integration intervals of $\rm 0.1\, Myr$, obtaining nearly identical results.\\

\noindent Assuming a Maxwellian velocity distribution and a black hole velocity $v_{\rm BH}$, the value of $\mathrm{\mathit{a}_{df}^{cless}}$ is determined by \citep{Chandrasekhar1943}: \begin{equation}\label{eq:trajectory_WBH}
\vec{a}_{\rm df}^{\rm \, cless}\,{=}\,-\frac{4 \pi G^2 \, \mathrm{M_{BH}} {\,\rho{(r\,{-}\,r_0)}} \ln \Lambda}{v_{\rm BH}^3} \left[ errf(\varGamma) - \frac{2 \varGamma}{\sqrt{\pi}}e^{-\varGamma^2}\right] \, \mathit{\hat{v}},
\end{equation}
with $\varGamma \,{=}\,|v_{\rm BH}|/\sqrt{2} \sigma $ where $\sigma$ is the dark matter velocity dispersion computed as $\sigma = \rm (\rm GM_{halo}/2R_{200})^{1/2}$, $errf$ is the error function and $\ln \Lambda$ is the Coulomb logarithm fixed to $3.1$ \citep{Escala2004,GualandrisMerritt2008,Blecha2008}. Finally, $\rho(r{-}r_0)$  is the mass density of the collisionless system enclosed within $r{-}r_0$. In our case $\rho(r{-}r_0)\,{=}\,\rho_{\rm DM}(r{-}r_0)\,{+}\,\rho_{\rm ICM}(r{-}r_0)$, i.e the mass of dark matter and the mass of stars in the inter-cluster medium (ICM or stellar halo). For the case of dark matter we assume a NFW profile \citep{NFW1996}\footnote{The concentration parameter as a function of redshift and subhalo mass has been computed by using the fits of \cite{Dutton2014}.} while for ICM an isothermal one.\\ 

\begin{figure} 
	\centering
	\includegraphics[width=1.0\columnwidth]{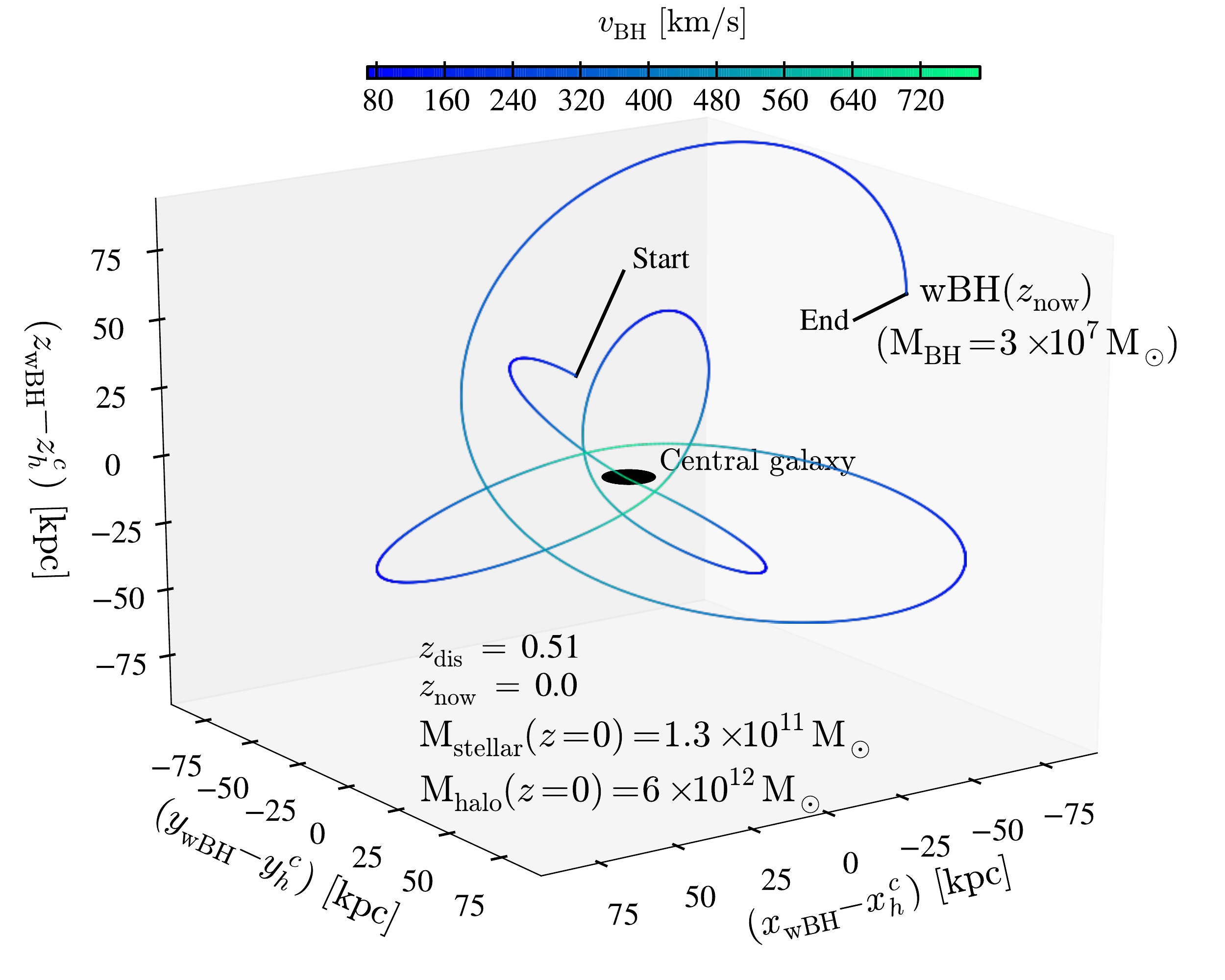}
    \caption{\textit{Orphan} wBHs orbit extracted from \LGalaxies run on top of MS subhalo merger trees. The example presents a $\rm {\sim}\,10^7\, M_{\odot}$ wBH formed at $z_{\rm dis}\,{=}\,0.51$ after the disruption of its host galaxy. The final ($z_{\rm now}\,{=}\,0$) halo and stellar mass of its central galaxy is $\rm 6{\times}10^{12} \, M_{\odot}$ and $\rm 1.3{\times}10^{11} \, M_{\odot}$, respectively. The color of the orbit encodes the modulus of the wBH velocity ($v_{\rm BH}$).}
    %Example of \textit{orphan} wBHs orbits extracted from \LGalaxies run on top of MS subhalo merger trees. \textbf{Upper panel}: Orbit of a $\rm {\sim}\,10^7\, M_{\odot}$ wBH formed at $z_{\rm dis}\,{=}\,0.51$ after the disruption of its host galaxy. The final ($z_{\rm now}\,{=}\,0$) halo and stellar mass of its central galaxy is $\rm 6{\times}10^{12} \, M_{\odot}$ and $\rm 1.3{\times}10^{11} \, M_{\odot}$, respectively. The color of the orbit encodes the modulus of the wBH velocity ($v_{\rm BH}$). \textbf{Lower panel}: The same as the upper panel but for a much more massive wBH ($\rm {\sim}\,10^8\, M_{\odot}$).}
\label{fig:Eamples_orbit_wBHs}
\end{figure}

\noindent Since galaxies are surrounded by hot gas which fills their host DM subhalos, we also take into account the dynamical friction caused by that gas. \cite{Escala2004} determined that the damping of the oscillation amplitude for an object moving inside a gas follows the equation:
\begin{equation}
\vec{a}_{\rm df}^{\, \rm gas}= -\frac{4 \pi G^2 \mathrm{M_{BH}} \rho_{\rm gas}(r - r_0) }{v_{\rm BH}^3} f(\mathcal{M}) \,\mathit{\hat{v}} \, ,
\end{equation}
\noindent where $\rho_{\rm gas}(r{-}r_0)$ is the mass density of the hot gas at position $r{-}r_0$, $\ln \Lambda$ is again the Coulomb logarithm fixed to $3.1$ and, $\mathcal{M}$ the \textit{Mach number} equal to $|v_{\rm BH}|/c_s$ with $c_s$ the sound speed computed as in \cite{Tanaka2009} and \cite{Choksi2017}: $c_s \,{\approx}\, 1.8 (1+z)^{1/2} (\mathrm{M_{halo}}/10^7\mathrm{M_{\odot}})^{1/3} (\Omega_{\rm M} h^2/0.14) \, \rm km/s$. The exact value of $f(\mathcal{M})$ is determined by:
\begin{equation}
  f(\mathcal{M})=\begin{cases}
    \frac{1}{2} \ln \Lambda \left[ errf(\frac{\mathcal{M}}{\sqrt{2}}) - \sqrt{\frac{2}{\pi}} \mathcal{M} e^{-\mathcal{M}^2/2}  \right] & \text{if $\mathcal{M} \leq 0.8$}\vspace{1mm}\\
    \frac{3}{2} \ln \Lambda \left[ errf(\frac{\mathcal{M}}{\sqrt{2}}) - \sqrt{\frac{2}{\pi}} \mathcal{M} e^{-\mathcal{M}^2/2}  \right] & \text{if $0.8\leq \mathcal{M} \leq 1.5$}\vspace{1mm}\\
    \frac{1}{2}\ln{(1-\mathcal{M}^{-2})} + \ln{\Lambda} & \text{if $\mathcal{M} > 1.5$}
  \end{cases}
\end{equation}

\noindent The integration of the wBH orbit stops when i) the black hole is re-incorporated in the galaxy (see Section~\ref{sec:Reincorporation}), ii) the recoil velocity of ejected BHs is larger than the subhalo escape velocity or iii) the black hole position exceed $\rm 3\,{\times}\,R_{200}$ ($\rm R_{200}$ is the subhalo virial radius) and it is still moving away from the galaxy.\\ 

As an example, in Figure~\ref{fig:Eamples_orbit_wBHs} we present orbit evolution of an \textit{orphan} wBHs extracted form \LGalaxies run on the MS subhalo merger trees. In particular, the  wBH has a  mass of $\rm {\sim}\,10^7\, M_{\odot}$ and it was formed at $z\,{\sim}\,0.5$ inside of a $\rm {\sim}\,10^{12}\, M_{\odot}$ subhalo. 

%\textbf{As an example, in Figure~\ref{fig:Eamples_orbit_wBHs} we present the orbits of two different \textit{orphan} wBHs extracted form \LGalaxies run on top of the MS subhalo merger trees. The first plot displays the 3D orbital evolution of a $\rm {\sim}\,10^7\, M_{\odot}$ wBH formed at $z\,{\sim}\,0.5$ inside of a $\rm {\sim}\,10^{12}\, M_{\odot}$ subhalo. The latter shows the same but for a much more massive wBH ($\rm {\sim}\,10^8\, M_{\odot}$) orbiting inside a $z\,{=}\,0$ subhalo of $\rm {\sim}\,10^{13}\, M_{\odot}$}. 

\subsubsection{Reincorporation of wBHs}\label{sec:Reincorporation}
We assume that a BH is reincorporated in a galaxy when it passes thought the galaxy, $r\,{<}\,R_{\rm gal}$\footnote{$R_{\rm gal}$ is the galaxy radius computed as the mass weighted average between the bulge, stellar disk and cold gas disk. In \cite{IzquierdoVillalba2019} we showed that the SAM, after including gas dissipation losses during the bulge size computation, is able to reproduce the galaxy radius of early and late type galaxies.}, with a velocity ($v_{\rm BH}$) smaller than the galaxy escape velocity ($v_{\rm esc}$), i.e: $v_{\rm BH}\,{<}\,v_{\rm esc}$. The exact value of $v_{\rm esc}$ is computed taking into account both bulge and disk (stellar and cold gas):
\begin{equation}\label{eq:escapevel}
    v_{\rm esc} \,{=}\, \sqrt{v^2_{\rm esc, Bulge} \,{+}\, v^2_{\rm esc, Stellar \, disk} \,{+}\,  v^2_{\rm esc, \,Cold \, disk}} \, ,
\end{equation}

In the case of the bulge component we assume a Hernquist profile \citep{Hernquist1990} with an escape velocity at $r_0$: \begin{equation}
v^2_{\rm esc, Bulge} \,{=}\, 2 \left( 1\,{+}\,\sqrt{2} \right) \,\frac{G \, \mathrm{M}_{\rm Bulge}}{r_{\rm bulge}},
\end{equation} where $ \rm M_{ Bulge}$ is the total bulge mass and $r_{\rm bulge}$ the bulge half-mass radius. Although, this profile can approximate classical bulges, it is a rough approximation for pseudobulges.\\ 

\noindent For stellar and cold gas disk we assume an exponential profile. In this case, the escape velocity at $r_0$ is determined by: \begin{equation}
v^2_{\rm esc, Stellar/Cold\, disk} \,{=}\, 3.36\,\frac{G \, \mathrm{M}_{\rm Disk}}{r_{\rm disk}},
\end{equation}
where $ \rm M_{Disk}$ is the total stellar/cold gas disk mass and $r_{\rm disk}$ its half-mass radius.\\

%\begin{figure} 
%	\centering
%	\includegraphics[width=0.65\columnwidth]{figures/Relations_All_Bulges_Poster.pdf}
%    \caption{Black hole mass - bulge mass scaling relation  at $z\,{=}\,0$. Upper panel represent the relation for all galaxies. Solid line is the median of the relation and shaded areas the 1$\sigma$-2$\sigma$ of the relation. Central and lower panel represent the same but for the population of classical bulges + ellipticals and pseudobulges, respectively. Dots are the data from \protect \cite{Kormendy2013} and dotted lines the fits of \protect \cite{Kormendy2013} and \protect \cite{Savorgnan2016}.}
%\label{fig:BH_Bulge_relation}
%\end{figure} 

%\begin{figure} 
%	\centering
%	\includegraphics[width=1.0\columnwidth]{figures/BH_Poster.pdf}
 %   \caption{Redshift evolution of the black hole mass function. We have added the observational results of \protect \cite{Marconi2004,Shankar2004,Shankar2013}}
%\label{fig:BH_mass_function}
%\end{figure} 

\subsubsection{Changing reference system during merger events} \label{section:Orbits_during_mergers}

\noindent The scenario described above complicates when galaxies merge. During the interaction, the satellite galaxy
%of the smaller halo (or the less bound one in the case of similar halo masses)
is first of all deprived of its host DM subhalo. In the following, we will refer to the ``subhalo merger time'' as the instant in which the satellite halo is no more identified within the underlying merger tree. Any wBHs which was hosted by the satellite galaxy now starts to be influenced by the potential of the larger central galaxy\footnote{The orbits of the wBHs moving around the central subhalo (galaxy) are also affected by the change of the dark matter potential as the differential equation which determines the orbit evolution depends on the halo mass at $r\,{<}\,r_{\rm wBH}$, see Eq~\eqref{eq:oscilationsWBH}.}. To track the new orbit, the position and velocity of the recently accreted wBHs, we need to change the reference system, so that the center is  given by the new central galaxy position and velocity \footnote{Notice that we assume that the galaxy position is always at the center of the host subhalo.}:
\begin{equation} 
\begin{split}
&x_{\rm wBH}^{\rm new}\,{=}\, \left( x_h^s + x_{\rm wBH} \right) - x_h^c  \, \, ;  \, \,
v_{\rm wBH}^{x, \rm new}\,{=}\, \left( v_h^{s,x} + v_{\rm wBH}^{\rm x} \right) - v_h^{c,x}, \\%  \,\hat{x}\\
&y_{\rm wBH}^{\rm new}\,{=}\, \left( y_h^s + y_{\rm wBH} \right) - y_h^c \, \, ;  \, \,
v_{\rm wBH}^{y, \rm new}\,{=}\, \left( v_h^{s,y} + v_{\rm wBH}^{\rm y} \right) - v_h^{c,y}, \\%
&z_{\rm wBH}^{\rm new}\,{=}\, \left( z_h^s + z_{\rm wBH} \right) - z_h^c \, \, ;  \,\, \,
v_{\rm wBH}^{z, \rm new}\,{=}\, \left( v_h^{s,z} + v_{\rm wBH}^{\rm z} \right) - v_h^{c,z}, \\%
\end{split}
\end{equation}
\noindent where $(x_h^s,y_h^s,z_h^s)$ and $(v_h^{s,x},v_h^{s,y},v_h^{s,z})$ are, respectively, the positions and velocity of the satellite subhalo at the moment of merger. $[\,(x_h^c,y_h^c,z_h^c) \,,\,(v_h^{c,x},v_h^{c,y},v_h^{c,z})\,]$ and $[\,(x_{\rm wBH},y_{\rm wBH},z_{\rm wBH}) \, , \, (v_{\rm wBH}^{\rm x},v_{\rm wBH}^{\rm y},v_{\rm wBH}^{\rm z})\,]$ represent the same as before but for the central subhalo and wandering BH, respectively.\\

%\noindent Despite pure radial orbits for \textit{ejected} wBHs is a reasonable supposition when both wBHs and their host halos evolve in isolation\footnote{We refer with this to all those \textit{ejected} wBHs that are still orbiting within the DM in which they were originated. No halo merger happened in their host halo since they were formed.}, the scenario might change when the latter merge with their central DM halo. In this case, the motion of \textit{ejected} wBHs inside the new host is unlikely to be purely radial. Thus, after an halo-halo merger we start tracking their 3D orbit following the methodology previously described. 

\noindent Although obtaining the exact value of $\,(x_{\rm wBH},y_{\rm wBH},z_{\rm wBH})$ and $(v_{\rm wBH}^{\rm x},v_{\rm wBH}^{\rm y},v_{\rm wBH}^{\rm z})$ for \textit{orphan} (or \textit{acquired ejected}) wBHs at the moment of the merger is simple because we follow their full 3D orbit, for \textit{inborn} \textit{ejected} wBHs is not trivial since we only follow their radial coordinate (see Section~\ref{section:ejectedWBHs}). In this case, we assume that they start the new orbit from a random location (in $\theta, \varphi$) in a sphere of radius $r$. Once $r$, $\theta$ and $\varphi$ are fixed, the Cartesian coordinates and velocities of the \textit{inborn} \textit{ejected} wBHs are determine by: \begin{equation} 
\begin{split}
&x_{\rm wBH}\,{=}\, |r|\sin{\theta}\cos{\varphi} \,\,\,\, ; \, \,\,\, v_{\rm wBH}^{x}\,{=}\, |v_{\rm BH}|\sin{\theta}\cos{\varphi},  \\%  \,\hat{x}\\
&y_{\rm wBH}\,{=}\, |r|\sin{\theta}\sin{\varphi} \,\,\,\, ; \, \,\,\, v_{\rm wBH}^{y}\,{=}\, |v_{\rm BH}|\sin{\theta}\sin{\varphi}, \\%\,\hat{y} \\
&z_{\rm wBH}\,{=}\, |r|\cos{\theta} \,\,\,\,\,\,\,\,\,\,\,\,\,\,\,\, ; \, \,\,\, v_{\rm wBH}^{x}\,{=}\, |v_{\rm BH}|\cos{\theta}. \\%\,\hat{z}
\end{split}
\end{equation}
where $r$ and $v_{\rm BH}$ are respectively the radial position and radial velocity of the \textit{ejected} wBH at the moment of the halo merger.\\

\section{Nuclear black holes} \label{sec:results_nuclear_BHs}

In this section we compare the predictions for nuclear black holes (i.e BHs located at the center of the host galaxy at the analyzed redshift) with  a variety of available  observational results. 
By checking the black hole mass function, the bolometric luminosity functions and spin values we prove that our black holes form a reliable population at $z\,{\leq}\,4$. We do not present results beyond $z\,{=}\,4$ since the model predictions are very sensitive to the exact seed mass assumed. In a future work, where a more careful seeding in \LGalaxies  is modeled, we will analyze the high-$z$ black hole population  (Spinoso et al. in prep.). On top of this, we show only the results for  BHs whose mass is larger than $\rm 10^6 \, M_{\odot}$, as at smaller masses the resolution of \texttt{Millenium} DM simulation does not allow us to draw reliable conclusions.

\subsection{Black hole mass assembly}

%It has been shown that in the local universe the mass of central black holes correlates with the bulge mass of their host galaxies \citep{KormendyKennicutt2004,HaringRix2004,McConnellANDMa2013,Kormendy2013}. This is the so-called \textit{scaling relation} and it has been interpreted as a evidence of a co-evolution between the BH and its galaxy. In upper panel of Fig.\ref{fig:BH_Bulge_relation} we compare the relation bulge-BH predicted by the model with the data of \cite{Kormendy2013}. The results are in good agreement with the observations, specially at high masses. In the same work of \cite{Kormendy2013}, the BH-Bulge relation was separately presented for classical- (including ellipticals) and pseudo- bulges. Even though \cite{Kormendy2013} found a well constrained relation for the former, it did not find a clear one for latter. The model predictions regarding the bulge morphology are presented in middle and lower panel of Fig.\ref{fig:BH_Bulge_relation}. As suggested by \cite{Kormendy2013}, a clear correlation is found for classical bulges (including ellipticals). In the case of pseudobulges, even though at low bulge masses ($\rm M_{bulge}\,{<}\,10^{10}\, M_{\odot}$) we find a clear trend, at larger ones it breaks down: the dispersion increases and the correlation flattens. As we will see in Section~\ref{sec:Recoild_and_scaling_Relation}, this is caused by the inclusion of BH recoil velocities in the model: by ejection of black holes the scaling relation of massive pseudobulges is blurred.\\

\begin{figure} 
	\centering
    \includegraphics[width=0.83\columnwidth]{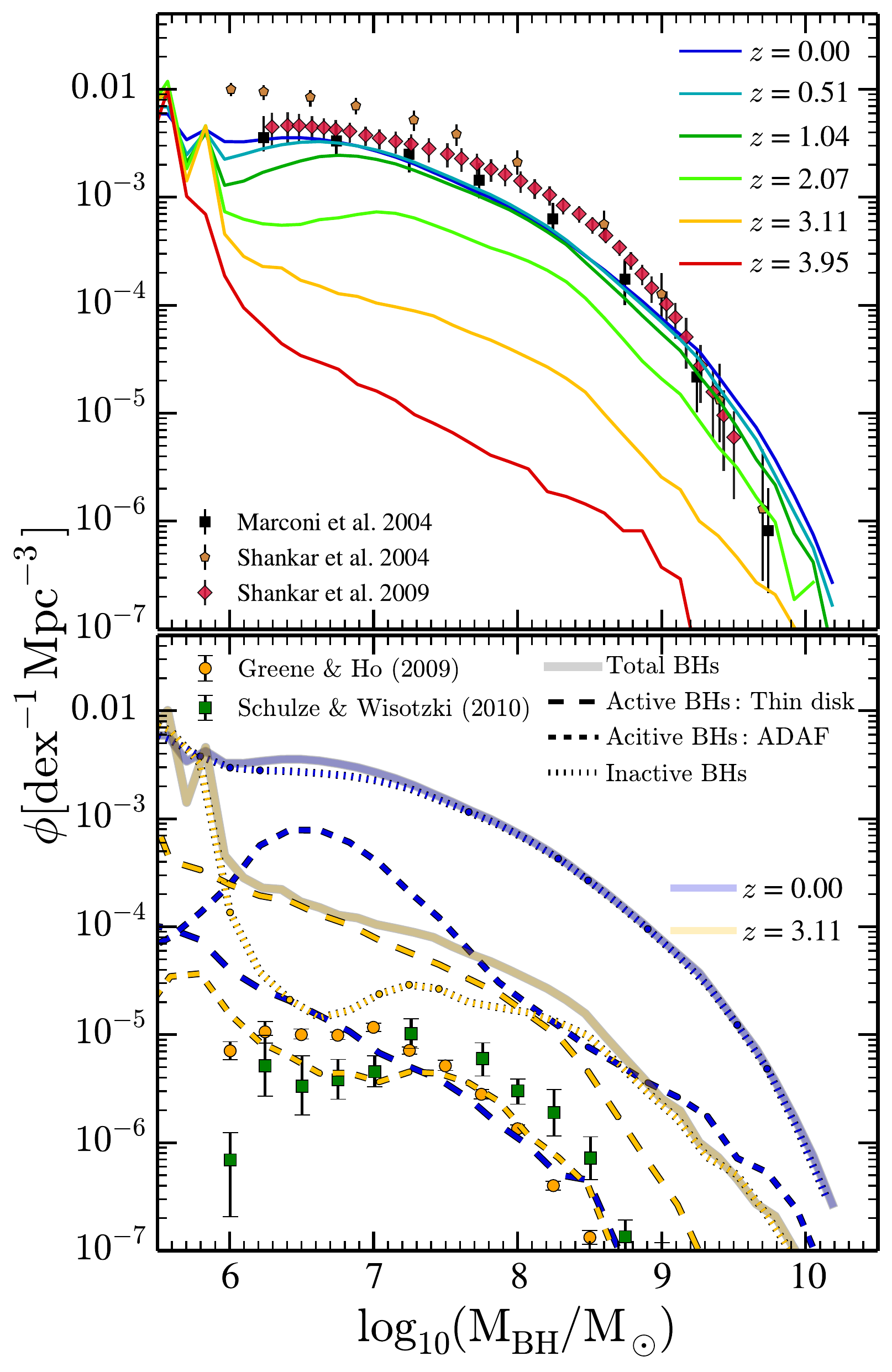}
    \caption{\textbf{Upper panel}: Redshift evolution of the black hole mass function compared to the observational results of \protect \cite{Marconi2004} and \protect \cite{Shankar2004,Shankar2009}. \textbf{Lower panel}: Black hole mass function at $z\,{\sim}\,0$ and $3$ (thick blue and yellow lines, respectively) divided between inactive (\fedd $\,{<}\,{10^{-4}}$, dotted lines) and active (\fedd $\,{>}\,{10^{-4}}$) black holes, where the latter are split in BHs accreting in the thin disk (\fedd${>}\,\rm \mathit{f}_{Edd}^{crit}$, dashed lines) and ADAF ( $10^{-4}\,{<}\,$\fedd${<}\,\rm \mathit{f}_{Edd}^{crit}$, short dashed lines) phase. The black hole mass function of active BHs at $z\,{\sim}\,0$ from \protect\cite{Greene2007} and \protect \cite{Schulze2010} are added for comparison.}
\label{fig:BH_mass_function}
\end{figure}

The evolution of the black hole mass function (BHMF) between $z\,{=}\,0$ and $z\,{=}\,4$ is shown in the upper panel of Fig.\ref{fig:BH_mass_function}.  There is a significant evolution from $z\,{\sim}\,4$ up to $z\,{\sim}\,2$, where the BHMF amplitude increases up to $\rm 2\,dex$. Conversely, at $z\,{<}\,1$ the BHMF does not evolve significantly. Such small evolution at low $z$ indicates that most of the local BHs were already assembled by $z\,{\sim}\,1$. Similar behavior is found in the theoretical work of \cite{MerloniANDHeinz2008} where a very weak evolution in the BHMF is seen below $z\,{\sim}\,1$ and, as we find here, most of it happens at small masses ($\rm {<}\,10^7\, M_{\odot}$). On the semi-analytical side, works such as \cite{Fanidakis2011,Griffin2018} or \cite{Marshall2019} show a stronger redshift evolution at $z\,{\lesssim}\,1$ where the normalization of the BHMF increases relatively fast at $\rm {>}\,10^8 \, M_{\odot}$ and decreases at $\rm {\lesssim}\,10^8 \, M_{\odot}$. On the other hand, the results of \cite{Hirschmann2012} show a similar weak evolution in the BHMF at $z\,{\sim}\,1$. %{\bf Silvia: would be good to mention how this compares with other theoretical results}. 
In the same figure, we have compared the predictions at $z\,{=}\,0$ with the observational constrains provided by \cite{Marconi2004} and \cite{Shankar2004,Shankar2009}. Even though the BHMF of the model is compatible with the observations, mainly with \cite{Marconi2004}, it displays a smaller amplitude when compared with \cite{Shankar2004,Shankar2009}. %{\bf[AS: But this is strictly true only for the Shankar observations, the results of the model seem to be quite consistent with Marconi 2004]}. 
As pointed out by \cite{Shankar2016}, this discrepancy might be caused by biases effects affecting the observations. \cite{Shankar2016} showed that, because of selection effects\footnote{For instance, they reported that local galaxies with black hole mass estimate with a dynamical approach are a biased subset of all galaxies, i.e at fixed stellar mass BHs display larger velocity dispersion that the bulk of the population, regardless of the exact morphological type.}, the normalization of the scaling relations used to link the BH mass with galaxy properties (bulge velocity dispersion, total stellar mass and bulge mass) is increased by a factor ${\gtrsim}\,3$ \citep[see also][]{Bernardi2007,Shankar2019}. The lower amplitude in the empirical relations pointed out by \cite{Shankar2019} yields smaller BH masses, lower black hole mass density and higher radiative efficiencies which would support rapidly spinning BHs and lower levels of gravitational wave emission, consistent with the current non-detection of this signal by pulsar timing array experiments \citep[see][]{Sesana2016}.\\ %{\bf detections - is this correct -S- also, shall we then remove Shankar 2013??} \\

%Lower panel of Fig.\ref{fig:BH_mass_function} presents the BHMF at $z\,{\sim}\,0$ and $3$ divided by active (\fedd $\,{>}\,{10^{-4}}$) and inactive BHs (\fedd $\,{<}\,{10^{-4}}$). As shown, the inactive population increases towards low redshifts, being the most massive BHs the ones that became non-active earlier. Whereas at $z\,{\sim}\,3$ inactive BHs have typical masses of $\rm {>}\,10^8\, M_{\odot}$, at $z\,{\sim}\,0$ they are characterized by $\rm {>}\,10^6\, M_{\odot}$. Regarding the type of accretion geometry, think disk (\fedd${>}\,\rm \mathit{f}_{Edd}^{crit}$) is the dominant one at $z\,{\sim}\,3$. On the other hand, at lower redshifts ADAF geometry ($10^{-4}\,{<}\,$\fedd${<}\,\rm \mathit{f}_{Edd}^{crit}$) takes more important and it becomes the principal one at $z\,{\sim}\,0$.\\ 

\begin{figure*}
	\centering
	\includegraphics[width=1.\columnwidth]{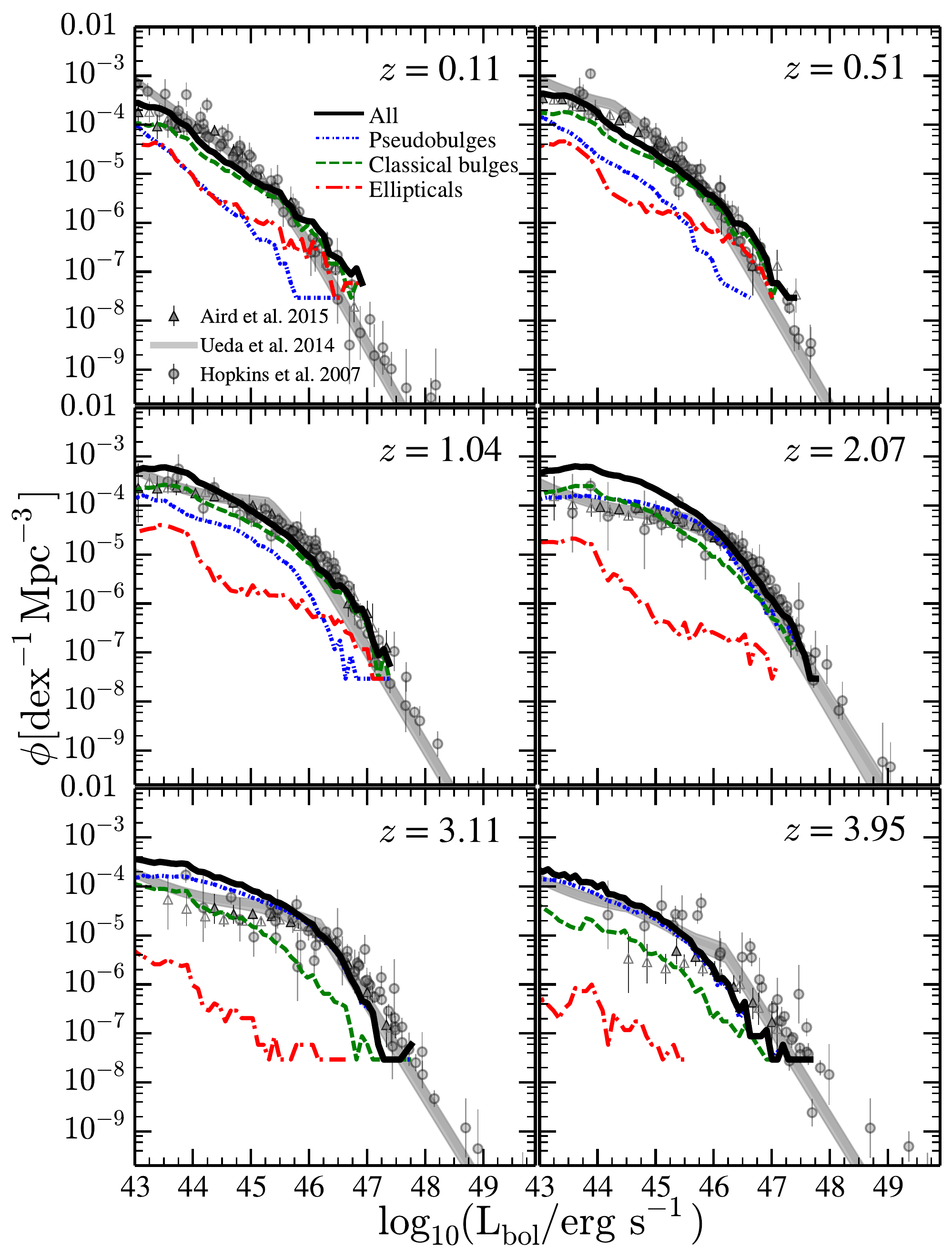}	
	\includegraphics[width=1.\columnwidth]{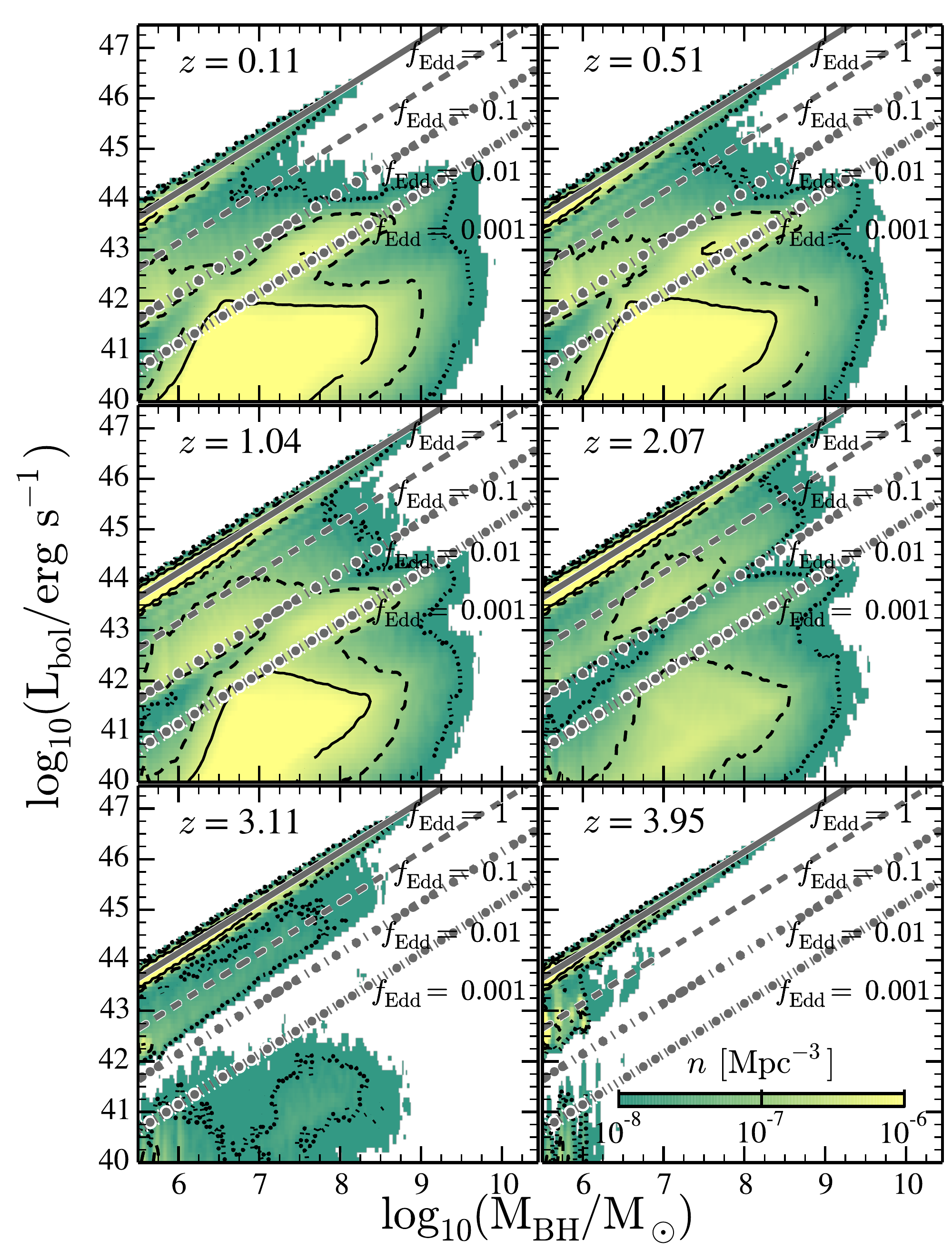}
    \caption{\textbf{Left panel}: Quasar bolometric luminosity functions ($\rm L_{bol}$) at $z\,{\approx}\,0.1, 0.5, 1.0, 2.0, 3.0, 4.0$. Solid lines represent the total quasar bolometric luminosity functions. Red dashed-dotted line, green dashed line, and blue dotted lines represent the same but for galaxies hosting respectively elliptical, classical bulge, and pseudobulge bulge structures. Luminosity functions are compared with the data of \protect\cite{Hopkins2007} (circles), \protect\cite{Aird2015} (triangles) and the fit of \protect\cite{Ueda2014} (shaded area). \textbf{Right panel}: Bolometric luminosity ($\rm L_{bol}$) - black hole mass ($\rm M_{BH}$) plane at $z\,{\approx}\,0.1, 0.5, 1.0, 2.0, 3.0$, and $4.0$. The color map encodes the number density ($n \, \rm [Mpc^{-3}]$) of objects, with solid, dashed and dotted contours indicating the regions with $n$ equal to $5{\times}10^{-7} \, \rm Mpc^{-3}$, $10^{-7} \, \rm Mpc^{-3}$, and $10^{-8} \, \rm Mpc^{-3}$, respectively. Diagonal lines represent 1, 0.1, 0.01, and 0.001 \fedd limits.} %The data of \protect\cite{Aird2015} has been converted from hard X-ray (filled triangles) and from soft X-rays (empty triangles) by using the bolometric correction provided in \protect\cite{Marconi2004}.}
\label{fig:LF}
\end{figure*}

The lower panel of Fig.\ref{fig:BH_mass_function} presents the BHMF of \textit{active} and \textit{inactive} BHs at $z\,{\sim}\,0$ and $3$. We consider as threshold between active and non-active the Eddington ratio \fedd ${\sim}\,{10^{-4}}$ \citep{Rosas-Guevara2016}, i.e below this value the emission of the BH is essentially undetectable against the emission of the host galaxy. Even though this threshold is quite arbitrary, we did not find significant differences in the results assuming a threshold between $10^{-3}\,{-}\,10^{-5}$. The inactive population increases from $z\,{\sim}\,3$ to $z\,{\sim}\,0$, with the most massive BHs becoming non-active at an earlier epoch:  while at $z\,{\sim}\,3$ inactive BHs have typical $\rm M_{BH} \, {>}\,10^8\, M_{\odot}$, at $z\,{\sim}\,0$, black holes with masses $\rm M_{BH}\,{>}\,10^6\, M_{\odot}$ are inactive. This behaviour is usually referred to as \textit{downsizing}, and it has been reported by observational works  \citep{Merloni2004,Heavens2004,Hasinger2005}, semi-analytical models \citep{Bonoli2009,Fanidakis2012,Hirschmann2012} and cosmological hydro-simulations \citep{Sijacki2009,Rosas-Guevara2016,Thomas2019}. In the same figure we divided the active population by accretion geometry. We assume that the emission of active BHs with \fedd${>}\,\rm \mathit{f}_{Edd}^{crit}$ can be described by the thin and radiatively-efficient Shakura-Sunyaev disk model \citep[][\textit{thin disk geometry}, hereafter]{ShakuraSunyaev1973}. On the other hand, the emission of active BHs with $10^{-4}\,{<}\,$\fedd${<}\,\rm \mathit{f}_{Edd}^{crit}$ are associated to a thick and radiatively-inefficient accretion disk to which we will refer as a \textit{Advection Dominated Accretion Flow} geometry (or just ADAF geometry, \citealt{Rees1982,NarayanYi1994}). As shown, regardless of the BH mass, $z\,{\sim}\,3$ active BHs are characterized by a thin disk geometry fuelled by Eddigton limited accretion flows (see Fig.~\ref{fig:LF}). This picture changes at $z\,{\sim}\,0$, where ADAF becomes the main accretion geometry for BHs with $\rm M_{BH}\,{>}\,10^6 M_{\odot}$. In this case, BHs are powered by the quiescent phase of cold gas accretion and the consumption of hot gas which surrounds its host galaxy (see right panel of Fig.~\ref{fig:LF}). We refer the reader to Section~\ref{sec:BHaccretion_model}, where a detailed description of how \fedd values are computed in our BH growth model. 

Finally, we have also explored the evolution of the black hole masses in the chaotic scenario proposed by \cite{King2005}\footnote{Following \cite{Sesana2014}, the chaotic scenario presented in \cite{King2005} is obtained fixing $F\,{=}\,1/2$ in Eq.\eqref{eq:npa}.}. For the sake of the present discussion, we only compare the main predictions of this \textit{chaotic} model to our findings. We have seen that, as a consequence of the different outcomes in the spin distribution, there is a faster BH assembly in the chaotic model than in the one explored in this work. Whereas the former display typically spin values of $a\,{\lesssim}\,0.1$, the latter is characterised by $a\,{\gtrsim}\,0.7$. As shown by Eq.\eqref{eq:eta}, the spin has an impact in the mass-to-energy conversion, implying longer timescales in the BH growth at larger spin parameters (see Eq.\eqref{eq:growth_rate}). %This spin effect is especially important for the case of $z\,{>}\,6$ BHs since they have grown up to $\rm M_{BH}\,{\gtrsim}\,10^9\,M_{\odot}$ by gas accretion in less than a $\rm 1\,Gyr$ \citep{Fan2001,Willott2007}. 
We have also checked that the fast growth undergone by the BHs in the chaotic scenario has an imprint in the faint end of $z\,{\lesssim}\,1$ bolometric luminosity functions (LFs). Since the BH growth in the chaotic scenario happens on a shorter time-scale than in the model presented here, the BHs tend to consume the gas reservoir faster, hence entering to the quiescent phase earlier. This implies that BHs at low $z$ do not have enough gas in order to effectively contribute to the faint-end of the bolometric luminosity function, reducing its amplitude.  
%Since the growth happens in a short time scale, the BHs consume the gas reservoir fast enough to enter in the quiescent phase at higher redshifts than in the model linked with the bulge. Thus, BHs run out of gas at low-$z$, loosing active BHs that contributed to the faint end of the AGN bolometric LF. 
In order to contrast this effect and match the observed faint-end of the AGN bolometric LF, the parameters $f_{\rm BH}^{\rm merger}$ and $f_{\rm BH}^{\rm DI}$ (Eq.\eqref{eq:QuasarMode_Merger}-\eqref{eq:QuasarMode_DI}) controlling the amount of gas accreted during mergers and DI, had to be increased. This resulted in a significant excess on the predicted BHMF at $\rm M_{BH}\,{>}\,10^8\,M_{\odot}$, making difficult its comparison to observations.\\

%To obtain a better match to the faint end of bolometric LF in the chaotic accretion it has been necessary to increase the gas accreted during mergers and DI (parameters $f_{\rm BH}^{\rm merger}$ and $f_{\rm BH}^{\rm DI}$ in Eq.\eqref{eq:QuasarMode_Merger}-\eqref{eq:QuasarMode_DI}) with the side effect of obtaining a larger amplitude in the BHMF at $\rm M_{BH}\,{>}\,10^8\,M_{\odot}$, making more difficult to match the observed BHMF.\\% {\bf[AS: It is not clear whether this is a problem or not. Is it more difficult to match the BH mass function in the chaotic scenario or not?]}\\

\subsection{The evolution of bolometric luminosity}

Since the discovery of the first quasar \citep{Schmidt1963}, it has been widely accepted that the quasar phase is triggered by the gas accretion onto BHs. The understanding of the quasar luminosity function at different redshifts is a crucial point for inferring the assembly history of BHs. They provide us with information about the BH growth rate, the nature of accretion disks and fundamental quantities such as BH spins and radiative efficiencies.
%{\bf Silvia: Let's start by saying why the Lbol is important.... we can discuss this. }
In the left panel of Fig.\ref{fig:LF} we present the evolution of the bolometric luminosity ($\rm L_{bol}$) function predicted by our model (in Appendix~\ref{Appendix:XRaysLFs} it is presented the same but for hard and soft X-rays luminosity). The predictions of our model are compared to the observational work of \cite{Hopkins2007,Aird2015} and \cite{Ueda2014}. A good agreement is achieved at any redshift bin. However, at $z<0.5$ our model slightly underestimates the number of objects in the range of $\rm 10^{44}{\lesssim}\,L_{bol}\,{\lesssim}\,10^{45}\, erg/s$. This might be caused by other feeding processes which are not accounted by our model. For instance, \cite{Volonteri2013} discussed that substantial  growth of BHs in ellipticals galaxies was due to the consumption of recycled gas of the evolving stellar population. In future works, we plan to include this extra growth channel in both elliptical and spiral galaxies. In the same figure, we have divided the luminosity functions in three types of bulge population: galaxies hosting a pseudobulge, classical bulge and elliptical structure. While at high-$z$ ($z\,{>}\,2$) the bolometric luminosity function is dominated by BHs accreting in pseudobulge structures, at lower redshifts classical bulges and elliptical galaxies are the main structures hosting active BHs. At $z\,{<}\,1$, pseudobulges host only faint AGNs ($\rm L_{bol}\,{\lesssim}\,10^{44} \, erg/s$). Since pseudobulges form in galaxies experiencing a quiet merger history \citep[see][]{IzquierdoVillalba2019}, our model points out that high-$z$ AGNs ($z\,{\gtrsim}\,2$) and a fraction (${\sim}\,10{-}20\%$) of intermediate luminous AGNs ($\rm L_{bol}\,{\sim}\,10^{44-45}\, erg/s$) at $z\,{\lesssim}\,1$ are mainly triggered by secular processes rather than galaxy encounters. This is in agreement with recent observational and theoretical results. For instance, the observational work of \cite{Allevato2011} pointed out that major or minor mergers alone are not able to reproduce the high bias factors of X-ray AGN selected sample. Similar results were obtained in \cite{Marian2019} for X-ray and optical selected AGNs at $z\,{\sim}\,2$. On the theoretical side, \cite{Martin2018}, by using the \texttt{Horizon-AGN} hydrodynamical simulation, found out that only $35\%$ of today's BH mass is directly attributable to merger-driven gas accretion. Similar results were found by \cite{Steinborn2018}, which using large-scale cosmological hydrodynamic simulations from the \texttt{Magneticum Pathfinder} set, concluded that mergers could not be the statistically prevalent fuelling mechanism for nuclear activity at $z\,{=}\,0-2$, except for very luminous AGNs, with $\rm L_{bol}\,{>}\,10^{46}\,erg/s$. From the semi-analytical models perspective, recent works of \cite{Lagos2008,Fanidakis2011,Griffin2018} and \cite{Marshall2019} found that, regardless of redshift, DIs are the main drivers of BH growth and AGN activity. Even though all these works use the same analytical prescription to detect instabilities in the galactic disk (see Eq.~\ref{eq:instability_criterium}), their implementation is slightly different from the one used here. For instance, the model of \cite{Fanidakis2011} (and \citealt{Griffin2018}) assumes that any disk instability event is able to destroy the galactic disk, transforming the galaxy into a pure spheroidal structure and triggering a burst of star formation. On the other hand, the works of \cite{Lagos2008} and \cite{Marshall2019} follows the approach presented here transferring to the galaxy spheroidal component only the amount of stellar mass needed to restore the disk stability. However, they do not distinguish between their triggering mechanism, as we do here.\\ %{\bf Silvia: But we have to say what is included in the DI in those works... }\\

To directly explore the relation between bolometric luminosity and BH mass, in the right panel of Fig.\ref{fig:LF} we show the black hole mass-bolometric luminosity  plane at six different redshifts.  At $z\,{\gtrsim}\,3$ there is a tight correlation between black hole mass and luminosity, as almost all  BHs are accreting at \fedd${=}\,1$. On the contrary, at $z\,{<}\,2$ most BHs are far from growing at the Eddigton limit,  and diverse combinations between BH mass and \fedd are present. For instance, whereas at $z\,{\sim}\,4$ all the quasars shining at $\rm L_{bol}\,{\sim}\,10^{45}\, erg/s$ are triggered by BHs of mass $\rm M_{BH}\,{\sim}\,10^7\, M_{\odot}$ with \fedd${=}\,1$, at $z\,{\lesssim}\,3$ the same luminosity is triggered by a wide range of BHs ($\rm M_{BH}\,{\sim}\,10^7-10^9$ $\rm M_{\odot}$) accreting at \fedd${\sim}\,1-\,0.01$. Only the highest luminosity ($\rm L_{bol}\,{\gtrsim}\,10^{46}\, erg/s$) are reached always by the same type of objects:  BHs of mass $\rm M_{BH}\,{\gtrsim}\,10^8\, M_{\odot}$ accreting at the Eddignton limit. Interestingly, the plane $\rm L_{bol}\,{-}\,M_{BH}$ displays two branches: one at \fedd${\gtrsim}\,10^{-3}$ and the other at \fedd${\lesssim}\,10^{-3}$. While the former is caused by cold gas accretion, the latter is originated by hot gas consumption. At $z\,{>}\,3$ these two regimes are  clearly separated, whereas at $z\,{\sim}\,2$ the branches start to blend. This is  because BHs accreting via the cold gas phase enter in the quiescent (or self-regulated) phase, characterized by low \fedd. Eventually, as we discussed in Fig.\ref{fig:LF} left panel, at $z\,{\sim}\,0$ the number density of active BHs is dominated by low \fedd values (or ADAF geometries) caused by both hot and cold gas accretion.

%Regardless redshift, very bright quasars, i.e , are always triggered by very massive black holes ($\rm M_{BH}\,{\sim}\,10^8\, M_{\odot}$) accreting at Eddington limit. On contrary, less bright quasars are triggered by different massive BHs at different accretion limits. For instance, quasars with $\rm L_{bol}\,{\sim}\,10^{45}\, erg/s$ at $z\,{\sim}\,3$ are triggered by BHs of $\rm M_{BH}\,{\sim}\,10^6\, M_{\odot}$ accreting at Eddington limit and more massive BHs ($\rm M_{BH}\,{\sim}\,10^6\, M_{\odot}$) consuming gas at low Eddington limit (\fedd${\lesssim}\,0.1$). As we decrease on redshift, the same range of luminosity is triggered by similar with different accretion limits. For instance, at . As is it shown in the figure, there is two different branches (more evident at $z\,{\sim}\,3$). These two branches are caused by the cold and hot phase accretion. The branch closer to Eddington limits is caused by the BH accretion of cold gas. On contrary XXX. By $z\,{\sim}\,2$ the two branches start to merger. This is caused because a BHs accreting in the cold gas phase enter in the quiescent (or self-regulated) phase, characterized by low Eddington limits. Eventually, as we discussed in Fig XXX, at $z\,{\sim}\,0$ the number density of active BHs are characterized by low \fedd values

\subsection{The evolution of black hole spin}

\begin{figure}
	\centering
	\includegraphics[width=1.0\columnwidth]{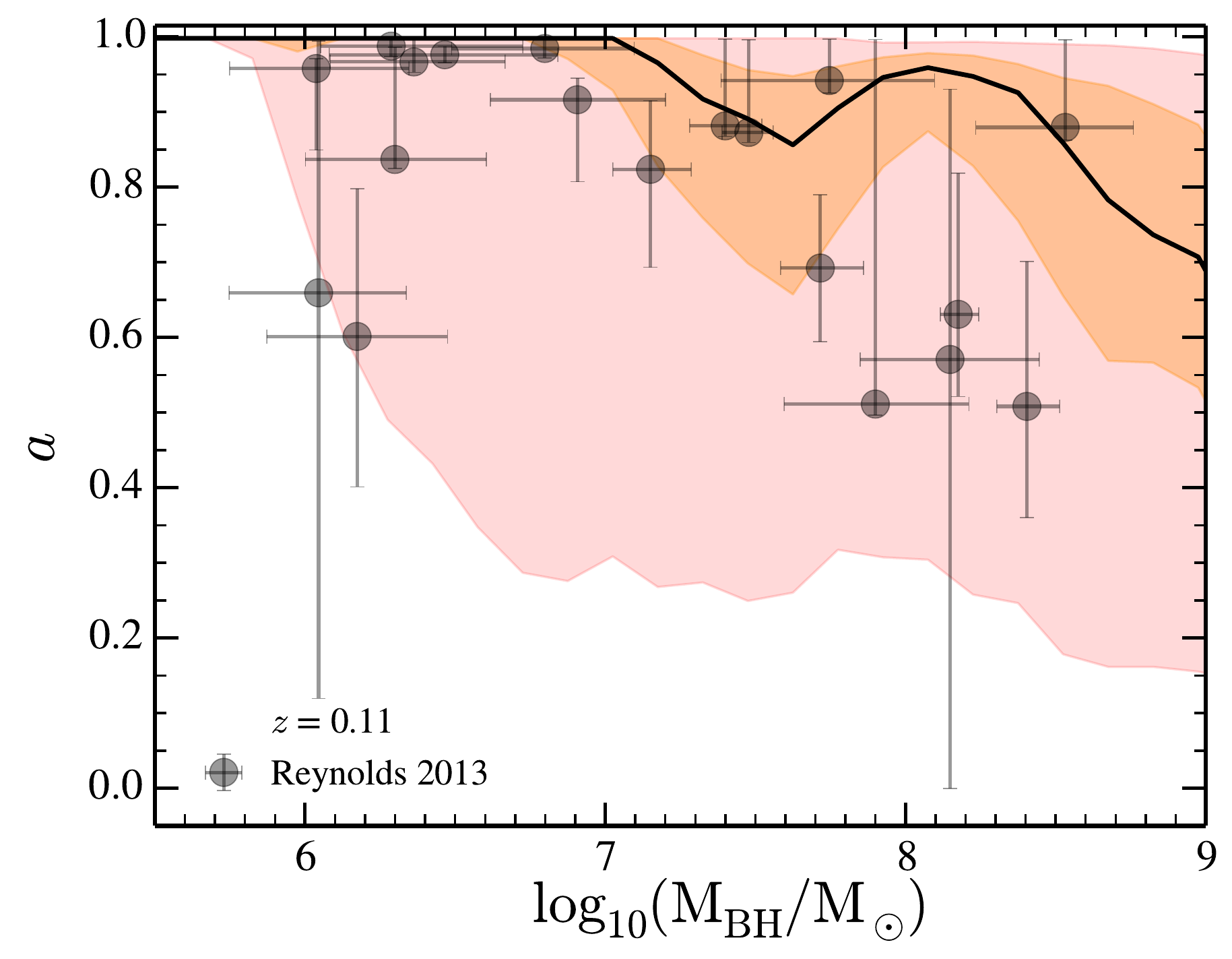}
   \caption{Black hole spin, $a$, as a function of black hole mass ($\rm M_{BH}$) at $z\,{=}\,0$ for black holes with hard X-ray luminosity $\rm L_{Hx}{>}10^{42} erg/s$. The solid black line represents the median value of the spin, whereas dark and light orange shaded areas display the $1\sigma$ and $2\sigma$ of the distribution. Black dots are the observational data of \protect\cite{Reynolds2013}.}
\label{fig:spin_z0}
\end{figure}

\begin{figure*}
	\centering
	\includegraphics[width=1.0\columnwidth]{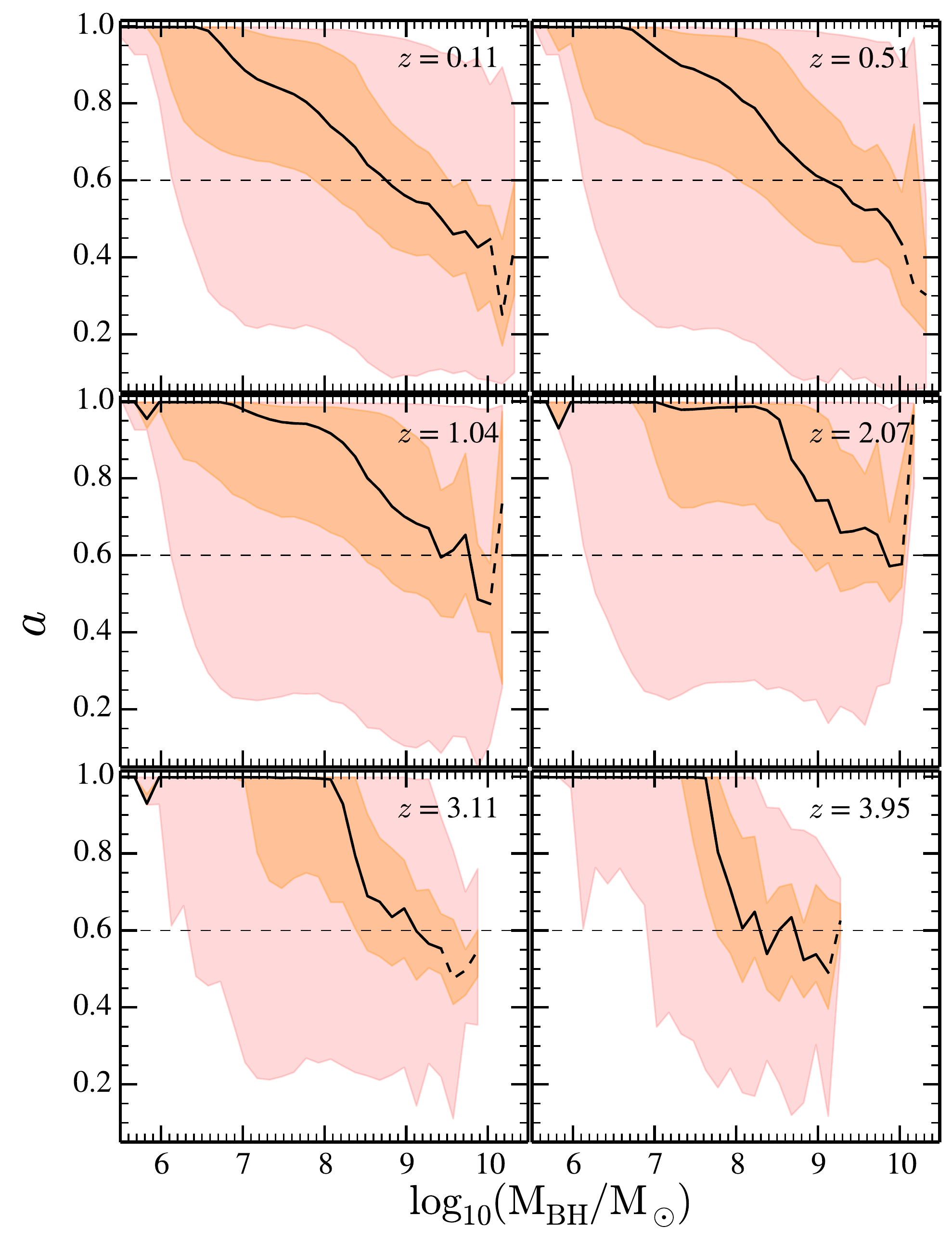}
	\includegraphics[width=1.0\columnwidth]{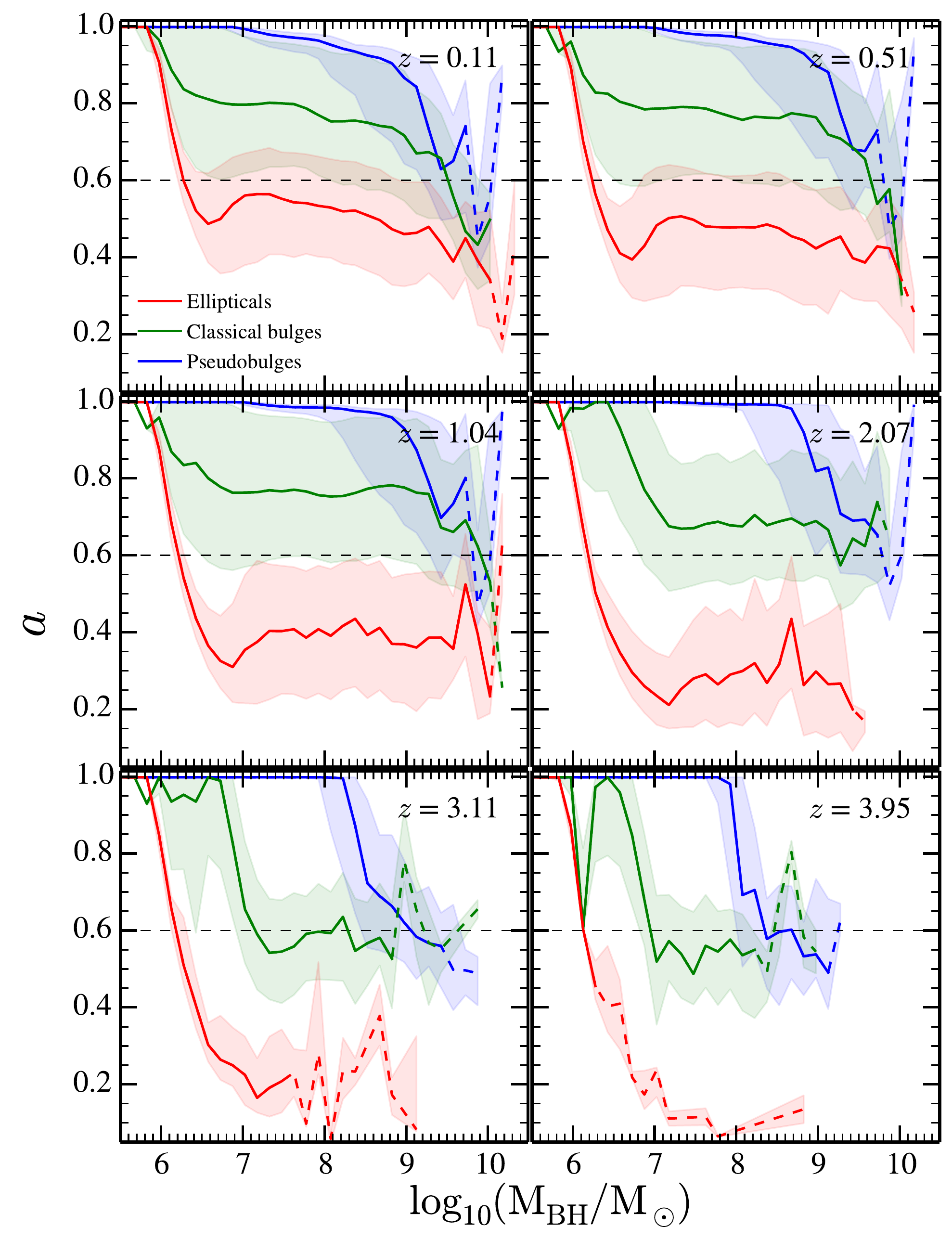}
    \caption{\textbf{Left panel}: Predicted black hole spins, $a$, as a function of black hole mass ($\rm M_{BH}$) and at different reshifts. Solid black lines represent the median value of the spin per black hole mass. Dark and light orange shaded areas display the $1\sigma$ and $2\sigma$ of the distribution. To guide the reader we have highlighted with dashed lines the value of $a\,{=}\,0.6$ (corresponding to an accretion efficiency ${\sim}\,0.1$). \textbf{Right panel}: The same as in the left panel but dividing the galaxies between classical bulges (green), pseudobulges (blue) and ellipticals (red). Shaded areas display the $1\sigma$ of the distribution. The large fluctuations at high masses are due to small sample statistics (in both panels dashed lines in the median relation corresponds to the black hole masses where the number of objects is smaller than 5).}
\label{fig:spin_BH}
\end{figure*}

%{\bf Silvia: some intro... }
Apart from mass, the spin parameter is the other fundamental property of BHs that has to be taken into account to build a complete picture of how black holes assemble their masses across cosmic time, given also the dependence of the radiation efficiency on the spin value.  %As discussed in Section~\ref{sec:Spin_evolution}, the spin has an imprint in the BH growth time scale implying a longer BH growth curve at larger spin value. 
From an observational point of view, spin measurements are still a challenge and the current estimates display large errors.  \cite{Reynolds2013} present a compilation of few BHs with reliable measurements of both spin and mass. However, all the spin values have been computed via X-ray spectroscopy of iron $\rm K\alpha$ line, biasing the results towards AGNs with bright hard X-rays luminosity (${\gtrsim}10^{42}\, \rm erg/s$, see Table 2 in \citealt{Sesana2014}). As a check of the spin predictions in \LGalaxies, we present in Fig.\ref{fig:spin_z0} a comparison of the model with \cite{Reynolds2013} data. In order to perform a fair comparison we have only selected in our sample the BHs with hard X-ray luminosity larger than $10^{42}\,\rm erg/s$ (see Appendix~\ref{Appendix:XRaysLFs} for hard and soft X-rays LFs). We see that the model predictions are overall consistent with the observations. \\

In the left panel of Fig.\ref{fig:spin_BH} we present the cosmological evolution of BH spin. The model generally predicts a rapidly spinning super-massive black hole population. While at $\rm M_{BH}\,{<}\,10^{6}\, M_{\odot}$ BHs tend to be maximally spinning, at $\rm M_{BH}\,{>}\,10^{6}\, M_{\odot}$ BHs display lower spin values \citep[see also][]{Sesana2014}, with average spin values descreasing with increasing BH mass. Also, the median spin values show a modest decrease with decreasing redshift, but only for BHs with $\rm M_{BH}\,{>}\,10^{6}\, M_{\odot}$. The different spins of small and high-mass BHs come from the ratio $|\vec{J_{d}}|/2|\vec{J}_{\rm BH}|$ during the gas consumption. At  $\rm M_{BH}\,{<}\,10^{6}\, M_{\odot}$, independently of the redshift and the nature of the accretion episode, the ratio $|\vec{J_{d}}|/2|\vec{J}_{\rm BH}|$ is always larger than one, resulting in a BH feeding characterized by $n_{Pa}\,{=}\,1$ (see Section~\ref{sec:Spin_evolution}). Conversely, for $\rm M_{BH}\,{>}\,10^{6}\, M_{\odot}$, the value of  $|\vec{J_{d}}|/2|\vec{J}_{\rm BH}|$ is not necessary larger than one and the precise number of $n_{Pa}$ depends on the galaxy bulge assembly. Indeed, this bulge dependence is seen in Fig.\ref{fig:spin_BH} when the BH population is divided according to the host bulge type (pseudobulge, classical bulge and elliptical). Regardless of the redshift, at fixed mass, BHs hosted in pseudobulges display larger spin values ($a\,{\gtrsim}\,0.9$) than the ones in classical bulges ($a\,{\sim}\,0.7$) and ellipticals structures ($a\,{\lesssim}\,0.4$). This is in agreement with the work of \cite{Orban2011}, which concluded, by observing local AGN, that BHs in local Narrow-line Seyfert I galaxies (with masses $\rm {\sim}\,10^6 M_{\odot}$) hosted in pseudobulge  structures need to be rapidly spinning in order to explain their duty cycles.
Notice that the model predicts a slight spin-up of BHs hosted in elliptical galaxies form $z\,{\sim}\,1$ ($a\,{\sim}\,0.3$) to $z\,{\sim}\,0$ ($a\,{\sim}\,0.4$). This is  because after the formation of the elliptical structure the galaxy undergoes successively minor mergers/\textit{smooth accretions} (a median value of $5$) which carry gas with large $v/\sigma$ towards the BH, ultimately causing its spin-up (see Figure 7 and 16 in \citealt{IzquierdoVillalba2019} where we show the number density of minor/\textit{smooth accretion} events and the typical redshift of the last minor merger).
This is along the line of recent observational work, pointing out that minor mergers (${\sim}\,8$ with mass ratio 1:10) contribute significantly to the build-up of massive elliptical galaxies at $z\,{\lesssim}\,1$ \citep{Boylan-Kolchin2006,Trujillo2011}.  %{\bf Silvia:mmm... this discussion should be for the number of mergers. Here we are talking about spin.}. %{\bf Silvia: David, the next paragraph is still a little confusing. For ellipticals things do not change at high masses. So, the only thing is that for classical and pseudobulges the spin decreases. We should discuss this. } 
The dependence on the bulge type is somewhat blurred for BHs with $\rm {>}\,10^8\, M_{\odot}$ inhabiting pseudo- and classical- bulge structures. This is because such massive BHs are hosted in massive galaxies ($ \rm M_{stellar}\,{>}\,10^{11} \, M_{\odot}$) whose bulge assembly is complicated, shaped by both mergers and disk instabilities. This intricate bulge assembly results in a complex evolution of the bulge $v/\sigma$, significantly deviating form the $v/\sigma$ values produced by DI or minor mergers alone. For instance, the bulge $v/\sigma$ in pseudobulges decreases after a minor merger while $v/\sigma$ in classical bulges increases after a DI. Other mechanisms that blur the spin-bulge dependence at $\rm {\gtrsim}\,10^8\, M_{\odot}$ are the \textit{gas-poor} BH-BH coalescences. In these cases the remnant BHs do not accrete gas and the spin change is driven by the BH coalescence. \cite{BertiVolonteri2008} showed that the spin evolution under only BH-BH mergers is driven towards $a{\sim}\,0.4-0.7$ (with some intrinsic dispersion).\\

\begin{figure*}
	\centering
	\includegraphics[width=1.0\textwidth]{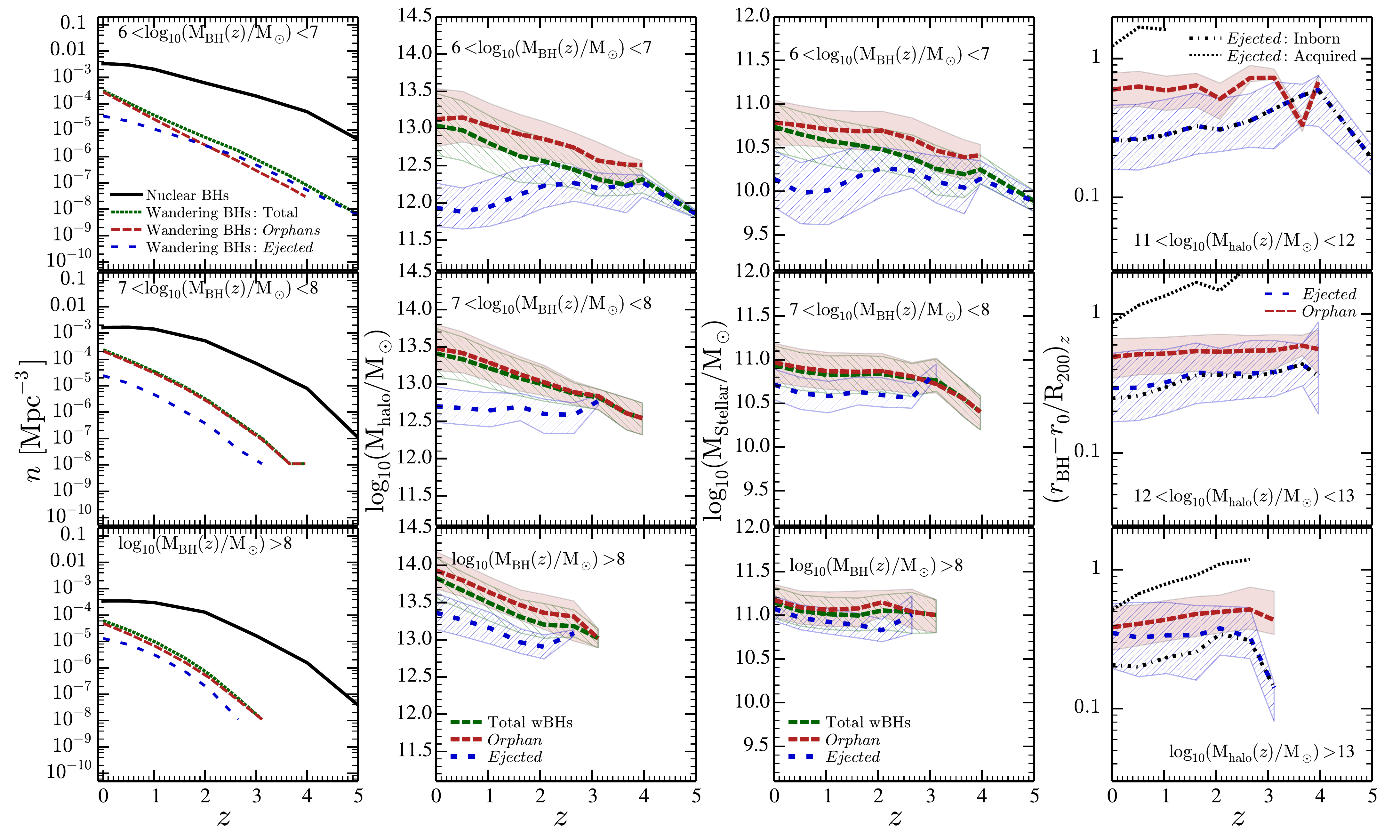}
    \caption{\textbf{Left panel}: Redshift evolution of the number density ($n$) of nuclear (thick black solid line) and wandering (thick short-dashed dark green line) BHs of in the ranges: $\rm 6\,{<}\,log_{10}(M_{BH}/M_{\odot})\,{<}\,7$ (upper panel), $\rm 7\,{<}\,log_{10}(M_{BH}/M_{\odot})\,{<}\,8$ (middle panel) and $\rm log_{10}(M_{BH}/M_{\odot})\,{>}\,8$ (lower panel). The population of wBHs has been divided into \textit{orphans} (long dashed dark red line) and \textit{ejected} (dotted dark blue line). \textbf{Central left panel}: Median subhalo mass in which wandering black holes are hosted at different redshifts. As in the left panel, each figure corresponds to a different BH mass bin. The wBH population has been divided into \textit{orphans} (long dashed dark red line) and \textit{ejected} (dotted dark blue line). The shaded areas indicate  the 1$\sigma$ dispersion. \textbf{Central right panel}: Same as the central left panel but for stellar masses. \textbf{Right panel}: Redshift evolution of the median positions of all wandering black holes with mass $\rm {>}\,10^{6}\, M_{\odot}$. The position, $r_{\rm BH}$, is given with respect to the subhalo center, $r_0$, and is normalized to the subhalo virial radius, $\rm{R}_{200}$. Each panel corresponds to a different subhalo mass bin. Blue short--dashed and red long--dashed lines represent the median positions of \textit{ejected} and \textit{orphan} wBHs respectively. The shaded areas indicate the 1$\sigma$ dispersion. We further split the ejected population into inborn and acquired, represented by the black short-dashed and blue dotted lines, respectively.} 
\label{fig:ndensity_Mhalo_position_wBHs}
\end{figure*}

During the last years several studies have addressed the evolution of BH spin by using semi-analytical models. For instance, \cite{Fanidakis2011} and \cite{Griffin2018} found out that the \textit{chaotic} scenario suggested by \cite{King2005} applied on \texttt{GALFORM} SAM (run on top of the \texttt{Millennium} and \texttt{P-Millenium} merger trees) yields a population of BHs characterized by low spin values ($a\,{\lesssim}\,0.4$) at any mass and redshift (see Figure 9 of \citealt{Fanidakis2011} and \citealt{Griffin2018}). Even though the chaotic scenario was able to reproduce the population of radio-loud AGNs in the local Universe, its outcome spin distribution makes difficult reconciling the model predictions with the recent claims of a rapidly spinning black hole population \citep{Orban2011,Reynolds2013,Trakhtenbrot2014,Shankar2016}. In the same work they also explored a \textit{prolonged} accretion scenario. Although in this case the model predicts BHs with $a\,{>}\,0.9$, the observed LFs are systematically above the observed one and its consistency with radio-loud AGNs in the local Universe is worse than in the chaotic scenario (see Figure 10 of \citealt{Fanidakis2011} and Figure 16 of \citealt{Griffin2018}). Other attempts have been done by \cite{Barausse2012} employing a SAM build on top of Press-Schechter merger trees. By using a more sophisticated approach consisting in feeding the BH respectively in a prograde and chaotic way during gas rich and gas poor mergers, they found a dichotomy in the spin distribution between rapidly ($a\,{\gtrsim}\,0.9$) and low ($a\,{\sim}0.2$) spinning BHs (see Figure 14 and 15 in \citealt{Barausse2012}). \cite{Sesana2014} updated the spin evolution of \cite{Barausse2012} by linking the BH growth with the assembly of the galactic bulge. As in this work, they found different spinning black holes hosted in elliptical and spiral galaxies. Finally, \cite{Volonteri2013} also studied the spin with a SAM based on Press-Schecter merger trees. Similar what we do here, they assumed that the BH quasar phase takes place in two different phases. Whereas the former is described by an Eddington limited growth with a spin evolution characterized by a prograde accretion, the latter displays low Eddington ratios with a chaotic spin evolution (even though they explored a coherent mode as well during this quiescent phase). They found a strong spin redshift evolution whereby $z>2$ BHs are characterized by large spin values ($a\,{>}\,0.8$) whereas $z<1$ BHs display small spins ($a\,{<}\,0.4$). However, the model did not include BH feeding through disk instabilities, a channel which seems to be fundamental in SAMs \citep{Fanidakis2012,Menci2014,Marshall2019}.

\section{Wandering black holes}\label{sec:res_wandering_BHs}
In this section we discuss our main results on the wandering black hole population. We first explore the frequency and location of the events  at different cosmological times. We then  focus on the local universe, investigating the characteristics of wBHs hosted in different kinds of galaxies.

\subsection{Wandering black holes across cosmic time}

%In Fig.\ref{fig:mass_density_WBHs} it is presented the mass density of wandering black holes as a function of redshift. The total mass in nuclear black holes is added for comparison. As shown, regardless redshift, the wBHs mass density is several orders of magnitude smaller the one in nuclear BHs. From $\rm 1\, dex$ at $z\,{\sim}\,0$, up to $\rm 2\,dex$ at $z\,{\sim}\,4$. Interestingly, when the contribution is divided between \textit{orphan} and \textit{ejected} wBHs, two different behaviors are seen. Whereas at $z\,{\gtrsim}\,3$ wBHs mass density is fully dominated by ejected black holes, at $z\,{\lesssim}\,2$ orphans BHs are the ones that take more importance. Even though Fig.\ref{fig:mass_density_WBHs} gives us an idea of the total mass in wBHs, it does not provide information about the relative contribution of wBHs at different masses. 

\begin{figure*}
	\centering
	\includegraphics[width=1.0\columnwidth]{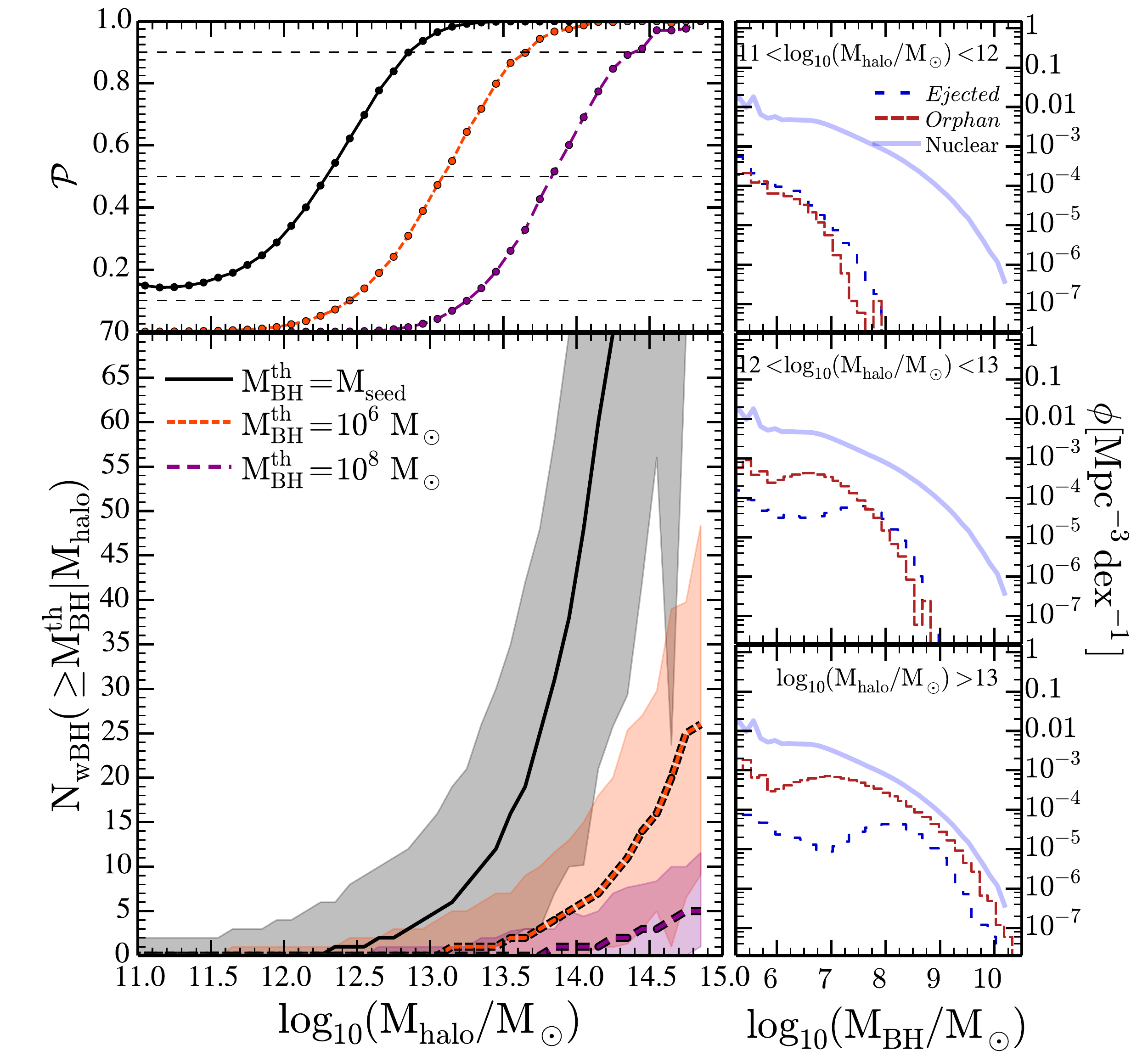}
	\includegraphics[width=0.99\columnwidth]{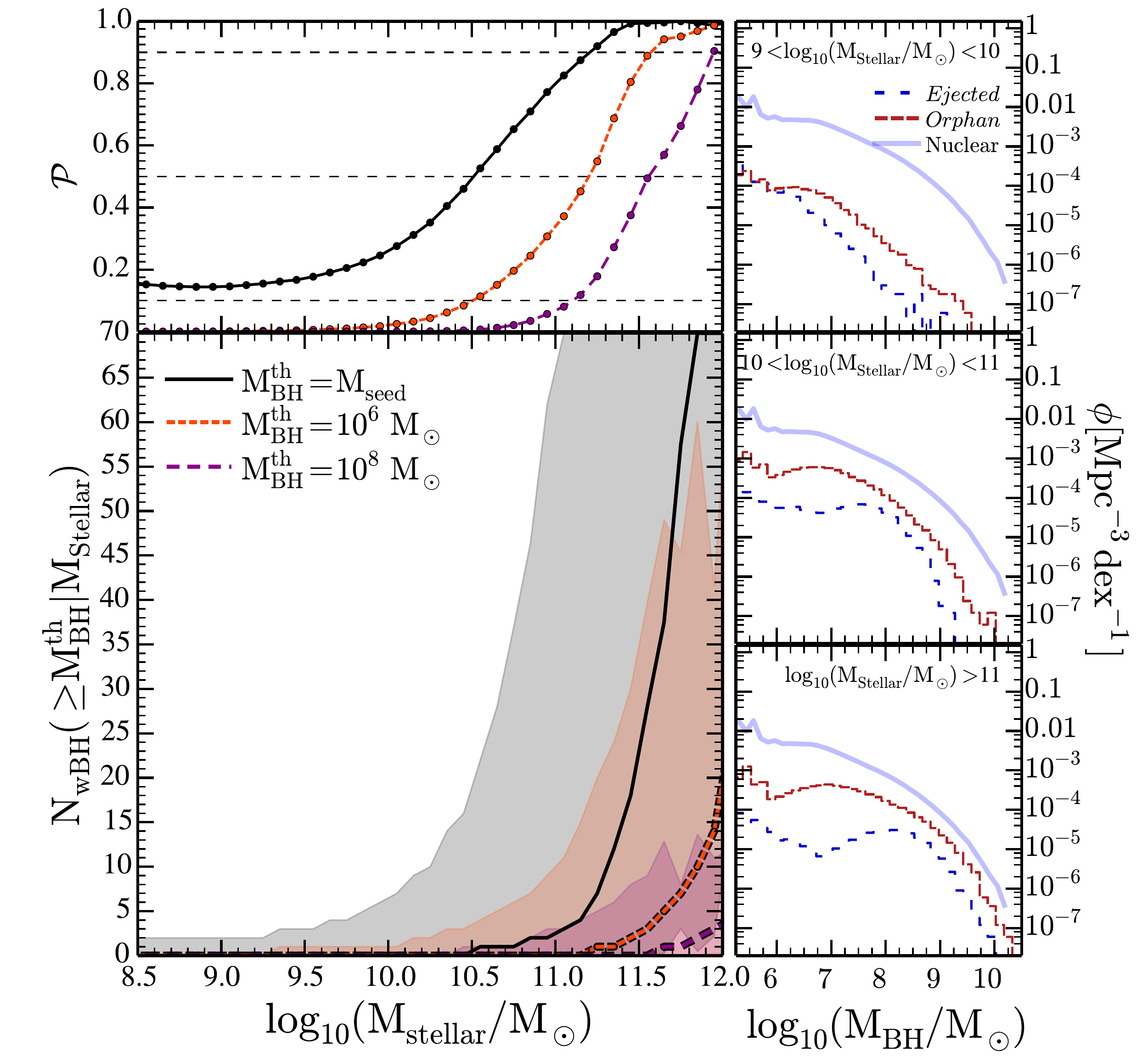}
    \caption{\textbf{Left panel}: Connection between the wandering black holes and their host subhalos at $z\,{=}\,0$.  \textit{Left upper panel}: Probability $\mathcal{P}$  at a given subhalo mass bin of finding at least one wBH above the mass threshold, $\rm M_{BH}^{th}$. Solid black line, short dashed orange line and long dashed purple line display three different $\rm M_{BH}^{th}$ thresholds: $\rm M_{seed}\,{=}\,10^{4}\, M_{\odot}$, $\rm 10^{6} \, M_{\odot}$ and $\rm 10^{8} \, M_{\odot}$, respectively. Horizontal dashed lines highlight the values $\mathcal{P} \,{=}\,0.1,0.5,0.9$ the \textit{Left lower panel}: Median number of wandering black holes, $\rm N_{wBH}$, per subhalo mass bin for a given $\rm M_{BH}^{th}$ value. Shaded areas represents the $3\sigma$ interval of the distributions. \textit{Right panels}: Wandering black hole mass function at three different subhalo mass bins. Solid blue, short red dashed and long blue dashed lines represent, respectively, the mass function for nuclear BHs, \textit{orphan} wBHs and \textit{ejected} wBHs. \textbf{Right panel}: Same as in the left panel but for stellar masses.}
\label{fig:Superplot_wBHS_halo}
\end{figure*}

The left panel of Fig.\ref{fig:ndensity_Mhalo_position_wBHs} shows the redshift evolution of the wBH number density ($n$). We show only results for  BHs whose mass is larger than $\rm 10^6 \, M_{\odot}$ since at smaller masses we can not draw solid conclusions given the resolution of the MS simulation, as previously discussed. Regardless of the BH mass, the number density evolves with redshift, reaching a maximum of ${\sim}\,10^{-3}\, \rm Mpc^{-3}$ at $z\,{\sim}\,0$. The lighter is the mass of the wBH the larger is the number density at high-$z$. For instance, while at $z\,{\sim}\,4$ black holes of $\rm M_{BH}\,{>}\,10^7 \, M_{\odot}$ have $n\,{\lesssim}\,10^{-6} \, \rm Mpc^{-3}$, black holes of $\rm 10^6\,{<}\,M_{BH}\,{<}\,10^7 \, M_{\odot}$ present $\rm 1\,dex$ larger number densities. For comparison, the figure displays the $n$ values of nuclear black holes. Regardless of mass and redshift, nuclear BHs are $\rm 1\,{-}\,3\,dex$ more numerous than wandering BHs, especially at $z\,{\gtrsim}\,4$. Concerning the different contribution between \textit{orphan} and \textit{ejected} BHs, we find that at $\rm M_{BH}\,{>}\,10^7 \, M_{\odot}$ the former dominate the number density of wBHs. On the other hand, at $\rm 10^6\,{<}\,M_{BH}\,{<}\,10^7 \, M_{\odot}$ the relative contribution of the two types of wBHs evolves with redshift; while \textit{ejected} wBHs display larger number density at $z\,{\gtrsim}\,1.5$, \textit{orphan} ones dominate at lower redshifts. This evolution is due to the fact that towards lower $z$ the  number of orphan galaxies increase, and, at the same time, the frequency of the \textit{major/minor} interactions decreases \citep[see, e.g., Figure 3.2 of][]{IzquierdoVillalba2019}. \\

%As we have shown, a population of very massive BHs ($\rm {>}\,10^8\, M_{\odot}$) is already present by $z\,{\sim}\,3$. 
In the two central panels of Fig.\ref{fig:ndensity_Mhalo_position_wBHs} the evolution of the median subhalo and stellar mass in which wandering black holes are hosted is presented. At $z\,{>}\,3$ wBHs with $\rm M_{BH}\,{>}\,10^8 \, M_{\odot}$  inhabit very massive subhalos and galaxies, with typical masses of $\rm M_{halo}\,{>}\,10^{13}\, M_{\odot}$ and $\rm M_{stellar}\,{>}\,10^{11}\,M_{\odot}$. Particularly, the host subhalo mass presents an increasing trend toward low $z$, displaying values from $\rm M_{halo}\,{\sim}\,10^{13} \, M_{\odot}$ at $z\,{\sim}\,4$ up to $\rm M_{halo}\,{\sim}\,10^{14} \, M_{\odot}$ at $z\,{\sim}\,0$. On the contrary, the stellar mass of their central galaxy hardly changes, maintaining a rather constant value of $\rm {\sim}\,10^{11}\, M_{\odot}$. WBHs with $\rm 10^7 \,{<}\,M_{BH}\,{<}\,10^{8}\, M_{\odot}$ display a similar behavior but are hosted in slightly less massive subhalos and galaxies. Finally, wBHs with $\rm 10^6\,{<}\,M_{BH}\,{<}10^7 \, M_{\odot}$ are typically placed in $\rm M_{halo}\,{\sim}\,10^{12.2}\, M_{\odot}$ and $\rm M_{stellar}\,{\sim}\,10^{10.5}\, M_{\odot}$. When the population of these wBHs is divided by formation scenario, we see that at $z\,{<}\,2$ \textit{ejected} wBHs tend to be hosted in less massive subhalo and galaxies ($\rm M_{halo}\,{\sim}\,10^{11.7}\, M_{\odot}$ and $\rm M_{stellar}\,{\sim}\,10^{10}\, M_{\odot}$, respectively) than at higher redshifts. However, the significance of this trend change is small.\\ %\textbf{Silvia: I would change this last two sentences. Once you have described the global trend, I would say:  when we divide the wBH population into orphan and ejected, we see that the general trend of increasing host halo mass with decreasing redshift is driven by the evolution of the typical halo masses hosting ejected BHs. ... wait, but looking at the left panel the orphans should actually dominate... so, why is that?}\\ %We interpret this as the result of the downsizing process that undergo the galaxy population \citep{Cowie1996} where small galaxies (with $\rm M_{BH}\,{<}\,10^7 \, M_{\odot}$) start to take the main role in the galaxy evolution and most massive ones star to quench and evolve in a very passive way until our days. \\

The position of  both \textit{ejected} and \textit{orphans} wBHs with $\rm M_{BH}\,{>}\,10^6\, M_{\odot}$ as function of redshift is presented in the right panel of Fig.\ref{fig:ndensity_Mhalo_position_wBHs}. The population has been divided in 3 different bins of subhalo masses while the distances (referred to the halo center, $r_{0}$) have been normalized by the subhalo virial radius, $\rm R_{200}$. As shown, regardless of subhalo mass, \textit{ejected} and \textit{orphan} wBHs at $z\,{>}\,0.8$ are located in different regions inside the subhalo. While the former reside at $\rm {\lesssim}\,0.3R_{200}$ (with a decreasing trend towards low $z$), the latter orbit  at $\rm {\sim}\,0.6-0.8\,R_{200}$ radii. On the contrary, at $z\,{<}\,0.8$ there are differences from a subhalo mass bin to another. While for $\rm M_{halo}\,{<}\,10^{13}\ M_{\odot}$ \textit{ejected} and \textit{orphans} wBHs still inhabit different subhalo regions, at $\rm M_{halo}\,{>}\,10^{13}\ M_{\odot}$ both types of wBHs are at similar distances from the subhalo center, $\rm {\sim} \, 0.3 \, R_{200}$. To understand this behavior, in Fig.\ref{fig:ndensity_Mhalo_position_wBHs} we divided the population of \textit{ejected} wBH in two sub-classes: the acquired and inborn ones. While the former have been incorporated from other galaxies after an subhalo merger, the latter were generated in-situ, expelled form the central galaxy of the subhalo after a recoil. A completely different behavior is seen for the two populations, regardless of the redshift. While inborn \textit{ejected} BHs are closer to the subhalo center than orphan BHs, the acquired \textit{ejected} wBHs populate similar regions, further form the subhalo center. Consequently, the change of trend of the ejected population in massive subhalos at low $z$ it is caused by the acquired \textit{ejected} wBHs, whose number becomes dominant with respect to the inborn ones.\\

\subsection{The environment of wandering black holes in the local universe}

We now focus on the local universe, exploring the typical halo and galaxy masses hosting wandering black holes as well as the frequency of wandering BHs for different galaxy type. As discussed before, we assume that wBHS are not accreting, thus not observable as active sources. In future works we will explore how simple assumptions for the growth of wBHs could translate into their observability for different host properties.\\

The mass function of wBHs at different subhalo and galaxy masses is presented in the the small panels of Fig.\ref{fig:Superplot_wBHS_halo}. For what concerns the subhalo mass, we note that  only the more massive subhalos host a significant population of wBHs, including some with mass larger than $\rm 10^7 \, M_{\odot}$. By dividing the wandering BHs by formation scenario we can see that for $\rm M_{halo}\,{\lesssim}\, 10^{13}\,M_{\odot}$ the \textit{ejected} population reaches larger masses than the orphan one, whereas at larger subhalo masses the trend is inverted. Indeed, the amplitude of the \textit{orphan} wBHs mass function increases with subhalo mass due to the increase of the average number of satellite galaxies\footnote{Typically, the average number of satellite galaxies increases with the subhalo mass in a power-law shape \citep[see e.g.][]{Berlind2003,Contreras2017}.}. Analogously, galaxies with the largest masses  are the ones which host the most massive wandering black holes, but already galaxies with $\rm M_{stellar}\,{\sim}\,10^{10}\, M_{\odot}$ can be surrounded by massive wBHs.\\
%Thus, massive halos are inhabited by a large number of \textit{oprhan} wBHs since they hosted more satellite galaxy which could be effectively disrupted, depositing their BHs in the halo.\\

The larger panels of Fig.\ref{fig:Superplot_wBHS_halo} present both the probability $\mathcal{P}$
%\footnote{The value of $\mathcal{P}$ is defined as $\rm \mathcal{P}(M_{halo,stellar})\,{=}\,N_{Galaxies}^{with\, wBHs}(M_{halo,stellar})/N_{Galaxies}^{Total}(M_{halo,stellar})$.} 
of finding at least one wBH above a given  mass threshold  $\rm M_{BH}^{th}$ (upper panel)  and the median number of wBHs ($\rm N_{wBHs}$) at a given subhalo mass (lower panel). Independently of the $\rm M_{BH}^{th}$, both $\mathcal{P}$ and $\rm N_{wBH}$ increase with the host subhalo and galaxy mass. However, the exact value of $\rm M_{BH}^{th}$ has an important effect in the values of $\mathcal{P}$ and $\rm N_{wBH}$. For instance, at $\rm M_{BH}^{th}\,{=}\,M_{seed}$ all subhalos above $\rm 10^{13}\,M_{\odot}$ (galaxies above $\rm 10^{10.75}\,M_{\odot}$) host at least one wBH, with a typical $\rm N_{wBH}\,{>}\,5$. However, only subhalos above $\rm 10^{14}\, M_{\odot}$  (galaxies above  $\rm 10^{11.5}\, M_{\odot}$) host a wBH with  $\rm M_{BH}^{th}\,{>}\,10^{8} M_{\odot}$.\\ %{\bf[AS:Check numbers in this last paragraph. For example, it seems to me that for $\rm M_{BH}^{th}\,{=}\,M_{seed}$ all  halos above $\rm 10^{13}\,M_{\odot}$ (not $10^{12.5}$) host a wBH.]}\\

\section{The imprint of gravitational recoil on the BH global properties}\label{sec:grav_rec_impr}

In this section we investigate the effects of gravitational recoil  in the properties of nuclear black holes across cosmic time. As we will show, long wandering phases can have a visible effect in the black hole occupation fraction and in the BH growth, in particular for certain galaxy types.

\subsection{Black hole occupation fraction} \label{sec:occupancyfraction}

\begin{figure}
	\includegraphics[width=1.0\columnwidth]{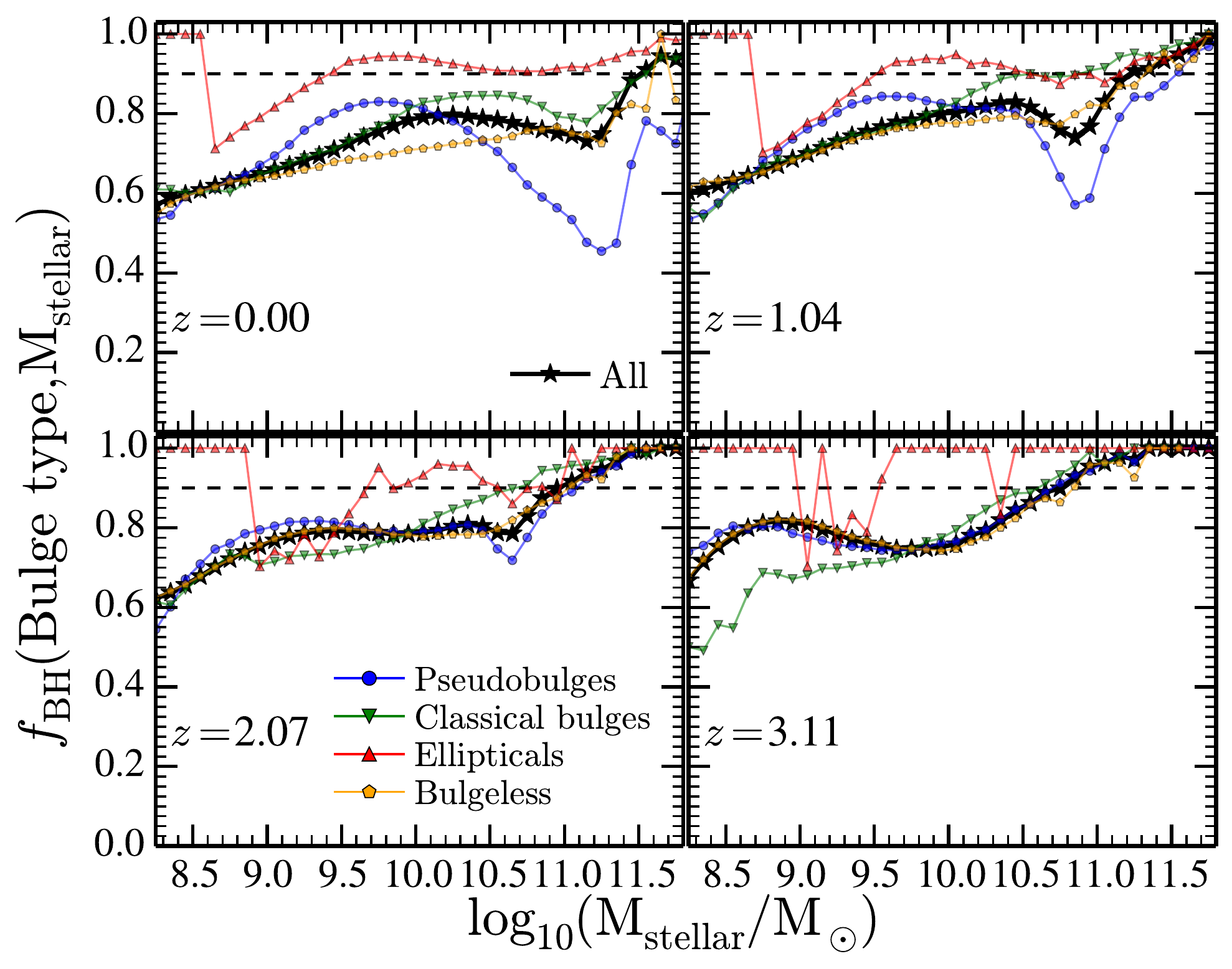}
	\centering
    \caption{Nuclear black hole occupation fraction, $f_{\rm BH}$, as a function of stellar mass and different bulge morphological types at $z\,{\sim}\,0,1,2, 3$. Black lines present the $f_{\rm BH}$ for all the galaxies. Blue, green, red and, yellow lines display the break-up into pseudobulges, classical bulges, ellipticals and bulgeless galaxies, respectively. Horizontal dashed line highlights $f_{\rm BH}\,{=}\,0.9$.}
\label{fig:galaxies_with_no_BH}
\end{figure}

The seeding procedure followed in this works (see Section \ref{sec:seeding_and_spin_init}) imply that all galaxies host a BH when they are initialized. But gravitational recoil after BH mergers can leave galaxies without a central BH, at least until the kicked BH loses the aquired energy and settles again in the galaxy center. Therefore, especially at later times, a fraction of galaxies is expected to be deprived of a central BH. In Figure \ref{fig:galaxies_with_no_BH} we show the nuclear BH occupation fraction as a function of galaxy stellar mass, and for different galaxy morphologies.  
 %, at different stellar masses, bulge morphology and redshift. %We highlight that the BH occupancy presented here should be considered as a lower limit given that our SAM predicts smaller $\rm B/T$  than the observed ones \citep[see][]{IzquierdoVillalba2019}, causing that the BHs could escape more easily from the galaxy potential. 
Regardless of the bulge morphology and redshift, $f_{\rm BH}$ decreases towards \textit{low stellar masses}: from ${\sim}\,80\%$ for $\rm M_{stellar}\,{\gtrsim}\,10^{11} M_{\odot}$ down to ${\sim}\,60\%$ for galaxies with $\rm M_{stellar}\,{\lesssim}\,10^{8.5} M_{\odot}$. This mass dependence is related to the  potential well of the galaxy: the smaller the galaxy, the lower the escape velocity, thus the higher the ejection probability for the central BH. Moreover, as we will see at the end of this section, for galaxies with $\rm M_{stellar}\,{<}\,10^9 M_{\odot}$, it is also much more probable  the ejection not only from the galaxy, but also from the subhalo. By dividing the galaxy population in different bulge morphological types, we find that, in massive galaxies with $\rm M_{stellar}\,{>}\,10^{10}\, M_{\odot}$, pseudobulges and bulgeless galaxies display lower occupation fractions than  both classical bulges and ellipticals. This is  because pseudobulges and bulgeless galaxies experience very few mergers during their evolution, thus it is more difficult for them to replenish their empty bulge after an ejection. Indeed, as we will see in Section~\ref{sec:Recoild_and_scaling_Relation} (Fig.\ref{fig:understanding_scatter_relation}), the reincorporation of BHs in a pseudobulge galaxy is not very common (${\lesssim}10\%$ of probability at any stellar mass) making it even more difficult for such structures to increase their nuclear BHs occupation fraction. %{\bf Silvia: but they can re-incorportate the ejected BH, no? Maybe we should say clearly that some of the nuclear BHs are not the re-incorportated ones, but they come from a merged galaxies. How frequent is that? } 
Besides mass and morphology, Fig.\ref{fig:galaxies_with_no_BH} shows a trend of decreasing $f_{\rm BH}$ %with decreasing redshift, especially 
towards low $z$. This is because BH-BH mergers are characterized by smaller values of $q$ (${\sim}\,0.01, 0.08, 0.1$ for pseudobulges, classical bulges and ellipticals, respectively) and poorer gas environments towards low $z$. For instance, at $z\,{\sim}\,2$, 80\% of BH-BH mergers in pseudobulges, classical bulges and elliptical galaxies are considered as wet mergers. On contrary, at $z\,{\sim}\,1.0$, $\rm {\sim}\,90\%, 60\%, 80\%$ of the BH-BH mergers in pseudobulges, classical bulges and elliptical galaxies are classified as gas poor mergers. These conditions increase the modulus of the kick velocity making it easier for a BH  to be ejected. In particular, the recoil velocity after a BH-BH coalescence in a  gas-rich environment is generally larger than in a gas-poor one since this latter is inefficient in leading the alignment between the two BH spins. On the other hand, small values of $q$ lead to larger kicks since the recoil value is proportional to the inverse of $(1\,{+}\,q)$ (see Eq.~\eqref{eq:recoil_velocity}). \\ %At $z\,{<}\,1$, $f_{\rm BH}$ displays relatively small changes caused by the decline in galaxy mergers. The main evolution is principally found at $\rm 10^{10}\,{<}\,M_{stellar}\,{<}\,10^{11}\, M_{\odot}$ due to the fact that these type of galaxies display ${\sim}\,2$ times more galaxy interactions than the rest ones \citep[see Figure 7 in][]{IzquierdoVillalba2019}. % whose number density is smaller than $\rm 10^{-4}\,Mpc^{-3}$ at $z\,{<}\,1$ in the whole MS box \citep[see Figure 7 in][]{IzquierdoVillalba2019}. \\ 
%{\bf[AS: Here I would put the discussion a bit differently, in fact it seeems to me that the occupation fraction overall decreases all the way from $z=3$ tot $z=0$, with no need to make a distinction at $z=1$.]}

\begin{figure}
	\centering
	\includegraphics[width=1.0\columnwidth]{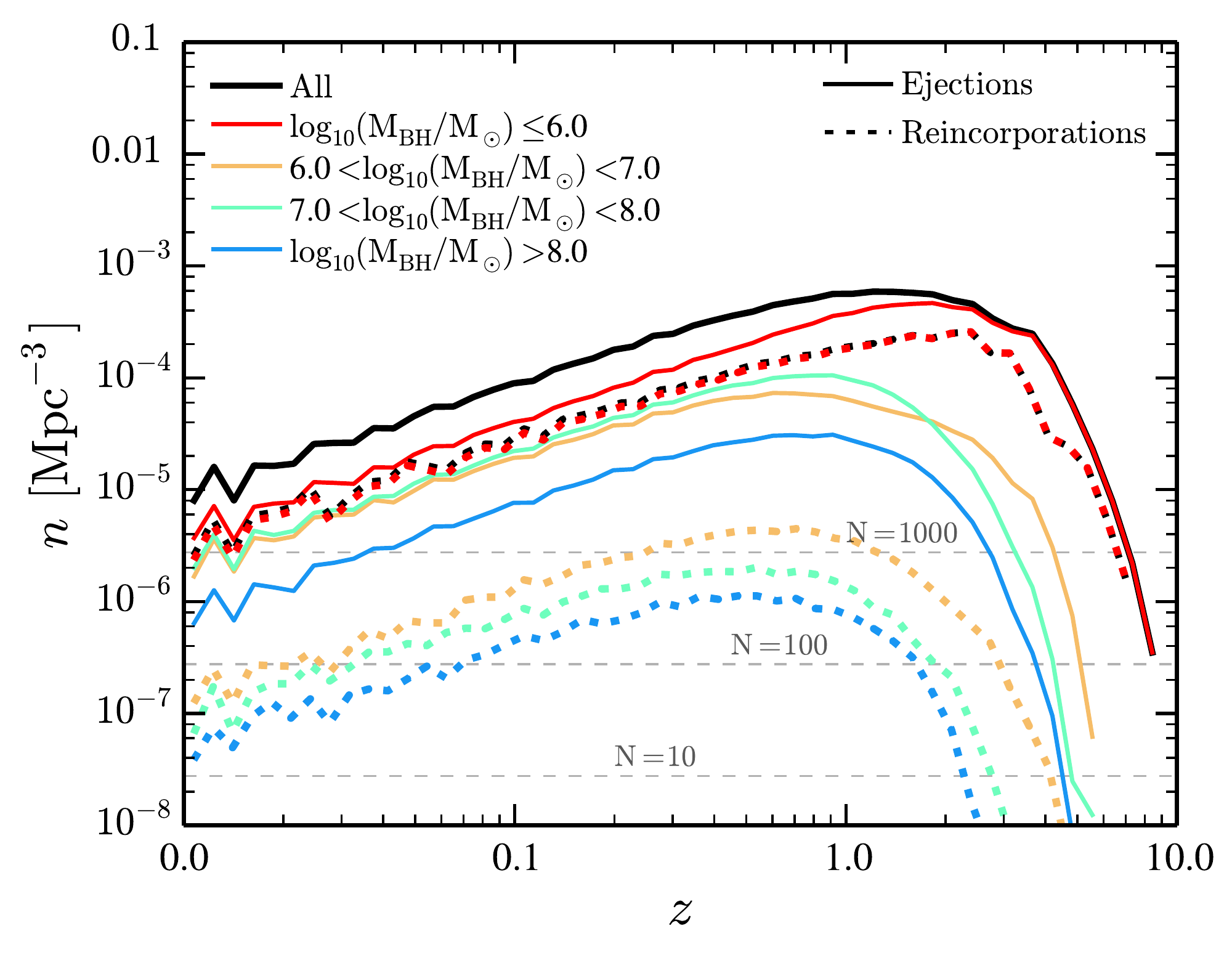}
    \caption{Redshift distribution of black hole ejections via gravitational recoils (solid lines) and reincorporation after a wandering phase (dashed lines). While black lines represent the predictions for the whole BH population, different colors are for different black hole mass bins as labeled in figure. Horizontal grey lines display the values of $n$ in which we have a number of objects ($\rm N \,{=}\,\mathit{n(z)}\,{\times}\,L_{Box}^3 $) equal to 10, 100 and 1000.} %{\bf[AS: I don't understand where those $N$ come from and how they are computed. Are those simply the correspinding number of BHs in the simulation? It is not clear.]}}
\label{fig:z_ejection}
\end{figure}

To explore in which moment the BH recoil velocities had major importance in driving ejections, Fig.\ref{fig:z_ejection} shows the comoving number density of BH ejections. Clearly, at $z\,{\sim}1.5$ the number density reaches a maximum, coinciding with the peak of the galaxy-galaxy merger frequency (see Figure 2 in \citealp{IzquierdoVillalba2019}). %{\bf Silvia: at a first glance, this seems to be in contrast with what we said before, that most ejections are at low-z. We should clarify this. What follows can help the discussion. We should stress that there are many factors at play}. 
By dividing the ejections in different bins of BH mass we find that the larger is the BH mass, the lower is the redshift of its ejection. While black holes of $\rm {<}\,10^6 M_{\odot}$ are typical ejected at $z\,{\gtrsim}\,2.5$, the ones with $\rm {>}\,10^8 M_{\odot}$ were \textit{kicked out} at $\,0.1{<}\,z\,{<}\,0.6$. This mass segregation is just a growth matter. At very  high-$z$ we do not find massive BHs ($\rm {>}\,10^8 M_{\odot}$) being ejected, since they did not have enough time to accrete gas and grow.\\ %Since the BH can escape from both galaxy and halo after a large gravitational recoil, we have explored which of two processes dominate the BHs ejections.\\%For avoiding too many figures in the paper, we do not include the plot.\\ % {\bf Silvia: this is not shown anywhere, right? We should clarify. I would include it after the next paragraph, where we talk about reincorporations}\\

Finally, in Fig.\ref{fig:z_ejection} we also show the number density of reincorporation as compared to the one of ejections. The latter is larger than former by a factor of $2\,{-}\,3$ depending on the exact redshift. This is because gravitational recoil would expel more easily a BH from both galaxy and subhalo, rather than only from the galaxy. In other words, gravitational kicks are likely to overcome at once both the subhalo and galaxy escape velocities rather than only the galaxy one. This directly reduces the effective number of BHs that can be can be reincorporated  in the parent galaxy. For instance, regardless of redshift, we have found that galaxies with $\rm M_{stellar}\,{<}\,10^{9}  M_{\odot}$, $\rm 10^{9}\,{<}\,M_{stellar}\,{<}\,10^{10} M_{\odot}$, $\rm 10^{10}\,{<}\,M_{stellar}\,{<}\,10^{11} M_{\odot}$ have, respectively, $2.5$, $2$ and, $1.5$ times \textit{more} galaxy+subhalo ejections than only galaxy ejections. On the other hand, in more massive systems ($\rm M_{stellar}\,{>}\,10^{11}M_{\odot}$) the trend is inverted and the former is $2$ times less common than the latter. A similar behavior is seen with subhalo mass.\\

%This is because ejections from both halo and galaxy after a gravitational recoil are more common than just a galaxy ejection, reducing the effective number of BHs that can be reincorporated  in the parent galaxy. For instance, regardless of redshift, we have found that galaxies with $\rm M_{stellar}\,{<}\,10^{9}  M_{\odot}$, $\rm 10^{9}\,{<}\,M_{stellar}\,{<}\,10^{10} M_{\odot}$, $\rm 10^{10}\,{<}\,M_{stellar}\,{<}\,10^{11} M_{\odot}$ have, respectively, $2.5$, $2$ and, $1.5$ times \textit{more} galaxy+halo ejections than only galaxy ejections. On the other hand, in more massive systems ($\rm M_{stellar}\,{>}\,10^{11}M_{\odot}$) the trend is inverted and the former is $2$ times less common than the latter. A similar behavior is seen with halo mass.\\ %Interestingly, when the reincorporation number density is divided by BH masses we can see that at $z\,{<}\,2$ BHs of $\rm {<}\,10^6 M_{\odot}$ it displays similar values than ejection. %{\bf Silvia: you should explain this better. Some are orphans, right?}. 
%On contrary, at any redshift BHs with $\rm {>}\,10^6 M_{\odot}$ display larger number densities of ejections than reincorporations. 
%{\bf Silvia: based on the discussion, it seems that dividing into stellar mass bins rather thank BH mass bins could be an option. What do you think?}

%at $z\,{>}\,1.0$ wandering BHs of $\rm M_{BH}\,{<}\,10^5 M_{\odot}$ are reincorporated more frequently. At lower redshift, BHs with $\rm 10^5 \,{<}\, M_{BH}\,{<}\,10^8 M_{\odot}$ have similar reincorporation number density than the ones with $\rm {<}\,10^5 M_{\odot}$.\\

\begin{figure*}
	\centering
	\includegraphics[width=1.0\textwidth]{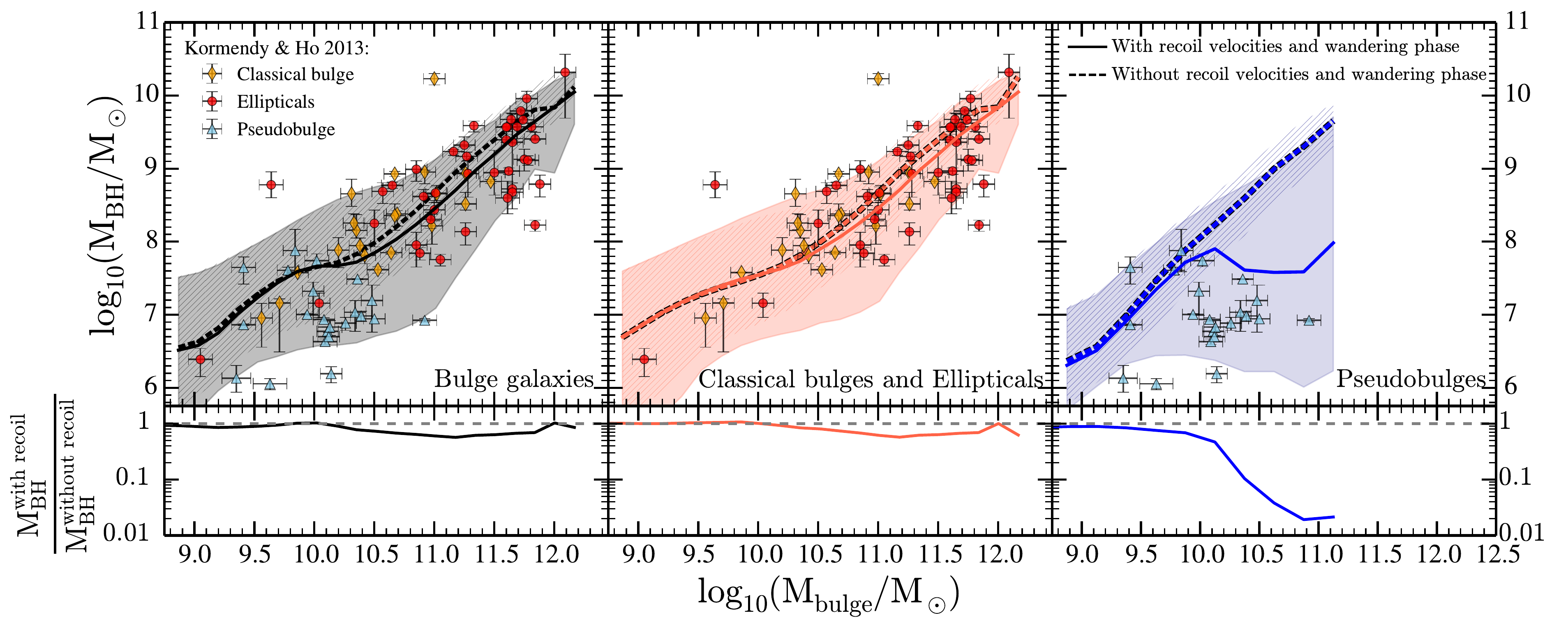}
    \caption{\textbf{Upper panels}: Black hole mass ($\rm M_{BH}$) - bulge mass ($\rm M_{Bulge}$) scaling relation in the local universe ($z\,{=}\,0$). Left panel represents the relation for the entire galaxy population, the central panel is for classical bulges and elliptical galaxies, and the right panel is for pseudobulges. Solid and dashed lines represent, respectively, the median of the relation when the SAM is run with and without including recoil velocities and wandering phase. Plain and dashed shaded area represent, respectively, the 2$\sigma$ of the relation with and without recoil velocities. Dots display the data from \protect \cite{Kormendy2013}. \textbf{Lower panels}: Ratio, at a fixed bulge mass, between the black hole mass in the model with recoil velocities and the black hole mass in the model without recoil velocities.}
\label{fig:scaling_reation_with_without_recoil}
\end{figure*}

\subsection{Effects on the scaling relations} \label{sec:Recoild_and_scaling_Relation}

As we have seen in the previous section, the population of nuclear black holes is affected by the physics of gravitational recoil included in the model. In this section we explore how this translates into predictions for the  bulge-black hole $z\,{=}\,0$ relation.\\

The left panel of Fig.\ref{fig:scaling_reation_with_without_recoil} shows the bulge-black hole scaling relation for \textit{all} galaxies when the SAM is run including or not both BH recoil velocities and wandering phase. Clearly, the scatter and the median relation are different in the two runs. The model displays a trend of hosting less massive BHs at fixed bulge mass in the run in which recoil velocities and wandering phase are taken into account. In the other two panels the population is divided by bulge morphological type: classical bulges and ellipticals (middle panel) and pseudobulges (right panel). While we still see some effects in the classical bulge population, the signatures of BH recoil are more pronounced for the pseudobulge population, where the median relation at $\rm M_{bulge}\,{\gtrsim}\,10^{10} M_{\odot}$ changes up to $\rm {\sim}\,2\,dex$ and the scatter increases up to $\rm 4 \, dex$.\\ 

These results clearly show that the inclusion of gravitational recoil physics is important in the predictions for the BH scaling relations, in particular in pseudobulge galaxies. Similar effects were found in the hydro-simulations of \cite{Blecha2011} and the analytical model of \cite{Gerosa&Sesana2015}. By running over 200 simulations of gaseous galaxy mergers with 60 different BH-BH merger configurations \cite{Blecha2011} found out that the inclusion of gravitational recoils leaves an imprint in the correlation between bulge velocity dispersion and the mass of the host BH. At fixed BH mass, the run with gravitational recoils displays large bulge velocity dispersion values. This deviation is caused by the ejection interrupting the BH growth while the galaxy continues its evolution. On the other hand, the work of \cite{Gerosa&Sesana2015} explored the repercussion of gravitational recoils in brightest cluster galaxies (BCGs). They pointed out that the ejections and subsequent replenishments of BHs after galaxy mergers increase the scatter of the BCGs black hole - bulge mass relation.\\%Thanks to the cosmological framework given by \texttt{Millennium} merger trees and the self-consistent BH spin evolution included in \LGalaxies in this work we have been able to extend the results of these works. %{\bf Silvia: I think here we should discuss a little more  the results  of Blecha and/or Gerosa, and mention how this work imroves on those}\\

\begin{figure}
	\centering
	\includegraphics[width=1.0\columnwidth]{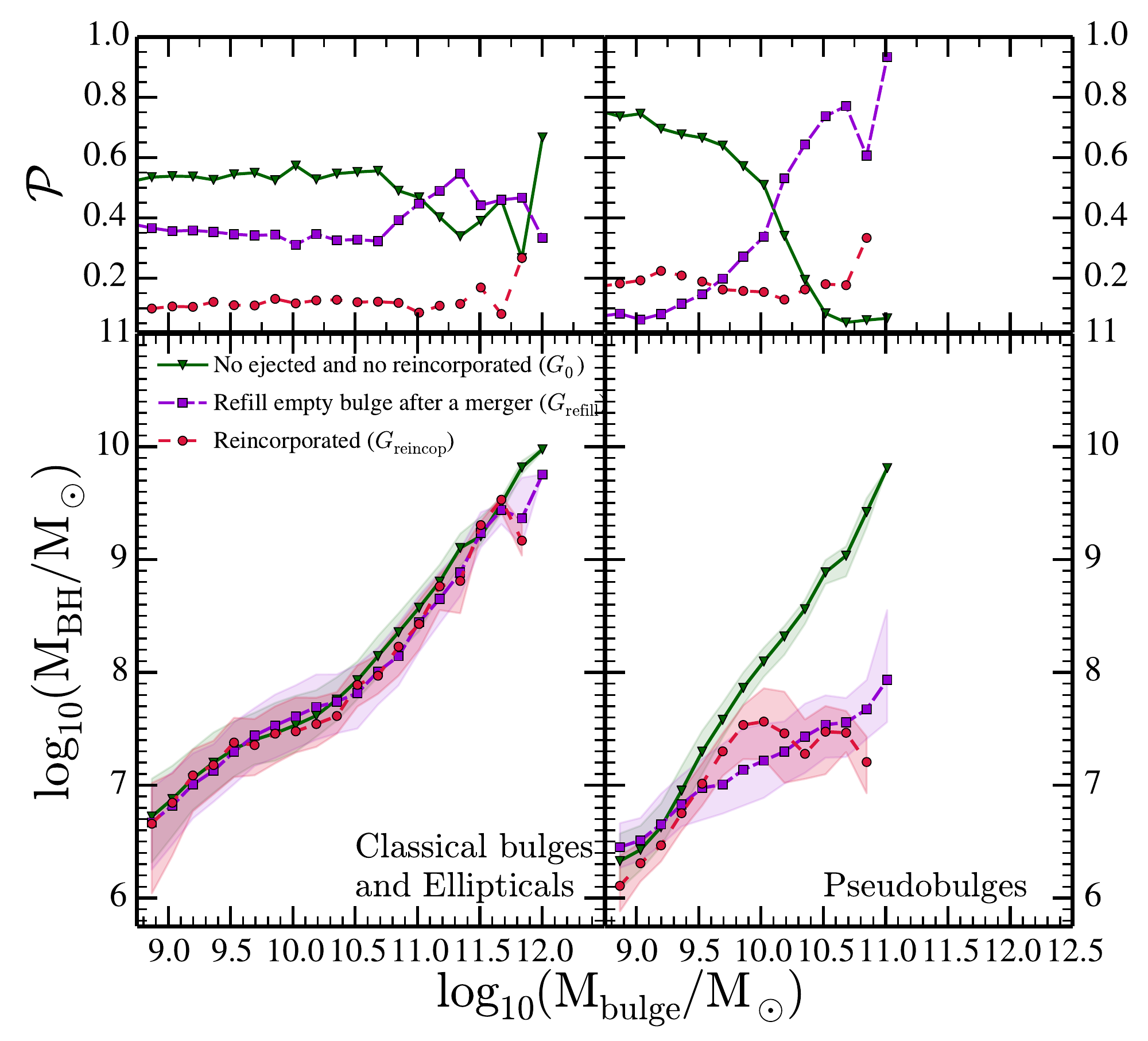}
    \caption{\textbf{Upper panels}: Probability, for a given bulge mass, of finding a nuclear BH which never underwent an ejection nor a wandering phase ($G_0$, green line), a nuclear BH which refilled an empty bulge after a merger (major, minor or \textit{smooth accretion}, $G_{\rm refill}$, violet line) and a BH which was reincorporated after a wandering phase ($G_{\rm reincop}$, red line)  \textbf{Lower panels}: Scaling relation for classical bulges plus ellipticals (left panel) and pseudobulges (right panel) when the nuclear BH inhabiting that bulge morphology is divided in the three previous types.}
\label{fig:understanding_scatter_relation}
\end{figure}

In Fig.\ref{fig:understanding_scatter_relation} we explore in more details the origin of the   deviations of the recoiled population with respect to the median relation. We again present   results for classical bulges (left panel) and pseudobulges separately (right panel). The median relation is divided into galaxies whose nuclear BHs never underwent an ejection nor a wandering phase ($G_0$), galaxies whose bulge did not host a central BHs but after a merger\footnote{major, minor or \textit{smooth accretion}.} it was refilled by one ($G_{\rm refill}$) and galaxies whose BHs were reincorporated after a wandering phase ($G_{\rm reincop}$). Regarding pseudobulges, galaxies belonging to $G_0$ follow a very similar relation as the model without recoil velocities (see Fig.\ref{fig:scaling_reation_with_without_recoil}). A similar trend is visible for the galaxies type $G_{\rm refill}$ at $\rm M_{bulge}\,{\lesssim}\,10^{10} M_{\odot}$. However, their trend breaks at larger bulge masses. Concerning galaxies of type $G_{\rm reincop}$, they show a much more flattened relation compared to the previous ones, regardless of bulge mass. This is because the merger (principally \textit{smooth accretions}) that filled the pseudobulge happened at relatively low $z$, typically $z\,{\lesssim}\,1.0$, with a galaxy hosting a BH that is $\rm {\sim}\,2 \, dex$ ($q\,{\sim}\,0.01$) lighter than the average BH that would inhabit the empty bulge. %In this way, the merger of an empty pseudobulge with a galaxy hosting a central BHs, causes, at fix bulge mass, the decrease of the relation around $\rm 2\,dex$.\\ 
As shown in the upper panel of Fig.\ref{fig:understanding_scatter_relation}, the relative contribution of $G_0$, $G_{\rm refill}$ and, $G_{\rm reincop}$ causes the change on the pseudobulge scaling relation with respect to the one without recoil velocities. Since galaxies $G_0$ are the dominant ones at $\rm M_{bulge}\,{<}\,10^{10} M_{\odot}$ we do not find any differences between the model with and without BH recoils. However, at $\rm M_{bulge}\,{>}\,10^{10.2} M_{\odot}$ the trend changes since galaxies type $G_{\rm reincop}$ take the main importance, causing the flattening in the relation, seen in Fig.\ref{fig:scaling_reation_with_without_recoil}.\\

Concerning classical bulges and ellipticals structures, all the three types $G_0$, $G_{\rm refill}$ and, $G_{\rm reincop}$ have similar bulge-BH mass relation than the relation without recoil (even though $G_{\rm reincop}$ is systematically below). This is caused by the fact that classical bulges and elliptical galaxies do not spend a lot of time without a BH: the reincorporation time of the BH is relatively low (${\lesssim}\,0.01 \, \rm Gy$) and since their formation is linked with the hierarchical growth of structures, they experience frequent mergers (with $q\,{\gtrsim}\,0.2-0.3$) in short time-scales. As it is shown in the upper panel of Fig.\ref{fig:understanding_scatter_relation}, galaxies $G_0$ dominate at any bulge masses. However, at $\rm M_{bulge}\,{>}\,10^{10.5} M_{\odot}$ the probability $\mathcal{P}$ of finding a galaxy $G_{\rm refill}$ increases up to $\mathcal{P}\,{\sim}\,0.4$. This causes the small change seen in the relation with and without recoil velocities in Fig.\ref{fig:scaling_reation_with_without_recoil}. We have also explored the $G_0$, $G_{\rm refill}$ and, $G_{\rm reincop}$ population at $z\,{=}\,2$ for both classical bulges (and ellipticals) and pseudobulges. In this case, at any mass and bulge type the $G_0$ population dominates. For instance, at $\rm M_{bulge}\,{\sim}\,10^{10.5}\, M_{\odot}$ $G_0$ counts for more than the $65\%$ and $80\%$ of the population of classical bulges and pseudobulges, respectively. This causes the median value in the runs with and without recoil velocities to be very similar, even though the population $G_{\rm refill}$ and $G_{\rm reincop}$ helps in increasing the scatter. \\

%We have checked that the BH merger ratio in classical bulges and ellipticals at $z\,{\lesssim}\,1$ is $q\,{\sim}\,0.2-0.3$, causing an smaller effect in the scaling relation than in the one for pseudobulges. Notice that the galaxies with a reincorporated BHs have very marginal effects in the classical and elliptical bulge population. Therefore, in both classical and pseudobulges the change and increase of the scatter in the scaling relation is caused by galaxies with BHs which refilled an empty bulge after a merger (major, minor or \textit{smooth accretion}).\\
 
In a test run, we have also checked how the bulge-black hole scaling relation behaves when in the model we do not assume any correlation between BH accretion and bulge assembly (see Eq.\eqref{eq:npa}). For that we have run \LGalaxies with the \textit{chaotic} scenario of \cite{King2005} which assumes an isotropic distribution for the angular momentum of the gas clouds\footnote{Following \cite{Sesana2014}, the chaotic scenario presented in \cite{King2005} is obtained fixing $F\,{=}\,1/2$ in Eq.\eqref{eq:npa}.}. We did not find significant differences, as the strength of the recoil velocities depends primarily on the capability of the gas surrounding the BBH in aligning the two BH spins at the moment of the merger, rather than on the spin value.  BH-BH coalescences at $z\,{\lesssim}\,1.5$ are mostly happening in gas poor environments, reducing the number of BH mergers with a spin alignment, independently on the spin model used.\\
%For doing the discussion short, we only discuss the main results. We have seen that the scaling relation in the chaotic scenario displays less spread than in the model used in this work. However, when the population is divided by bulge morphology, the pseudobulge scaling relation predicted by the chaotic model displays a very similar shape than the one presented in Fig.\ref{fig:scaling_reation_with_without_recoil}. This is caused by the combination of the quiet merger history that characterize the pseudobulge population and the low $z$ of the last interaction ($z\,{\sim}\,0.5$) which causes that the BH-BH coalescence happen in a gas poor environment, increasing the modulus of the recoil velocity.\\}

\begin{figure}
	\centering
	\includegraphics[width=1.0\columnwidth]{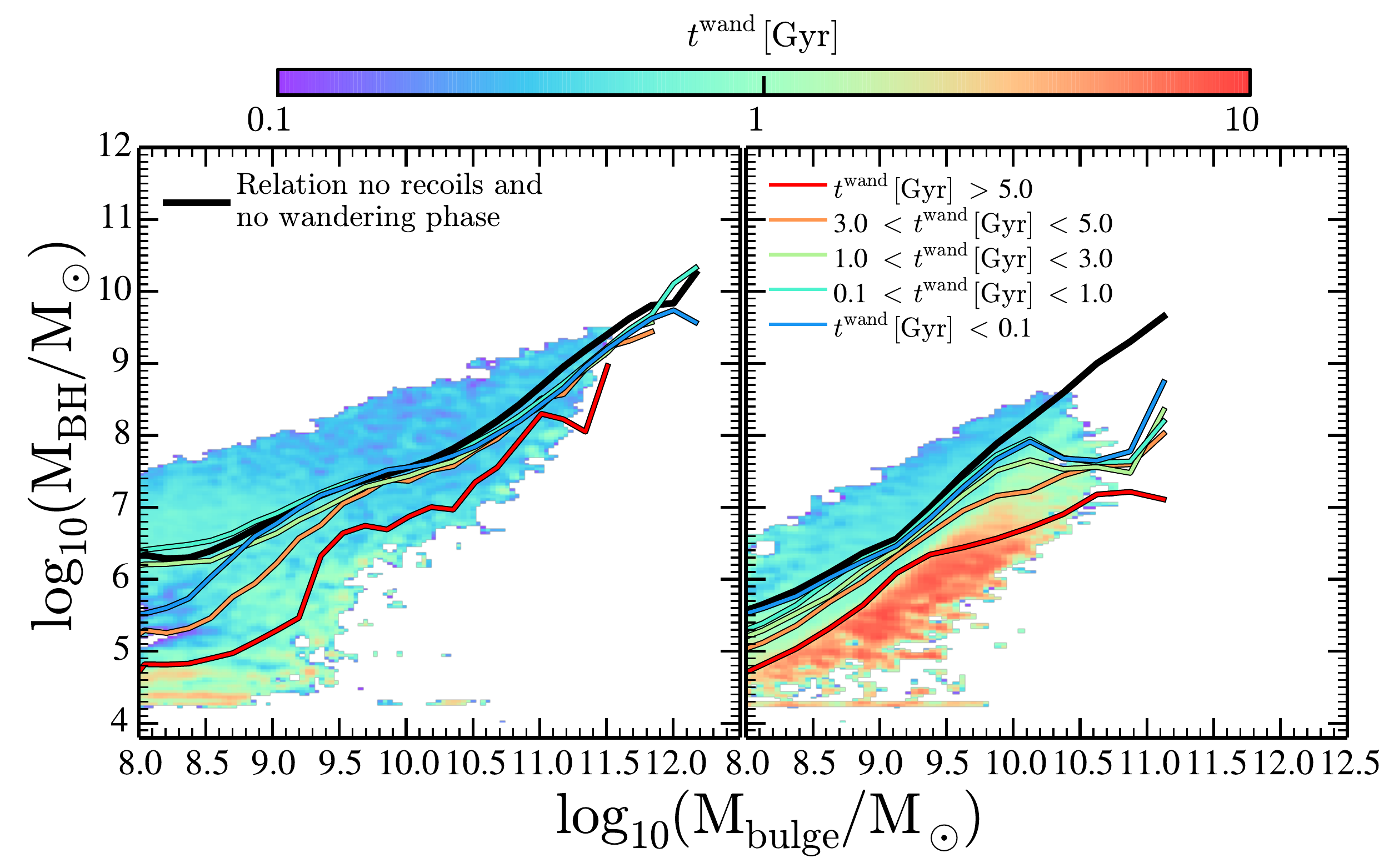}
    \caption{Bulge - black hole mass relation for classical bulges plus ellipticals (left panel) and pseudobulges (right panel) when \textit{only} nuclear black holes which underwent a wandering phase are selected. The color map indicates the median value of the time spent in the wandering phase ($t^{\rm wand}$) per bin of bulge and black hole mass. In each panel, the solid black line represents the scaling relation when no recoil velocities and wandering phase are included, whereas colored lines represent the median relation for nuclear BHs in a given bin of wandering phase, as labeled in figure.}
\label{fig:wandering_phase_effect_relation}
\end{figure}

Finally, even though we have showed that reincorporated BHs have small impact in the bulge - black hole mass relation, in Fig.\ref{fig:wandering_phase_effect_relation} we have explored the effect of the BH reincorporation after a wandering phase in the scaling relations. In the figure it is shown the scaling relation for only classical bulges (including ellipticals) and pseudobulges whose central black hole was reincorporated after undergoing a wandering phase. Independently of the bulge morphology, there is a segregation by time in the wandering phase ($\rm \mathit{t}^{wand}$): BHs which underwent longer wandering phase are systematically below the scaling relation, as growth in the model is only allowed when BHs are centrals (see Section~\ref{sec:wandering_BHs}). This trend is more clear in the pseudobulge population. 

\section{Summary and conclusions}\label{sec:conclusion}

In this paper we have studied the mass assembly, spin evolution and environment  of both nuclear and wandering supermassive black holes. To do that, we have updated with new physical prescriptions  the \cite{IzquierdoVillalba2019} version of \LGalaxies semi-analytical model (SAM).  The new prescriptions have been applied to the merger trees of the \texttt{Millennium} simulation, although galaxies have been initialized using the outputs of the higher-resolution simulation  \texttt{MillenniumII}.\\ %and \texttt{Millennium II} merger trees.\\%, allowing us to trace the most luminous and massive black holes and to improve the resolution limitations of the \texttt{Millennium}.\\ 

%and spin evolution of nuclear black holes and the properties that characterize the wandering black hole population

%Taking advantage of the SAM we have included different physical recipes to track  the black hole growth and spin evolution

%In this paper we have presented a model of black hole growth and spin evolution embedded inside of the last version of \LGalaxies semi-analytical model \citep[SAM,][]{Henriques2015}, in the variant presented in \cite{IzquierdoVillalba2019}. The model, applied on top of the \texttt{Millennium} and \texttt{Millennium II} merger trees, we have served us to explore the .\\

%The model is applied on top of the \texttt{Millennium} and \texttt{Millennium II} merger trees. While the former is ideal to trace the most luminous and massive black holes, the latter allows us to improve the resolution limitations of the \texttt{Millennium}.\\ 

%\davcoment{Add the fact that we use MS and we use the properties of the MSII to initialize the MS halos!}

The starting point of our BH model is the gas accretion from a \textit{hot} and \textit{cold} phase. While the former is linked with the consumption of part of the hot gas atmosphere which surrounds the galaxy, the latter is triggered by cold gas accretion right after a galaxy disk instability (DI) and/or a galaxy merger. %%Following \cite{Marulli2008} and \cite{Bonoli2009}, we have coupled the BH accretion rate during the cold phase with a quasar-light curve divided in a first stage of rapid growth truncated at Eddington limit, followed by a quiescent regime of slow accretion rates which imitates a self-regulated growth phase. 
During both phases of gas accretion, we track the evolution of BH spin ($a$)   using the approach presented in \cite{Dotti2013} and \cite{Sesana2014}. We linked the number of prograde transient accretion disk (BH spin-up) with the degree of coherent motion in the bulge. In particular, we have assumed that DIs, which lead to the bar/pseudobulge formation in our model, are the processes that increase the coherence of the bulge kinematics. On the other hand, mergers, which lead to the assembly of classical bulges and ellipticals, bring  disorder to the bulge dynamics. These assumptions result in a correlation between the predicted BH spin values and the host galaxy morphology.  Regardless of redshift and BH or bulge mass, galaxies with a pseudobulge structure host nuclear BHs with larger spin values than both classical and elliptical galaxies. In particular, at $\rm M_{BH}\,{>}\,10^6 M_{\odot}$ pseudobulges, classical bulges and elliptical galaxies host respectively BHs with a typical spin of $a\,{\sim}\,0.9$, $0.7$ and $0.4$. At lower BH masses, as a consequence the implemented model, all the three bulge morphology host maximum spinning BHs. %%Besides, given the complex evolution of bulge galaxies, we find the general trend of BH spinning down towards low-$z$ caused by the decrease of DI number density and the increase of mergers \citep[see][]{IzquierdoVillalba2019}. Indeed, at $z\,{\sim}\,0$ the population of massive BHs is dominated by galaxies hosting classical bulge structure.\\
The model also follows the formation of binary black hole (BBH) systems by assuming that the lifetime of a BBH (from the formation to the coalescence) inversely correlates with the BH merger ratio and the galaxy gas fraction at the moment of the galaxy encounter. This simple assumption will be further improved in future works. After coalescence, the remnant has a spin value fully determined by the \cite{BarausseANDRezzolla2009} analytic expression.\\ 
%{\bf Silvia: I would include here the rest of the model, discussing first recoil (given that we just talked about BBH merger, and then orphan BHs. We can then summarize all the predictions in few bullet points. What do you think? }%%However, since in the model we do not track the orientation of the spin but only the modulus, we have to assumed that in gas rich BH mergers (\textit{wet} mergers), the dynamical friction between BHs and gas finally aligns (with a random offset of $0\degree-10\degree$, see \citealt{Dotti2010}) both BHs. On contrary, in poor gas BH mergers (\textit{dry} mergers) both BHs do not reach an alignment and the orientations are random \citep{Barausse2012}. 
%Even though the mass increase by BH mergers is secondary compared with the growth by gas accretion, the model points out that BH-BH coalescence play an important role in shaping the spin of supermassive black holes hosted in elliptical galaxies. In particular, caused by gas-poor BH mergers, BHs of $\rm M_{BH}\,{>}\,10^8 M_{\odot}$ hosted in elliptical galaxies undergo a spin up ($a\,{\gtrsim}\,0.4$), in agreement with \cite{Fanidakis2011}. According to our model, this spin up takes place at $z\,{\lesssim}\,1$.\\

Due to conservation of linear momentum, in the instant of the BH-BH coalescence the propagation of gravitational waves
imparts a recoil velocity to the remnant black hole. %Using the \cite{Lousto2012} equations, the model computes the magnitude of the recoil velocity. %by taking into account both the mass and spin of the two interacting BHs. 
If the modulus of the recoil velocity, computed according to \cite{Lousto2012} equations, is larger than the escape velocity of the host galaxy, the model assumes that the black hole is kicked from its host and incorporated in the DM subhalo as a \textit{wandering black hole}. Here we tag these BHs as \textit{ejected} wBH. Since it is not clear which is the amount of mass in accretion disk that can be retained or accreated by a BH after the recoil, we have neglected the fact that the ejected BH can retain an accretion disk. The model also tracks   the formation of another type of wandering black holes, \textit{orphan} wBH, which are originated after the complete disruption of their host galaxy via tidal forces.
Independently of their origin, wBH orbits are tracked by using numerical integration, taking into account the host subhalo and galaxy properties for the computation of  gravitational acceleration and dynamical friction. In particular, the integration of the wBH orbit stops when i) the black hole is re-incorporated in the galaxy, i.e., when the BH passes through the galaxy center with a velocity smaller than its escape velocity, ii) the recoil velocity of ejected BHs is larger than the subhalo escape velocity or iii) the black hole position exceed $3$ times the hosts subhalo virial radius and it is still moving away from the galaxy.\\

Turning to the model predictions, we find a good consistency with the observed local black hole mass function (BHMF), spin values, BH-bulge mass relation and quasar (bolometric, soft and hard X-rays) luminosity functions. The model predicts a stalling in the BHMF at $z\,{\sim}\,1$, where no significant change up to $z\,{=}\,0$ is shown. This indicates that most of the local BHs were already assembled by $z\,{\sim}\,1$. By dividing the BHs between inactive and active we have found  that the former population increases form $z\,{\sim}\,3$ to $z\,{\sim}\,0$, with the most massive BHs becoming non-active earlier. %While at $z\,{\sim}\,3$ inactive BHs have typical $\rm M_{BH} \, {>}\,10^8\, M_{\odot}$, at $z\,{\sim}\,0$, they are characterized by $\rm M_{BH}\,{>}\,10^6\, M_{\odot}$. 
The accretion geometry of the active population displays also a redshift evolution. Regardless of  BH mass, $z\,{\sim}\,3$ active BHs are characterized by a thin disk geometry fuelled by Eddigton-limited accretion flows. ADAF, instead, becomes the main accretion geometry at $z\,{\sim}\,0$  for BHs with $\rm M_{BH}\,{>}\,10^6 M_{\odot}$, powering BHs during the quiescent phase of cold and hot gas accretion. On top of this, the luminosity functions of the model indicate that accreting BHs are hosted in different bulge morphologies at different cosmological times. While at $z\,{>}\,2$ the high end of the luminosity function is  dominated by BHs accreting in pseudobulge structures, at lower redshifts classical bulges and elliptical galaxies are the main structures hosting the most powerful active BHs. Since in the SAM the pseudobulge formation is detached form the merger framework \citep[see][]{IzquierdoVillalba2019} our model points out that $z\,{\gtrsim}\,2$ AGNs %and intermediate luminous AGNs ($\rm L_{bol}\,{\sim}\,10^{44-45}\, erg/s$) at $z\,{\lesssim}\,1$ 
are mainly triggered by secular processes instead of galaxy encounters, in agreement with recent observational and theoretical results.\\

%,i.e the population of BH outside of galaxies in bound orbits within the dark matter halos. 
%We have also explored the model predictions about the wandering black hole (wBH) population. %For that, we have taken into account both, disruption of satellite galaxies (\textit{orphan} wBH) and BH ejections via gravitational recoils (\textit{ejected} wBH) as the two mechanisms leading the formation of this type of population. By numerical integration, we have been able to track the orbits of the wBHs, including both BH reincorporation and the complete runaway of the BHs from both galaxy and halo system. 
%Since in this work the BH seeding procedure is still rough, we have only focused on wBHs whose mass is larger than $\rm 10^{6}\,M_{\odot}$. {\bf Silvia: I would cut this previous sentence}.
One of the main novelties of the present study is a thorough exploration of the population of wandering black holes (wBHs), either {\it orphan} or {\it ejected}. 
When looking at their mass density, we have found an increasing trend towards low $z$, reaching a maximum of $\rm 10^{-4}\,Mpc^{-3}$ at $z\,{\sim}\,0$, which is considerable but still $\rm 1\,{-}\,3\, dex$ smaller (depending the BH mass) than that of nuclear black holes. Regarding the spatial distribution, \textit{ejected} and \textit{orphan} wBHs occupy different regions inside the subhalo. While the former reside at $\rm {\lesssim}\,0.1R_{200}$, with a decreasing trend towards low $z$, the latter inhabit the $\rm {\gtrsim}\,0.2-0.4\,R_{200}$ regions. Concerning the environments of wBHs in the local universe, the model predicts that subhalos of $\rm M_{halo}\,{<}10^{13} \,M_{\odot}$ rarely host a wBH with $\rm {>}\,10^6 \, M_{\odot}$. But if they do, the wBH is typically formed after a gravitational recoil, i.e. of the {\it ejected} type. The picture is inverted at $\rm M_{halo}\,{>}10^{13} \,M_{\odot}$, where it is relatively more common to find a wBH of $\rm M_{BH}\,{>}\,10^6 \, M_{\odot}$ (typical ${\sim}\,5$) principally formed after the disruption of satellite galaxies, i.e. of the {\it orphan} type. We find the same  tendency when looking at wBH statistics against the host stellar mass. \\

Besides being an important channel of wandering black hole formation, we have found that gravitational recoils also affect the co-evolution between the black hole and the host galaxy. In particular, they cause a systematic depletion of nuclear BHs towards low redshift and stellar mass. While the former dependence is because low-$z$ BH-BH coalescences are characterized by poorer gas environments than the ones at higher-$z$, condition which increase the modulus of the kick velocity, %{\bf[AS: This is not correct. Kicks are the largest for $q=1$, so low $q$ mergers should decrease the probability of the remnant being ejected. It is true, however, that gas poor mergers are expected to host BBHs with random spin orientations, which tend to increase kicks. So I would say there is a tradeoff between the two effects. at low $z$, $q$ is smaller, but the fact that most mergers are dry, still causes the typical $v_{kick}$ to be larger, thus causing more deplation.]} 
the latter is simply caused by the fact that low-mass galaxies have smaller potential wells than massive ones, making it more difficult for them to retain the BH after a kick. By dividing the galaxy population according to bulge type, we have found that at $\rm M_{stellar}\,{\gtrsim}\,10^{10} \, M_{\odot}$ both classical bulge and elliptical structures tend to display larger occupancy fractions than pseudobulges and bulgeless galaxies. This behavior is due to the fact that  pseudobulges are detached form the merger framework, making more difficult for them to replenish their empty bulge after an ejection. On the other hand, at $\rm M_{stellar}\,{\lesssim}\,10^{10} \, M_{\odot}$ classical bulges are more affected by the gravitational recoils, having more than the 40\% of empty bulges. By running the SAM with and without gravitational recoils and wandering phase we have also explored the imprints of gravitational recoils in the $z\,{=}\,0$ bulge-BH scaling relation. We have found that both the median of the relation and its scatter are different in the two runs. 
There is a trend of hosting less massive BHs at fixed bulge mass in the run in which recoil velocities and wandering phase are taken into account. A similar tendency is found when the population is divided into classical bulges, elliptical galaxies and pseudobulges. In particular, the latter structures are the ones which suffer a more pronounced effect, where the median relation at $\rm M_{bulge}\,{\gtrsim}\,10^{10} M_{\odot}$ changes $\rm {\sim}\,2\,dex$ and the scatter increases. % up to $\rm 4 \, dex$ %{\bf[AS: This is not completely fair though. Usually for 'scatter' one refers to the 1$\sigma$ uncertanty (or 68\% confidence interval, whereas it is my understanding that the 4dex shown in figure 11 are 2$\sigma$ regions.]}. 
Finally, we have also explored the effect of  BH reincorporation after a wandering phase in the BH-bulge relation. Regardless of bulge morphology, the larger is the time of the BH in the wandering phase before its reincorporation, the larger is its offset form the median BH-bulge relation.\\

Despite the model presented here represents a considerable step forward in modeling the evolution of the population of black holes across cosmic time, still more effort is needed to construct a full coherent framework that can produce new quantitative predictions that can be tested against current and future observations. For example, including a physical model for the expected activity of wBH will lead to useful predictions about the number of active wBHs that may be observed in the local Universe and their more likely location with respect to their host galaxy/group/cluster. In the near future, new instrumentation will make it possible to constrain several aspects of the cosmic BH formation and evolution. The \textit{Laser Interferometer Space Antenna} \citep[LISA,][]{eLISAConsortium2013} will constrain the nuclear BH merger rate and their spin properties, thus also providing precious information for the theoretical modeling of the {\it kicked} wBH population; \textit{James Webb Space Telescope} \citep[JWT,][]{Gardner2006} and \textit{Advanced Telescope for High-ENergy Astrophysics} \citep[ATHENA,][]{Nandra2013} will allow to pierce deeper in the cosmos, probing progressively further and lower mass accreting BHs. Only a concerted interplay of theory and observations will unfold the full history of the cosmic, nuclear and wandering, supermassive black hole population.

\section*{Acknowledgements}
We thank the anonymous referee for the helpful comments which improved the quality of the manuscript. The authors thank Lucio Mayer for useful discussions and comments. D.I.V. and D.S. acknowledge the support from project \textit{AYA2015-66211-C2-2 MINECO/FEDER, UE} of the Spanish Ministerio de Economia, Industria y Competitividad and the Aragón Government through
the Reseach Groups E16\_17R. D.I.V. particularly thanks the grant \textit{Programa Operativo Fondo Social Europeo de Arag\'{o}n 2014-2020 (Construyendo Europa desde Arag\'{o}n}). A.S. is supported by the ERC through the CoG grant 'B Massive', grant number 818691. Y.R.G acknowledges the support of the European Research Council through grant number ERCStG/716151. S.B. and Y.R.G. acknowledge PGC2018-097585-B-C22, MINECO/FEDER, UE of the Spanish Ministerio de Economia, Industria y Competitividad. This work used the 2015 public version of the Munich model of galaxy formation and evolution: \LGalaxies. The source code and a full description of the model are available at http://galformod.mpa-garching.mpg.de/public/LGalaxies/. This work has made used of CEFCA's Scientific High Performance Computing system which has been funded by the Governments of Spain and Aragón through the Fondo de Inversiones de Teruel, and the Spanish Ministry of Economy and Competitivenes (MINECO-FEDER, grant AYA2012-30789).
%\bibliographystyle{mnras}
%\bibliography{references}

\bibliography{references}
\bibliographystyle{mnras} 

\appendix

\section{Ejections via Gravitational recoil and three body scattering}\label{Appendix:Ejection_Recoil_Three_Body}

\begin{figure}
	\centering
	\includegraphics[width=1.0\columnwidth]{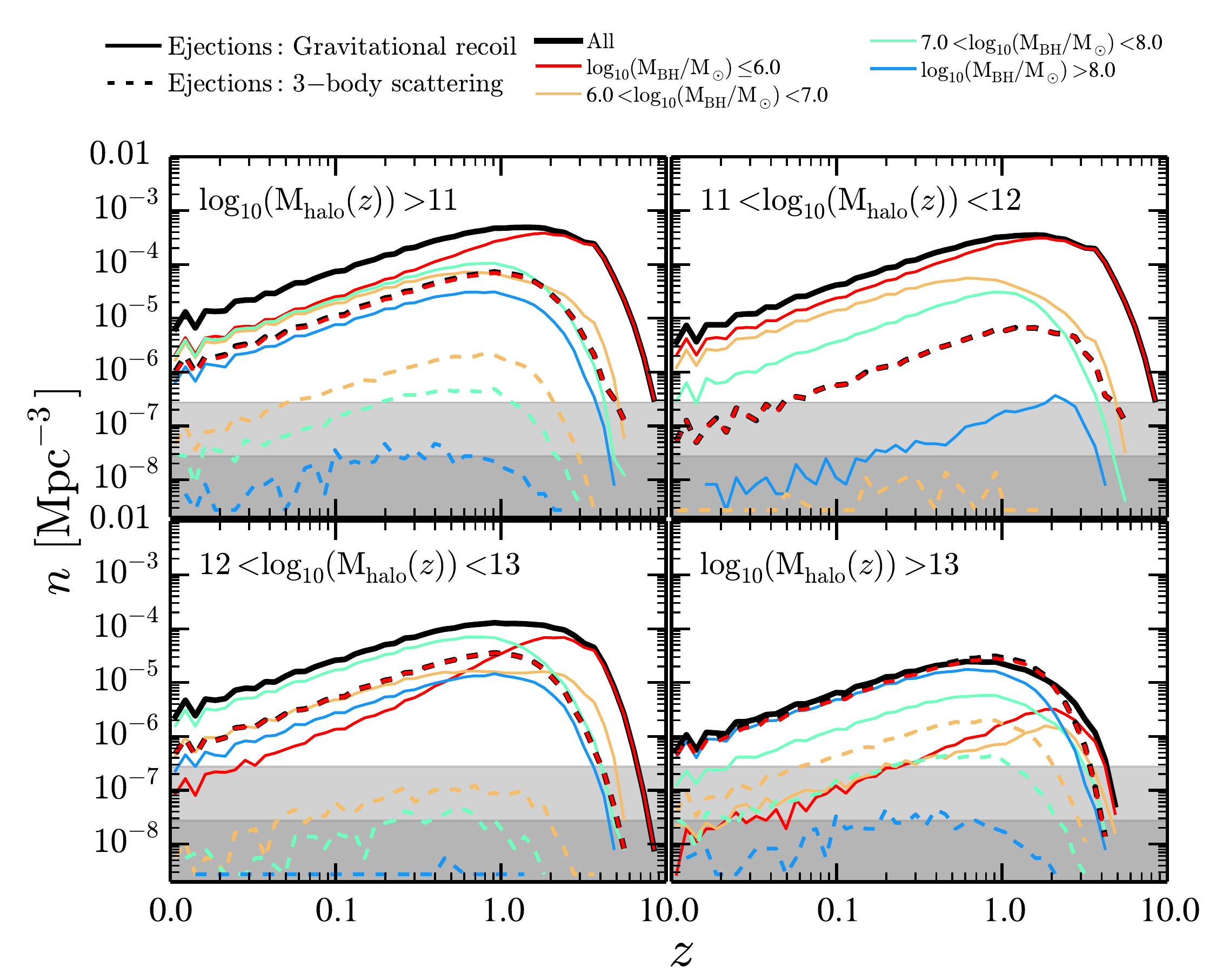}
    \caption{Number density ($n$) of nuclear black holes that are ejected from the nuclear part of the galaxy. Each panel represents a bin of subhalo mass ($\rm M_{halo}$). While solid lines display the ejections caused by gravitational recoils, dashed ones display the same but for BHs that were ejected via 3-body scattering. Colors represents different bins of BH mass. With shaded areas it is highlighted the value of $n$ in which we have less than 100 (faint grey) and 10 (dark grey) events.} 
\label{fig:n_density_ejection_scatteringWBHs}
\end{figure}

As we have discussed in Section~\ref{section:BH_delay_merger} nuclear BHs can be expelled from the galaxy nuclear region trough both gravitational recoils and recoils after 3-body scattering. While the former is led by the final BH-BH coalescence, the latter is the consequence of a complex interaction between a binary BH system and an intruder BH which causes the ejection of the less massive BH. In this appendix, we show that the ejections after a gravitational recoil are more common that the ejections via 3-body scattering. Thus, the population of \textit{ejected} wBHs is fully dominated by gravitational recoiled BHs. To prove so, in Figure~\ref{fig:n_density_ejection_scatteringWBHs} we present the evolution of the number density ($n$) of ejections trough gravitational ejection and 3-body scattering. As shown, at $\rm M_{halo}\,{>}\,10^{11}\,M_{\odot}$ ejections after gravitational recoils dominate the ejected population, being $1\,{-}\,2 \rm \, dex$ more frequent at any redshift. As we can see, the ejection after 3-body scattering is dominated by low mass BHs ($\rm M_{BH}\,{<}\,10^6 \, M_{\odot}$) at any redshift and subhalo mass. On the other hand, the ejection of BHs of $\rm M_{BH}\,{>}\,10^8\, M_{\odot}$ after a 3-body scattering is relatively rare, with less than 300 events at $z\,{<}\,2$. These events happens mainly in $\rm M_{halo}\,{>}\,10^{13} \, M_{\odot}$, which are the ones hosting the galaxies with the most massive BHs in the SAM. 

%\section{Distribution of BH-BH merger delay}\label{Appendix:BHBHDelay}

\section{X-ray luminosity functions}\label{Appendix:XRaysLFs}
\begin{figure}
	\centering
	\includegraphics[width=1.0\columnwidth]{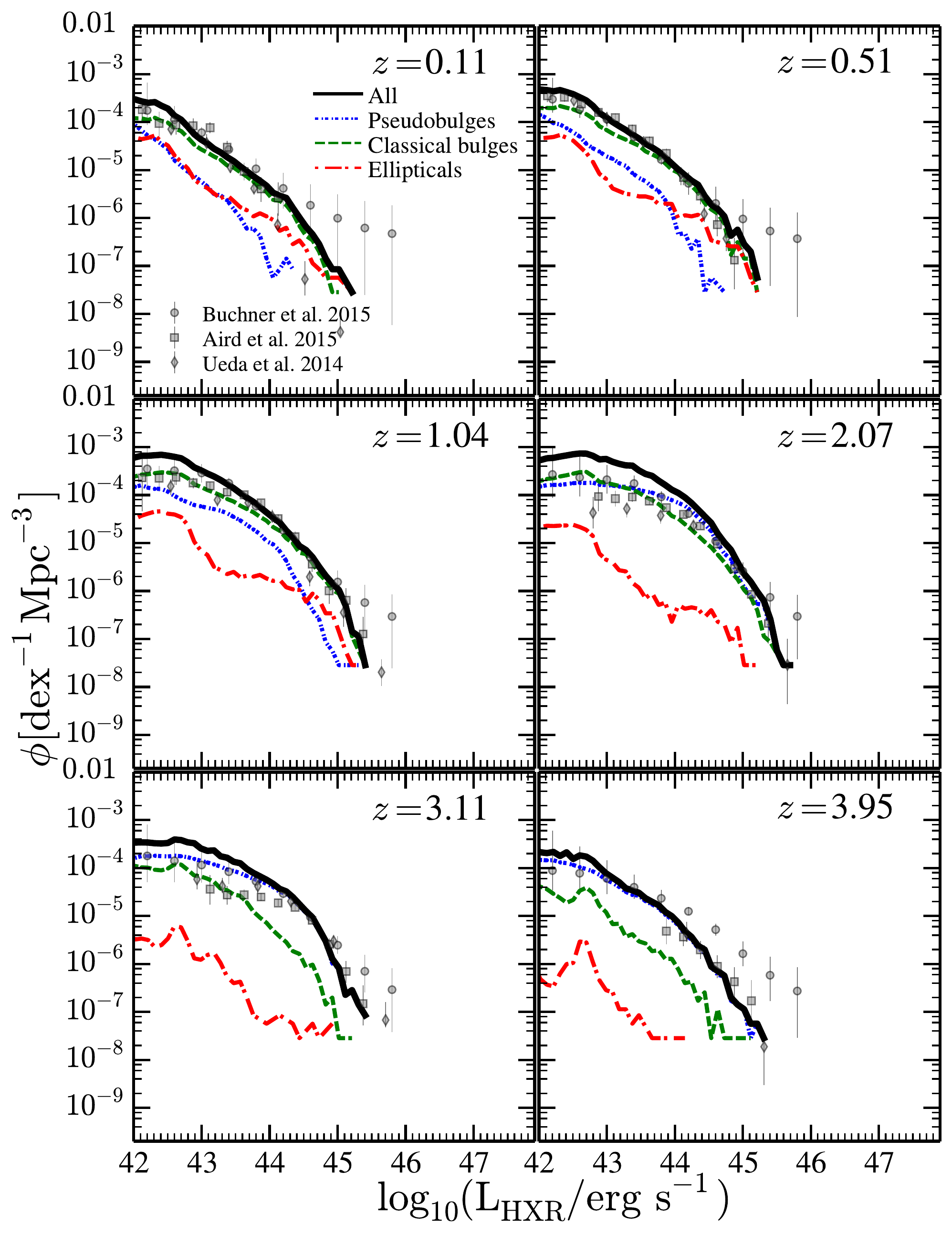}
    \caption{X-ray luminosity functions at $z\,{\approx}\,0.1,0.5,1.0,2.0,3.0,4.0$. In solid black line we present the predicted \textit{hard} X-ray (2-10 keV) luminosity functions. Red dashed-dotted line, green dashed line, and blue dotted line represents the luminosity functions for galaxies hosting, respectively, elliptical, classical bulge,and pseudobulge bulge structure. Each luminosity functions is compared with the compilation of \protect\cite{Buchner2015,Aird2015,Ueda2014} (circle, square and, diamond points, respectively).}
\label{fig:LF_Xrays_HARD}
\end{figure}

\begin{figure}
	\centering
	\includegraphics[width=1.0\columnwidth]{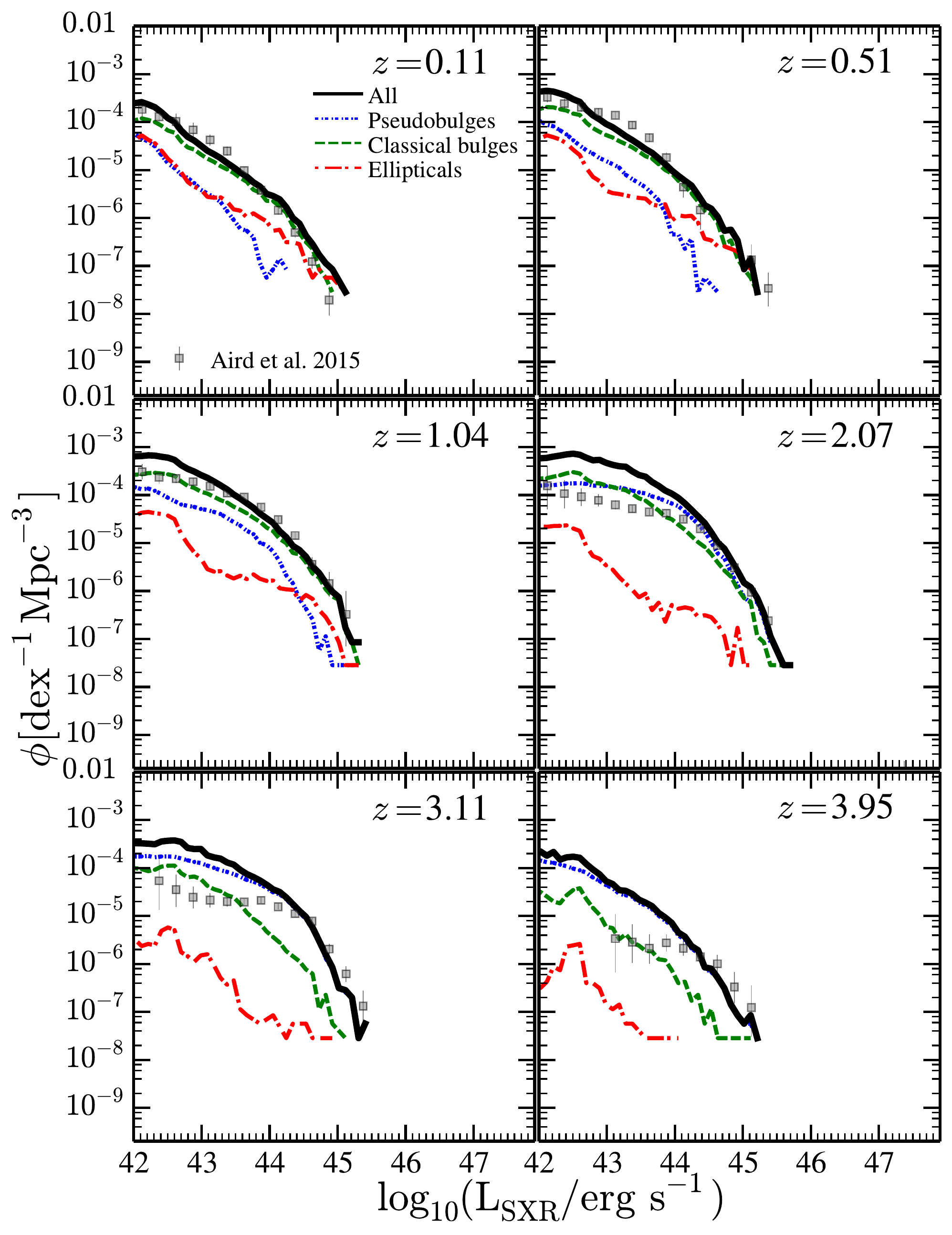}
    \caption{X-ray luminosity functions at $z\,{\approx}\,0.1,0.5,1.0,2.0,3.0,4.0$. In solid black line we present the predicted \textit{soft} X-ray (0.5-2 keV) luminosity functions. Each luminosity functions is compared with the compilation of \protect\cite{Buchner2015,Aird2015,Ueda2014} (circle, square and, diamond points, respectively).}
\label{fig:LF_Xrays_SOFT}
\end{figure}

The redshift evolution of X-ray luminosity functions is presented in Fig.\ref{fig:LF_Xrays_HARD} and Fig.\ref{fig:LF_Xrays_SOFT}. Soft ($\rm 0.5\,{-}\,2 \,keV$) and hard ($\rm 2\,{-}\,10 \,keV$) X-ray luminosity have been computed by using the bolometric corrections derived in \cite{Marconi2004}:
\begin{eqnarray}
\rm log_{10}\left(L_{Hx}/L_{bol}\right) \,{=}\, -1.54 - 0.24\mathcal{L} - 0.012\mathcal{L}^2 + 0.0015\mathcal{L}^3\\
\rm log_{10}\left(L_{Sx}/L_{bol}\right) \,{=}\, -1.64 - 0.22\mathcal{L} - 0.012\mathcal{L}^2 + 0.0015\mathcal{L}^3
\end{eqnarray}
where $\rm \mathcal{L}\,{=}\,log_{10}(L_{bol}/L_{\sun}) -12$, $\rm L_{Hx}$ is the hard X-ray luminosity and $\rm L_{Sx}$ soft X-ray luminosity. Predictions are compared to the observational works of \cite{Ueda2014} \cite{Aird2015} and \cite{Buchner2015}. Given the uncertainty in modelling the fraction of obscured AGNs, we prefer to compare the simulations to \cite{Aird2015} observed soft X-ray luminosity functions for which the obscured fraction has already been taken into account. The fraction of obscured objects in the hard X-ray band is thought to be relatively small, so for this work we consider that there is no obscuration at hard X-ray wavelengths. As we can see, good agreement between observations and model is reached. In the same figures we have divided the population between elliptical, classical bulges and pseudobulges. As we found with the bolometric luminosity, at $z\,{\gtrsim}\,2$ the luminosity functions are dominated by BHs accreting in pseudobulges structures. On contrary, at $z\,{\lesssim}\,1$  classical bulges and elliptical galaxies are the structures that preferentially host AGNs and quasars.\\

%\bibliography{paper}
%\bibliographystyle{apalike}
% Don't change these lines
\bsp	% typesetting comment
\label{lastpage}
\end{document}